\documentclass[11pt]{article}
\usepackage[utf8]{inputenc}
\usepackage{caption}
\usepackage{braket,slashed,bm}
\usepackage{simplewick}
\usepackage{jheppub}
\usepackage{array,multirow}
\usepackage{amsmath} \DeclareMathAlphabet{\mathpzc}{OT1}{pzc}{m}{it}
\usepackage[normalem]{ulem}
\usepackage{calligra}
\usepackage{floatrow}
\DeclareMathAlphabet{\mathpzc}{OT1}{pzc}{m}{it}
\usepackage{mathrsfs}
\usepackage{tikz}

\usepackage{subcaption}
\usepackage{lscape}
\usepackage{graphicx}
\usepackage{booktabs}
\newfloatcommand{capbtabbox}{table}[][\FBwidth]

\graphicspath{{./Figures/}}

\def\beq{\begin{equation}}
\def\eeq{\end{equation}}
\def\bea{\begin{eqnarray}}
\def\eea{\end{eqnarray}}
\def\nn{\nonumber \\}
\def\hyp{\mathsf{y}}

\renewcommand{\d}{\delta}

\def\gcb{{\overline g_{1}}}
\def\gcw{{\overline g_{2}}}

\def\ckin{C_{H,\text{kin}}}
\newcommand{\Lagr}{\mathcal{L}}
\renewcommand{\to}{\rightarrow}

\DeclareMathOperator{\re}{Re}

\begin{document}
\title{More accurate $\sigma(\mathcal{G} \,\mathcal{G}\rightarrow h)$,
$\Gamma(h \rightarrow \mathcal{G} \,\mathcal{G}, \mathcal{A} \mathcal{A}, \bar{\Psi} {\Psi})$
and Higgs width results via the geoSMEFT}

\author[a]{Adam Martin,}

\author[b]{and Michael Trott}

\affiliation[a]{Department of Physics, University of Notre Dame, Notre Dame, IN, 46556, USA}

\affiliation[b]{Walter Burke Institute for Theoretical Physics,
California Institute of Technology, Pasadena, California, USA
and Perimeter Institute for Theoretical Physics, Waterloo, ON, Canada}

\abstract{We develop Standard Model Effective Field  Theory (SMEFT) predictions of $\sigma(\mathcal{G} \,\mathcal{G}\rightarrow h)$,
$\Gamma(h \rightarrow \mathcal{G} \,\mathcal{G})$, $\Gamma(h \rightarrow \mathcal{A} \mathcal{A})$ to incorporate
full two loop Standard Model results at the amplitude level, in conjunction with dimension eight SMEFT corrections.
We simultaneously report consistent $\Gamma(h \rightarrow \bar{\Psi} {\Psi})$ results including leading QCD corrections
and dimension eight SMEFT corrections.
This extends the predictions of the former processes $\Gamma, \sigma$
to a full set of corrections at
$\mathcal{O}(\bar{v}_T^2/\Lambda^2 (16 \pi^2)^2)$ and $\mathcal{O}(\bar{v}_T^4/\Lambda^4)$, where $\bar{v}_T$ is the electroweak
scale vacuum expectation value and $\Lambda$ is the cut off scale of the SMEFT.  Throughout, cross consistency between the
operator and loop expansions is maintained by the use of the geometric SMEFT formalism.
For $\Gamma(h \rightarrow \bar{\Psi} {\Psi})$, we include results at $\mathcal{O}(\bar{v}_T^2/\Lambda^2 (16 \pi^2))$
in the limit where subleading $m_\Psi \rightarrow 0$ corrections are neglected.
We clarify how gauge invariant SMEFT renormalization counterterms combine
with the Standard Model counter terms in higher order SMEFT calculations when the Background Field Method is used.
We also update the prediction of the total Higgs width in the SMEFT to consistently include some of these higher order perturbative
effects.}

\maketitle
\section{Introduction}

In this paper we calculate production and decay results of the Higgs boson
in the Standard Model Effective Field Theory (SMEFT) to $\mathcal{O}(\bar{v}_T^2/\Lambda^2 (16 \pi^2)^2)$
and $\mathcal{O}(\bar{v}_T^4/\Lambda^4)$.\footnote{Here $\bar{v}_T$ is the electroweak
scale vacuum expectation value, including
higher dimensional operator corrections \cite{Alonso:2013hga}, and $\Lambda$ is the cut off scale of the SMEFT.
 $\mathcal{G}$ is the canonically normalised
gluon in the SMEFT. $\mathcal{A}$ is the canonically normalized
photon field in the SMEFT.}
SMEFT perturbations to the SM predictions of $\sigma(\mathcal{G} \mathcal{G} \rightarrow h)$, $\Gamma(h \rightarrow \mathcal{A} \mathcal{A})$ need to be
characterized to higher orders in the effective field theory (EFT) due to the relative sensitivity that these (SM loop induced) processes carry to SMEFT corrections, compared to other
(typically SM tree level) processes \cite{Manohar:2006gz,Passarino:2012cb,Passarino:2016pzb}.
This introduces a relevant numerical sensitivity
to the treatment of higher order corrections and interference effects in these processes in future (and current \cite{Ellis:2020unq,Almeida:2021asy,Ethier:2021bye,ATLAS:2022xyx})
global SMEFT fits. In addition, we report results in the $m_\Psi \rightarrow 0$ limit for subleading corrections
up to dimension eight, and including one loop QCD corrections, for $\Gamma(h \rightarrow \bar{\Psi} {\Psi})$.

To develop such SMEFT results
in a reproducible manner, a clear calculation scheme needs to be adopted for perturbative
 ($\hbar/16 \pi^2$), and SMEFT operator ($1/\Lambda$) corrections.
To systematically calculate in the SMEFT, with fully defined $\mathcal{O}(\bar{v}_T^4/\Lambda^4)$ corrections
for the mass eigenstate fields requires a characterisation of
the low n-point interactions that define key experimental quantities (mass, mixing angles,
coupling and canonically normalized fields).
The geoSMEFT was defined in Ref.~\cite{Corbett:2019cwl,Helset:2020yio,Hays:2020scx,Corbett:2021cil}
and is a compact formalism that accomplishes this task using field space geometry.
These geometries are reflective of the underlying field redefinition invariance present in the SMEFT.
We use the geoSMEFT (see Appendix \ref{setup}) to define $\mathcal{O}(\bar{v}_T^4/\Lambda^4)$ corrections in this work.

In Ref.~\cite{Corbett:2021cil}, it was emphasised that the geoSMEFT -- as it is a background field independent
formulation of the SMEFT expansion -- encourages a class of perturbative corrections
to be calculated in a specific background field independent manner using the Background Field Method (BFM) approach
to gauge fixing.
The operator and loop expansions are not formally independent in SMEFT calculations
-- due to scheme dependence introduced by defining conventions in leading order results.
The geoSMEFT and the BFM are, in this sense, fundamentally linked
when theoretical self consistency is demanded to subleading order(s).
The scheme dependence is unsurprising in principle, as higher order perturbative corrections always carry
a significant scheme dependence. However in the case of the SMEFT, scheme dependence is not simply numerical.
Formulated at the Lagrangian level, the SMEFT is based on the
freedom to redefine the theoretical description with operators being removed or introduced by
field redefinitions (or appropriate use of the Equations of Motion).
As such, scheme dependence in the SMEFT is also associated with operator basis dependence,
and the specific parameter dependence present in a calculation. This scheme dependence
is also present in the geoSMEFT, despite its background field independence,
when the field space connections, metrics, etc are expanded out to a
particular order in $1/\Lambda$ in a particular operator basis.

In this paper, we extend/replace and update results in Ref.~\cite{Corbett:2021cil} by adding a class of  $\mathcal{O}(\bar{v}_T^2/\Lambda^2 (16 \pi^2)^2)$
corrections to $\sigma(\mathcal{G} \,\mathcal{G} \rightarrow h)$,  $\Gamma(h \rightarrow \mathcal{A} \mathcal{A})$
and $\Gamma(h \rightarrow \mathcal{G} \,\mathcal{G})$.\footnote{$v$ is sometimes used to denote
the vacuum expectation value in the SM, and the bare version of this parameter is $v_0$.
The inferred vacuum expectation value will necessarily be $\bar{v}_T$ when higher dimensional operators
are present, or  $v$ when such operator corrections are not present experimentally perturbing measurements.
As such, our use of $\bar{v}_T$ and $v$ is interchangeable in most results below.} This upgrades these results
to include a {\em full} set of self-consistent and cross-consistent $\mathcal{O}(\bar{v}_T^2/\Lambda^2 (16 \pi^2)^2)$ and $\mathcal{O}(\bar{v}_T^4/\Lambda^4)$
corrections at the observable level. We also report corrections to $\Gamma(h \rightarrow \bar{\Psi} {\Psi})$
up to dimension eight and including the leading QCD corrections in the $m_\Psi \rightarrow 0$ limit for subleading corrections.
Finally, we also update the calculation of the total Higgs width
in the SMEFT to include a full set of these corrections.

\section{Framework of the calculation}
Consider the perturbation due to a SMEFT
operator to a dimensionless SM amplitude in an on shell process (such as
$\sigma(\mathcal{G} \,\mathcal{G} \rightarrow h)$ or $\Gamma(h \rightarrow \mathcal{A} \mathcal{A})$):
\bea
\mathcal{A} = \langle \cdots \rangle_{SM} + \sum_i C^{(6)}_i \,  \langle \cdots \rangle_{\bar{v}_T^2/\Lambda^2} + \cdots
\eea
For three-particle on shell processes, such as $1 \to 2$ decays or $2 \to 1$ production, derivative terms $\mathcal O(\partial^{2}/\Lambda^{2})$ are trivial within the SMEFT expansion so all corrections scale with  $\bar{v}_T^2/\Lambda^2$. Thus, each amplitude has  a series of SMEFT corrections
\bea
 \Sigma_n \langle \cdots \rangle_{\bar{v}_T^{2n}/\Lambda^{2n}}
\eea
associated with operators in $\mathcal{L}^{(4+2n)}$.

\subsection{Terms retained in the calculation(s)}
Due to a proliferation of superscripts and subscripts indicating the various
expansions present in these calculations, we introduce a more schematic notation.
Amplitudes are expanded as
\bea
\mathcal \mathcal{A} = \sum_{i,j} \langle \cdots| \cdots \rangle^i_{(v^2/\Lambda^2)^j}.
\eea
We generally use $\Delta$ to indicate a loop correction
while a power of $\delta$ is used to indicate a SMEFT perturbation $\propto 1/\Lambda^2$
for more condensed notation, and to track the scaling of cross terms in the expansions.
In this work, we focus on improving the treatment of $\langle \mathcal{F} \mathcal{F}|h \rangle^{0,1,2}$,
and $\langle h| \mathcal{F} \mathcal{F} \rangle^{0,1,2}$,
where $\mathcal{F} = \{\mathcal{G}, \mathcal{A}\}$
compared to Ref.~\cite{Corbett:2021cil}.
Each of the terms for the amplitudes in this work scale as
\begin{align*}
\langle \mathcal{F} \mathcal{F}|h \rangle^1_{SM} &\sim \Delta, & \quad  \langle \mathcal{F} \mathcal{F}|h \rangle^2_{SM} &\sim \Delta^2, \\
\langle \mathcal{F} \mathcal{F}|h \rangle^0_{\mathcal O(v^2/\Lambda^2)} &\sim \delta, & \quad   \langle \mathcal{F} \mathcal{F}|h \rangle^0_{\mathcal O(v^4/\Lambda^4)} &\sim \delta^2, \\
\langle \mathcal{F} \mathcal{F}|h \rangle^1_{\mathcal O(v^2/\Lambda^2)} &\sim \delta \, \Delta & \quad \langle \mathcal{F} \mathcal{F}|h \rangle^2_{\mathcal O(v^2/\Lambda^2)} &\sim \Delta^2 \, \delta,
\end{align*}
and so on. Cross terms when the amplitude is squared scale as
\bea
\langle \mathcal{F} \mathcal{F}|h \rangle^1_{SM} \, \langle \mathcal{F} \mathcal{F}|h \rangle^1_{\mathcal O(v^2/\Lambda^2)} \propto \Delta^2 \delta.
\eea
In this work, we include the corrections $\langle \mathcal{F} \mathcal{F}|h \rangle^2_{SM}$
and $\langle \mathcal{F} \mathcal{F}|h \rangle^1_{\mathcal O(v^2/\Lambda^2)}$
as defined above. The first term leads to corrections of the order
\bea
\langle \mathcal{F} \mathcal{F}|h \rangle^2_{SM} \langle \mathcal{F} \mathcal{F}|h \rangle^0_{\mathcal O(v^2/\Lambda^2)} \propto \Delta^2 \delta,
\eea
which should be retained for consistency at the amplitude squared level
when $\langle \mathcal{F} \mathcal{F}|h \rangle^1_{SM}$ $ \times \langle \mathcal{F} \mathcal{F}|h \rangle^1_{\mathcal O(v^2/\Lambda^2)}$
terms are retained, as in Ref.~\cite{Corbett:2020ymv}. We retain the terms that scale as $\delta$, $\Delta$, $\delta^2$, $\Delta^2$, $\delta \, \Delta$
in the amplitude expansion in this work. Note that $\Delta, \Delta^2$ terms are pure SM terms. We retain the SM cross terms of order
$\Delta^2$, $\Delta^3$  in the amplitude squared. As well as terms of order $\delta \, \Delta$, $\delta \, \Delta^2$, $\delta^2$, $\delta^2 \, \Delta$
for the SMEFT corrections in the amplitude squared. All other higher order terms are consistently dropped.

Note that
when constructing the interference term, one could choose to numerically retain
the corrections of the order
\bea
\langle \mathcal{F} \mathcal{F}|h \rangle^2_{SM} \langle \mathcal{F} \mathcal{F}|h \rangle^1_{\mathcal O(v^2/\Lambda^2)} \propto \Delta^3 \delta.
\eea
If this choice is made to improve numerical accuracy for some Wilson coefficient dependence,
then a consistent calculation at the amplitude squared level should also retain
the finite and scheme dependent interference terms following from $\epsilon/\epsilon$ cancelations that are also generated by
\bea
\langle \mathcal{F} \mathcal{F}|h \rangle^1_{\mathcal O(v^2/\Lambda^2)} \langle \mathcal{F} \mathcal{F}|h \rangle^2_{SM}
\propto \Delta^3 \delta.
\eea
We report a series of results below retaining different classes of terms to make the numerical impact
of the different sets of corrections clear (see Eqn.~3.17 and Eqn.~3.50). Also note that this class of terms is the same order as
\bea
\langle \mathcal{F} \mathcal{F}|h \rangle^0_{\mathcal O(v^2/\Lambda^2)} \langle \mathcal{F} \mathcal{F}|h \rangle^3_{SM}
\eea
corrections which are also neglected.
This class of corrections is particularly sensitive
to the combination of the SMEFT counterterms and the SM counterterms in a consistent calculation scheme.
See further discussion in Appendix \ref{combiningschemes}.

\section{Analytic results}\label{sigmaggh}

 To define the perturbative corrections to next to leading order (NLO),
the infrared/ultraviolet (IR/UV) divergences present in the perturbative expansions have to be canceled/subtracted
in some calculational scheme.
Combining these results with SMEFT perturbations requires some care.
We reiterate and incorporate these results to fix notational conventions.

\subsection{$\sigma(\mathcal{G} \mathcal{G} \rightarrow h)$}
We define the full amplitude for $\mathcal{G} \,\mathcal{G} \to h$ as \cite{Corbett:2021cil}
\begin{align}
\mathcal \mathcal{A}_{\mathcal{G} \mathcal{G}h} = \langle \mathcal{G} \mathcal{G}|h \rangle^1_{SM}+\langle \mathcal{G} \mathcal{G}|h \rangle^2_{SM} + \langle \mathcal{G} \mathcal{G}|h \rangle^0_{\mathcal O(v^2/\Lambda^2)} + \langle \mathcal{G} \mathcal{G}|h \rangle^1_{\mathcal O(v^2/\Lambda^2)} +  \langle \mathcal{G} \mathcal{G}|h \rangle^0_{\mathcal O(v^4/\Lambda^4)} + \cdots
\end{align}
The two loop SM contributions to this amplitude are $\langle \mathcal{G} \mathcal{G}|h \rangle^2_{SM}$.
The relevant results for $\langle \mathcal{G} \mathcal{G}|h \rangle^2_{SM}$ are known
and reported in Refs.~\cite{Dawson:1990zj,Graudenz:1992pv,Spira:1995rr,Harlander:2005rq,Gehrmann:2005pd,Anastasiou:2009kn,Anastasiou:2006hc}.

The first careful study of interference with $\langle \mathcal{G} \mathcal{G}|h \rangle^0_{\mathcal O(v^2/\Lambda^2)}$
effects was reported in Ref.~\cite{Manohar:2006gz}.
Results for $\langle \mathcal{G} \mathcal{G}|h \rangle^1_{\mathcal O(v^2/\Lambda^2)}$ are reported in many
works, including Refs.~\cite{Deutschmann:2017qum,Grazzini:2018eyk}, in different calculation schemes than used here.
Renormalization results to dimension eight have started to appear in Refs.~\cite{Chala:2021pll,Helset:2022tlf,Helset:2022pde}
enabling $\mathcal{O}(\Delta \, \delta^2)$ results to be developed, and recently results of this order were reported in
Ref.~\cite{Asteriadis:2022ras}, also in a different scheme than used here.

One of the central points of this paper, is the need to combine input parameter extractions,
and observables in a consistent calculational scheme up to $\mathcal O(\delta^2)$
and $\mathcal O(\Delta^2 \delta)$. We provide significant calculational details for our
results to be reproducible including these corrections.

\subsubsection{$\langle \mathcal{G} \mathcal{G}|h \rangle^{1}_{SM}$ and $\langle \mathcal{G} \mathcal{G}|h \rangle^{2}_{SM}$ results for $\sigma(\mathcal{G} \,\mathcal{G}\rightarrow h)$}
The top quark leading contribution to the SM amplitude is expanded in perturbation theory as
\cite{Ellis:1975ap,Georgi:1977gs,Shifman:1979eb,Dawson:1990zj,Harlander:2005rq,Gehrmann:2005pd,Anastasiou:2006hc,Anastasiou:2020qzk}
\begin{eqnarray}
\langle \mathcal{G} \mathcal{G}|h \rangle^1_{SM}
= i \, \delta_{ab} \,  K_{ab} \, \frac{1}{\bar{v}_T^{(0)}} \, \left(- \frac{s}{\hat{\mu}^2}\right)^{-\epsilon}
\left(\frac{\alpha_s^0 \, S^\epsilon \, \hat{\mu}^{-2 \epsilon}}{4 \pi} M_{t,SM}^{(0)} \right).
\end{eqnarray}
Here, $a,b$ are the gluon colors with $\epsilon$ polarization vectors, $\gamma_E$ is the Euler-Mascheroni constant, and
we have shifted to a $\rm \overline{MS}$ renormalization introducing $\mu^{2} \rightarrow \hat{\mu}^2/S =  \hat{\mu}^2 e^{\gamma_E}/(4 \pi)$
where $S^\epsilon =  (4 \pi)^\epsilon e^{- \epsilon \, \gamma_E}$ to simplify finite terms. The factor $K_{ab}$ is
\begin{eqnarray}
K_{ab} &\equiv& \epsilon_a(p_1) \cdot \epsilon_b(p_2) \, s/2 - p_1 \cdot \epsilon_b(p_2) \, p_2 \cdot \epsilon_a(p_1), \\
&=& - \langle h| h \mathcal{G}^{\mu\nu} \mathcal{G}_{\mu \nu} | \epsilon_a \epsilon_b \rangle^0/4.
\end{eqnarray}
where $p_{1,2}$ are the incoming gluon momenta with $s = (p_1+ p_2)^2 \equiv m_h^2$ and $\mathcal{G}_{\mu \nu}$ is the field strength tensor of the canonically normalized gluon field.

The normalized, leading order partonic cross section in the SM then depends on ($z = \hat{m}_h^2/s$) as
\bea
\sigma^{SM}_{LO}(\mathcal{G} \mathcal{G} \rightarrow h;z) \equiv \frac{\hat{\sigma}^{SM}_{LO}(\mathcal{G} \mathcal{G} \rightarrow h)}{1- \epsilon} \,
\, z \,\delta(1- z)
\eea
where $\langle \mathcal{G} \mathcal{G}|h \rangle^0_{SM}$ starts at one loop, so
$\sigma^{SM}_{LO}(\mathcal{G} \mathcal{G} \rightarrow h;z)$ scales as $\Delta^2$
and will be denoted $\Delta^2 \sigma^{SM}_{LO}(\mathcal{G} \mathcal{G} \rightarrow h;z)$
to emphasize this fact. Stated another way,
\bea
\Delta^2 \hat{\sigma}^{SM}_{LO}(\mathcal{G} \mathcal{G} \rightarrow h;z)&\equiv&
\frac{\pi}{4} \lim_{\epsilon \rightarrow 0}
\left|C^{SM}_{h \mathcal{G} \mathcal{G}}\right|^2,
\eea
where $C^{SM}_{h \mathcal{G} \mathcal{G}}$ is the Wilson coefficient of the operator $h \mathcal{G}_{\mu \nu}^a \mathcal{G}_{\mu \nu}^a$
with normalization
\bea
\Delta C^{SM}_{h \mathcal{G} \mathcal{G}} = - \frac{\alpha_s^{(r)}}{\bar{v}^0_T \, 16 \pi} \, \left(- \frac{s}{\hat{\mu}^2}\right)^{-\epsilon} M_{t,SM}^{(0)}.
\eea
The corresponding cross section in the SMEFT has a modified Wilson coefficient, given by
\bea
C^{SMEFT}_{h \mathcal{G} \mathcal{G}} = \Delta C^{SM}_{h \mathcal{G} \mathcal{G}} + \frac{\tilde{C}^{(6)}_{HG}}{\bar{v}^0_T}
 + \cdots.
\eea

An expression for $M_t^{(0)}$ is given in Ref.~\cite{Anastasiou:2006hc} and is numerically\footnote{A factor of four has been
absorbed into this expression comparing to $M_{LO}^0$ in Ref.~\cite{Anastasiou:2006hc}.}
in the $m_t \rightarrow \infty$ limit \cite{Anastasiou:2020qzk}
\begin{eqnarray}
\Delta C^{SM,m_t \rightarrow \infty}_{h \mathcal{G} \mathcal{G}} &=& - \frac{\alpha_s^{(r)}}{\bar{v}^0_T \, 16 \pi} \, \left(\frac{\hat{m}_t^2}{\hat{\mu}^2} \right)^{-\epsilon} \, M_{t,SM}^{(0),m_t \rightarrow \infty}, \nonumber \\
 &=& - \frac{\alpha_s^{(r)}}{\bar{v}_T^0 \, 16 \pi} \, \left[-\frac{4}{3} \left(1 + \frac{\pi^2}{12} \epsilon^2  -  \epsilon \, L_{\hat{m}_t} + \frac{1}{2} L^2_{\hat{m}_t} \epsilon^2 + \mathcal{O}(\epsilon^3) \right)\right],
\end{eqnarray}
where
$L_{m} = \log \left(m^2/\hat{\mu}^2\right)$.
The numerical term in this expression in the square brackets is related to the function commonly defined  and used
in the literature $A_{1/2}(\tau_t) = - 1.37664$ in the exact top mass limit,
where $\tau_t = 4 \hat{m}_t^2/\hat{m}_h^2 = 7.59871$.
Similarly, Ref.~\cite{Anastasiou:2020qzk} gives the exact higher order expressions to build up
\begin{eqnarray}\label{NLOresult}
\langle \mathcal{G} \mathcal{G}|h \rangle^2_{SM} &=&
 i \, \delta_{ab} \,  K_{ab} \, \left[\left(- \frac{s}{\hat{\mu}^2}\right)^{-\epsilon}
\frac{\alpha_s^0 \, S^\epsilon \hat{\mu}^{-2 \epsilon}}{4 \pi}\right]^2 \, \frac{1}{\bar{v}^0_T}\, M_{t,SM}^{(1)},
\end{eqnarray}
where $\langle \mathcal{G} \mathcal{G}|h \rangle^2_{SM}$ scales as a $\Delta^2$
perturbation and
\begin{eqnarray}\label{NLOresultdecompose}
M_{t,SM}^{(1)} =  M_{UV} + M_{UV,m} + M_{IR} + M_{fin} + M_{fin,s}  \log \left(- \frac{s}{\hat{\mu}^2} \right).
\end{eqnarray}
Each of the terms in the decomposition in Eqn.~\eqref{NLOresult} given in
Eqn.~\eqref{NLOresultdecompose} is defined in
Ref.~\cite{Anastasiou:2020qzk} and previous literature using a variety of calculation schemes;
$M_{UV,m}$ corresponds to UV poles and related finite terms canceled by UV mass renormalization,
$M_{UV}$ corresponds to the remaining UV renormalization of the NLO result, and
$M_{fin}$ and $M_{fin,s}$ correspond to finite NLO terms, with the later multiplying the complete scale dependence
in $M_{t,SM}^{(1)}$. Finally,
$M_{IR}$ corresponds to finite terms related to IR driven cancelations between these NLO contributions to
$\sigma(\mathcal{G} \,\mathcal{G} \to h)$ and $\sigma(\mathcal{G} \,\mathcal{G} \to h \mathcal{G})$.
Results in the literature must be modified into the background field method (BFM) to combine consistently with a
BFM based SMEFT calculation and counterterms (i.e. when using the geoSMEFT to define $1/\Lambda^n$ corrections).
We report the required modifications in the following sections.

\subsubsection{NLO finite terms}

We organise the NLO contributions by defining
\bea
\langle \mathcal{G} \mathcal{G}|h \rangle^{2,F}_{SM} \equiv
i \, \delta_{ab} \,  \frac{K_{ab}}{\bar{v}^0_T} \, \left[\left(- \frac{s}{\hat{\mu}^2}\right)^{-\epsilon}
\frac{\alpha_s^0 \, S^\epsilon \hat{\mu}^{-2 \epsilon}}{4 \pi}\right]^2 \, \left(
M_{t,SM}^{(1)} - M_{UV} - M_{UV,m} - M_{IR} \right)
\eea
so that the UV and IR subtractions and cancelations, which have an intricate interplay in these results
are separately considered.  The $\langle \mathcal{G} \mathcal{G}|h \rangle^{1,F}_{SM}$ renormalized and IR subtracted
finite terms (so defined) are related to matching and running in the EFT. In the $m_t \rightarrow \infty$ limit,
the corresponding subset of terms is
\bea\label{twoloopmatch2}
\langle \mathcal{G} \mathcal{G}|h \rangle^{2,F}_{SM} =
\frac{\alpha^{(r)}_s}{4 \pi}
\, \left[11 + c_1 \, \epsilon + (- \beta_0 + c_2 \, \epsilon) \log \left(- \frac{\hat{m}_h^2}{\hat{\mu}^2}\right) \right]
\langle \mathcal{G} \mathcal{G}|h \rangle^1_{SM,\epsilon \rightarrow 0},
\eea
where $\beta_0 = 11 N_c/3 - 2 n_F/3$, $N_c=3$ and \cite{Anastasiou:2020qzk}
\bea
c_1 &=& \left[- \frac{\pi^2 \, \beta_0}{12} + 28 \, \log(z)+ 12 \, \zeta_3 - \frac{40}{3} \right], \\
c_2 &=& \left[- \frac{1}{2}  \beta_0 \, \log\left(\frac{-s}{\mu^2}\right)- 2 \beta_0 \, \log(z)+ 8 \right].
\eea
Here $\log(z) = \log (- s/m_t^2)/2$. The $11 \alpha_s/4 \pi$ factor in Eqn.~\eqref{twoloopmatch2} is recognised as the two loop matching contribution to the $m_t \rightarrow \infty$
effective operator \cite{Dawson:1990zj}.  This non-log term was included in Ref.~\cite{Corbett:2020ymv}. The log term is an additional contribution present
that is not captured in the two loop matching contribution.\footnote{We thank Babis Anastasiou for confirming some typos in the literature result of
Ref.~\cite{Anastasiou:2020qzk} that are corrected for here.} This log dependence also is consistent with naive expectations
as the direct matching contribution at two loops needs to be augmented with log terms
due to running in the EFT.  To further consistently
improve these results beyond Ref.~\cite{Corbett:2020ymv}
we must also improve the finite terms resulting from the UV and IR subtracted
cancelation between $M_{t,SM}^{(1)}$ and the IR contributions from the process
$\sigma(\mathcal{G} \,\mathcal{G} \to h \mathcal{G})$.\footnote{The hat superscripts have a dual meaning, indicating a input parameter for the higgs mass and the use of $\rm \overline{MS}$ renormalization
for the renormalization scale.}

Interference of $\langle \mathcal{G} \mathcal{G}|h \rangle^0_{\mathcal O(v^2/\Lambda^2)}$ with $\langle \mathcal{G} \mathcal{G}|h \rangle^{2,F}_{SM}$ leads to the contributions
\bea\label{polelimitNLOfinite}
\frac{\Delta^2 \delta \sigma(\mathcal{G} \,\mathcal{G} \to h)_F}{\Delta^2 \hat{\sigma}^{SM}_{LO}(\mathcal{G} \,\mathcal{G} \to h;z)}
&=&
\frac{\alpha_s}{2 \pi}\left(11 - \beta_0 \, L_{\hat{m}_h}  \right)
\frac{\tilde{C}^{(6)}_{HG}}{\bar{v}^0_T \, \Delta C^{SM}_{h \mathcal{G} \mathcal{G}}}, \nonumber \\
&=& 6 \, \left(11 - \beta_0 \, L_{\hat{m}_h} \right)
\tilde{C}^{(6)}_{HG}.
\eea
The $\epsilon$ terms in $c_1, c_2$ interfere and generate constant finite terms in the
$\langle \mathcal{G} \mathcal{G}|h \rangle^2_{SM} \langle \mathcal{G} \mathcal{G}|h \rangle^1_{\mathcal O(v^2/\Lambda^2)}$
interference with the renormalization of the leading order cross section.
These contributions are
\bea\label{polelimitNLO2finite}
\frac{\Delta^3 \delta \sigma(\mathcal{G} \,\mathcal{G} \to h)_F}{\Delta^2 \, \hat{\sigma}^{SM}_{LO}(\mathcal{G} \,\mathcal{G} \to h;z)}
&=&
- \frac{3 \beta_0 \,\alpha_s}{2 \pi} \left({\rm Re}[c_1] + {\rm Re}[c_2] \, L_{\hat{m}_h}
+ \frac{3 \, \pi^2 \, \beta_0}{2}\right)
\tilde{C}^{(6)}_{HG}.
\eea

\subsubsection{UV divergences}
The renormaliation of the SM result has the remaining contributions
\bea\label{twoloopmasterUV1}
\mathcal{M}_{UV} + \mathcal{M}_{UV,m} =  \left(-\frac{s}{\hat{\mu}^2}\right)^{-\epsilon}
\left(\frac{Z_{m_t}^2}{Z_{m_h^2}} \tau_t \frac{\partial}{\partial \tau_t} \right)
 Z_g^2 \, Z_{\hat{\mathcal{G}}} \, \frac{Z^{1/2}_{\hat{h}}}{Z_v^{1/2}} \,
 i \, \delta_{ab} \,  K_{ab} \, \frac{1}{\bar{v}_T^{(r)}} \frac{\alpha_s^{(r)}}{4 \pi} \, M^{(0)}_{t,SM}.
\eea

The result in Eqn.~\eqref{twoloopmasterUV1} has one overall power of $\left(-s/\hat{\mu}^2\right)^{-\epsilon}$
due to the conventional choice in Refs.~\cite{Anastasiou:2006hc,Anastasiou:2020qzk}, followed here, to organize the calculation
in such a way that we factorize the complete $\mu$ dependence in $M_{t,SM}^{(1)}$ into the terms with "fin" superscripts.

Here we have modified the notation of Ref.~\cite{Anastasiou:2006hc} to make the mass dimensions
of the corrections clearer.\footnote{$\langle \mathcal{G} \mathcal{G}|h \rangle^1_{SM}$ is a function of a dimensionless ratio
in the bare masses (with $0$ superscripts)
$\tau_t^0 = 4 (m_t^{(0)})^2/(m^{(0)}_h)^2$. To make the mass dimensions of the corrections clearer and extend to the SMEFT more easily, we choose
to take a derivative with respect to $\tau_t$ more explicit.
The correction comes about due to Taylor expanding the perturbative corrections within $Z_{m_t}^2$ in the SM,
which reduces to past results for the SM, once correcting for a
typo in Ref.~\cite{Anastasiou:2006hc} in Equation 7.6, which is missing a factor of $m_t$ in the numerator.
The notation agrees in the mass dimensions with Ref.~\cite{Anastasiou:2020qzk}.}
In the SM, the non-vanishing counter terms (proportional to the QCD coupling) are
$Z_g,Z_{\hat{\mathcal{G}}},Z_{m_t}$ -- $Z_{m_h^2}$ does not have QCD corrections in the SM.
The neglect of $Z^{1/2}_{\hat{h}}$, $Z_v^{1/2}$ is trivial when only considering their lack of one loop
QCD corrections, but when considering EW corrections the use of the background field method
introduces important differences compared to alternate schemes. In the background field
method, EW corrections from $Z^{1/2}_{\hat{h}}$, $Z_v^{1/2}$ exactly cancel, including finite terms
-- a helpful simplification.

In Ref.~\cite{Anastasiou:2006hc,Anastasiou:2020qzk,Deutschmann:2017qum}
a $\rm \overline{MS}$ renormalization scheme is chosen so that the mass counter term is effectively given by
\bea
 \Delta Z_{m_t} &=& - \frac{\alpha^{(r)}_s}{4 \pi} C_F \frac{3}{\epsilon},
\eea
with $C_F = (N_c^2-1)/2 N_c$. We adopt this $\rm \overline{MS}$ renormalization for the top quark mass in this work.
Note that in the BFM, the fermion fields are not modified and the counterterm is the same and gauge independent.
To use the results in Ref.~\cite{Anastasiou:2020qzk} we need to account for the finite terms in the
renormalization of the leading order result.
The explicit form of the finite terms due to mass renormalization is given
by
\bea
\mathcal{M}_{UV,m} = \frac{6}{\epsilon} C_F \, \left(-\frac{s}{\hat{\mu}^2}\right)^{\epsilon} (m_t^0)^2 \frac{\partial}{\partial (m_t^0)^2}
\left(\left(\frac{(m_t^0)^2}{-s}\right)^{-\epsilon}\langle \mathcal{G} \mathcal{G}|h \rangle^1_{SM} \right)
\eea
which leads to a pure finite term, even in the $m_t \rightarrow \infty$ limit, that is effectively a matching contribution
to the leading order operator $C_{HG}^{(6)}$. The resulting correction is given by
\bea\label{mlimitren}
\frac{\Delta^2 \delta \sigma(\mathcal{G} \,\mathcal{G} \to h)_{ren,m}}{\Delta^2 \, \hat{\sigma}^{SM}_{LO}(\mathcal{G} \to h;z)}
&=&  36  \times C_F  \, \tilde{C}^{(6)}_{HG}.
\eea
The form of $Z_g$ and $Z_{\hat{\mathcal{G}}}$ depend on the scheme and gauge chosen.
In Ref.~\cite{Anastasiou:2006hc,Anastasiou:2020qzk,Deutschmann:2017qum},
the combination of $Z_g^2 Z_{\hat{\mathcal{G}}}$ leads to the effective renormalization
to cancel the poles in the matrix element
\bea
Z_g^2 \, Z_{\hat{\mathcal{G}}} \, \left(-\frac{s}{\hat{\mu}^2}\right)^{-\epsilon} \,
i \, \delta_{ab} \,  K_{ab} \, \frac{1}{\bar{v}_T^{(r)}} \, \frac{\alpha_s^{(r)}}{4 \pi} \,  M^{(0)}_{t,SM}
= - \left[\frac{\alpha^{(r)}_s}{4 \pi}\right]^2 \frac{\beta_0}{\epsilon} (-\frac{s}{\hat{\mu}^2})^{-\epsilon}
i \, \delta_{ab} \,  K_{ab} \, \frac{1}{\bar{v}_T^{(r)}} \, M^{(0)}_{t,SM}. \nn
\eea
When considering the calculation in the $m_t \rightarrow \infty$ limit,
a composite operator is present. In the unbroken phase of the theory,
the operator is $H^\dagger H \mathcal{G}^{\mu \nu} \mathcal{G}_{\mu \nu}$.
The composite operator renormalization is performed after the gluon
wavefunction renormalization is subtracted.
In Ref.~\cite{Deutschmann:2017qum}, and related works, both the gluon field and the composite operator
are not further renormalized due to the calculational scheme chosen, so this subtlety is rather irrelevant.
The full cancelation of the UV pole comes from the renormalization
of the strong coupling as a result.

In the BFM, the relationship between the QCD coupling and wavefunction renormalization is
$\mu^{2 \epsilon} \, Z_g^2 Z_{\hat{\mathcal{G}}} \equiv 1$, including finite terms.
On the other hand, the composite operator, in this case, gets its own renormalization counter term
\cite{Alonso:2013hga}
\bea
Z_{HG} = 1 - \frac{\beta_0 \,\alpha_s}{4 \pi \, \epsilon} + \cdots.
\eea
which leads to the same net subtraction of UV poles. In the BFM, this renormalization occurs with the
SM matching contribution to the composite operator interfering
with $\langle \mathcal{G} \mathcal{G}| h \rangle^0_{\mathcal O(v^2/\Lambda^2)}$
and the $C_{HG}$ Wilson coefficient itself.
This renormalization is given by
\bea
\langle \mathcal{G} \mathcal{G}| h \rangle^0_{\mathcal O(v^2/\Lambda^2)} \rightarrow Z_{HG} \, \frac{\tilde{C}^{(6)}_{HG}}{\bar{v}_T} \langle \mathcal G^{\mu \nu} \mathcal G_{\mu \nu} h \rangle _0.
\eea

If the choice is also made, as in Ref.~\cite{Deutschmann:2017qum}, to normalize the operator $\tilde{C}^{(6)}_{HG}$ by factors
of $(g_s^0)^2$ explicitly, then the renormalization of the composite operator
can again vanish, and a further renormalization due to the extra factor of the strong coupling is introduced,
again leading to the same net counter term being introduced.
This subtlety potentially introduces some confusion when comparing results in the literature if different normalizations,
and calculation schemes are not carefully defined.

The UV pole canceled by these counter terms also (accidentally) cancels against
an IR contribution with opposite sign. This renormalization introduces a contribution to the cross section
\bea\label{polelimitren}
\frac{\Delta^2 \delta \sigma(\mathcal{G} \,\mathcal{G} \to h)_{ren}}{\Delta^2 \, \hat{\sigma}^{SM}_{LO}(\mathcal{G} \to h;z)}
&=& - 6 \beta_0 \left[\frac{1}{\epsilon} +1 - L_{\hat{m}_t} \right] \, \tilde{C}^{(6)}_{HG}.
\eea
The finite term as $\epsilon \rightarrow 0$ comes from the $\epsilon$ dependence in the SM amplitude interfered with.

\subsubsection{IR divergences}
For the IR divergences, it is well known that a universal form is present in
the pole structure for a renormalized one loop amplitude for the production of a
Higgs boson from two massless gauge bosons in the SM. The \cite{Kunszt:1994np,Catani:1996vz,Catani:2000ef}
dipole subtraction scheme allows one to write $\mathcal{M}_{t,IR}^{(1)}$ with
a universal (scheme dependent) set of IR poles as \cite{Anastasiou:2006hc,Anastasiou:2020qzk}

\bea\label{twoloopmasterIR}
\mathcal{M}_{t,IR}^{(1)} = \frac{-e^{\epsilon \, \gamma_E}}{\Gamma(1- \epsilon)}
\left[\frac{2 N_c}{\epsilon^2} + \frac{\beta_0}{\epsilon} \right] M_{t,SM}^{(0)}.
\eea
in $\rm \overline{MS}$.
The number of active flavors is $N_f = 5$. By definition, the IR physics before heavy states are integrated out
is the same as that in the SMEFT with a fixed set of matched Wilson coefficients. The SMEFT contains additional local contact
operator corrections to the SM interaction terms that modify the UV. In principle, the presence of higher order local contact operators
can modify the IR radiation field present
compared to the SM with a point like higgs particle, leading to further modifications of this result
at higher orders. This is the SMEFT multipole expansion, reflecting possible Higgs substructure, see discussion in Ref.~\cite{Jenkins:2013fya,Elvang:2016qvq}.
In practice, this does not occur in the SMEFT
to the level of precision we are interested in calculating in this paper.

The interference of the remaining NLO contributions to $\mathcal{G} \,\mathcal{G} \to h$
\bea
i \, \delta_{ab} \,  \frac{K_{ab}}{\bar{v}^0_T} \, \left[\left(- \frac{s}{\hat{\mu}^2}\right)^{-\epsilon}
\frac{\alpha_s^0 \, S^\epsilon \hat{\mu}^{-2 \epsilon}}{4 \pi}\right]^2 \, M_{IR},
\eea
with the tree level insertion of $\tilde{C}^{(6)}_{HG}$\footnote{Here we introduced
the notation $L_+ = L_{\hat{m}_h} + L_{\hat{m}_t}$.} gives the subtraction scheme dependent terms
\bea
\frac{\Delta^2 \delta \sigma(\mathcal{G} \,\mathcal{G} \to h)_{sch.}}{\Delta^2 \, \hat{\sigma}^{SM}_{LO, \epsilon \rightarrow 0}(\mathcal{G} \,\mathcal{G} \to h;z)}
&=& 6\left[ - \frac{6}{\epsilon^2}
+6 \frac{L_+}{\epsilon} - \frac{6}{\epsilon} - 14 + 3 \pi^2 - 3 L_+^2 + 6 L_+  + \beta_0 L_{\hat m_h}
 \right] \tilde{C}^{(6)}_{HG}. \nn
 \eea
Note that the IR poles are the same in the renormalization scheme used in Ref.~\cite{Anastasiou:2006hc,Anastasiou:2020qzk,Deutschmann:2017qum}
and in the BFM.  The corresponding SMEFT expression differs from the SM in finite terms as the $\epsilon$ expansion
of $M_{t,SM}^{(0), m_t \rightarrow \infty}$ is not squared.
The log structure and constant terms differ in the SMEFT and the SM,
even though the IR pole structure is the same, as the higher order terms in $\epsilon$ coming from the
SM top sub-loop function are different.

\begin{figure}[ht!]
\includegraphics[width=0.5\textwidth]{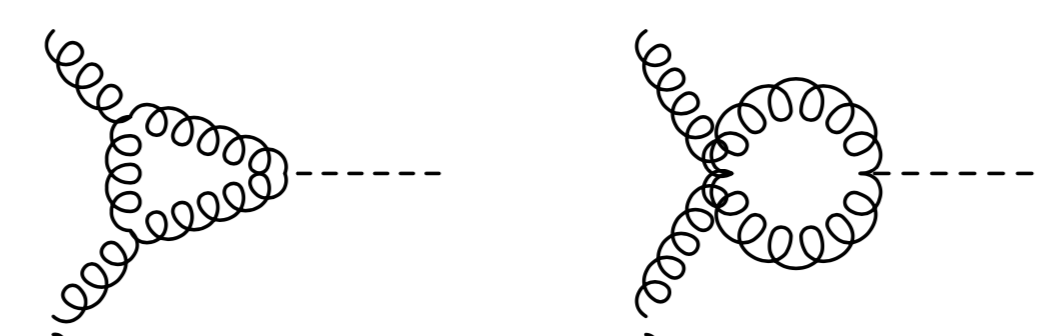}
\caption{QCD one loop contribution to $\mathcal{G} \,\mathcal{G} \rightarrow h$.}
\label{fig:2}
\end{figure}

Adding up all terms gives the NLO results from Fig.~\ref{fig:2}
of order $\Delta^2 \delta \sigma(\mathcal{G} \,\mathcal{G} \to h)$
\bea\label{combinedresultUV}
\frac{\Delta^2 \delta \sigma(\mathcal{G} \,\mathcal{G} \to h)}{\Delta^2 \, \hat{\sigma}^{SM}_{LO, \epsilon \rightarrow 0}(\mathcal{G} \,\mathcal{G} \to h;z)}
&=& 6\left[ - \frac{6}{\epsilon^2} - \frac{\beta_0}{\epsilon}
+6 \frac{L_+}{\epsilon} - \frac{6}{\epsilon} + \beta_0 \, L_{\hat{m}_t}
 + 3 \, \pi^2 +5  - \beta_0 - 3 L_+^2 + 6 L_+\right] \tilde{C}^{(6)}_{HG}, \nonumber \\
\eea

Here we have suppressed common factors of $\delta (1- z)$ in the numerator and denominator.
The $\epsilon$ poles Eqn.~\eqref{combinedresultUV} are all of IR origin.
These poles cancel against poles in $\mathcal{G} \,\mathcal{G} \to h \mathcal{G}$ in the limit
that the final state gluon is soft/colinear
for any local contact operator of the form $h \mathcal{G}_{\mu \nu}^a \mathcal{G}_{\mu \nu}^a$.
There are finite term differences between the SMEFT and the SM involved in this IR cancelation.

\begin{figure}[ht!]
\includegraphics[width=0.75\textwidth]{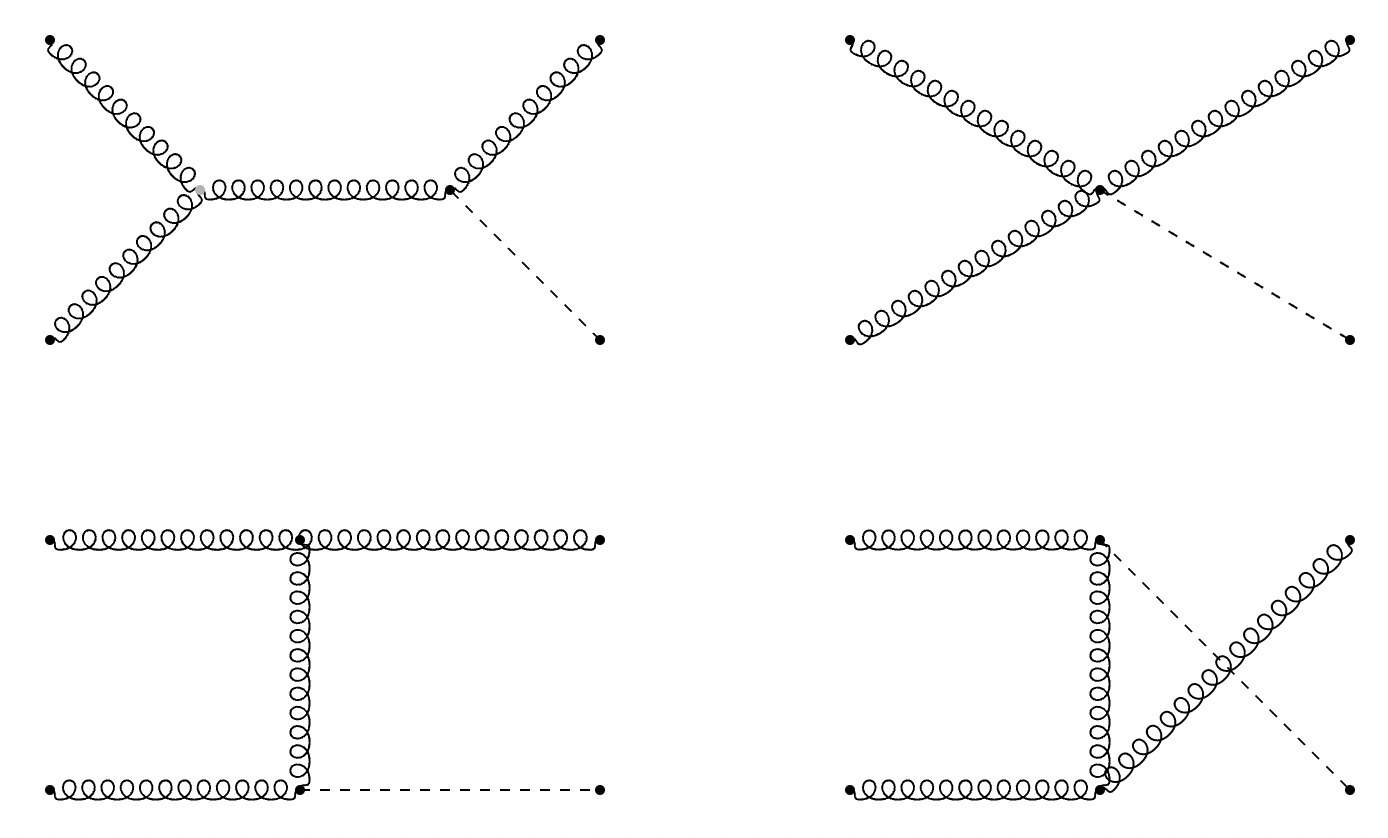}
\caption{$\mathcal{G} \,\mathcal{G} \rightarrow h \,\mathcal{G}$ Required to cancel IR divergences
in the two loop matrix element for $\mathcal{G} \,\mathcal{G} \rightarrow h$.}
\label{fig:3}
\end{figure}

The $\mathcal{G} \,\mathcal{G} \to h \mathcal{G}$ amplitude squared is shown in Fig.~\ref{fig:3}
and is a modification of results in Ref.~\cite{Dawson:1990zj}
\begin{align}\label{eq:ggg}
|\mathcal A(\mathcal{G} \,\mathcal{G} \to h \mathcal{G})|^2 = (4 \pi) \, 384 \,\alpha_s^{(0)} |C^{SMEFT}_{h \mathcal{G} \mathcal{G}}|^2 \frac{(\hat{m}^8_h + s^4 + t^4 + u^4)(1-2\epsilon) + \frac 1 2 \epsilon\, (\hat{m}^4_h + s^2 + t^2 + u^2)^2}{s\, t\, u},
\end{align}
where $C_{hGG}$ is the coefficient of $\langle h\, | \mathcal{G} \,\mathcal{G}\rangle^0$ and $s,t,u$ here are the usual Mandelstam variables
for this $2 \rightarrow 2$ process. Expanding out to the linear in $\tilde{C}_{HG}$ interference term
\begin{align}\label{eq:ggg2}
\Delta \delta |\mathcal A(\mathcal{G} \,\mathcal{G} \to h \mathcal{G})|^2 = \frac{768 \pi \alpha_s^{(0)}}{\bar{v}_T^0} 2 {\rm Re} \left(\frac{\Delta C^{SM}_{h \mathcal{G} \mathcal{G}}}{\mu^{2\epsilon}} \tilde{C}_{HG}\right) \frac{(\hat{m}^8_h + s^4 + t^4 + u^4)(1-2\epsilon) + \frac 1 2 \epsilon\, (\hat{m}^4_h + s^2 + t^2 + u^2)^2}{s\, t\, u}.
\end{align}
In the $m_t \rightarrow \infty$ limit, this becomes (after renormalizing)\footnote{Here we are dividing by a $1/2$ that we explain below.}

\bea\label{eq:ggg3}
\Delta \delta |\mathcal A(\mathcal{G} \,\mathcal{G} \to h \mathcal{G})|^2 &=&
\left(\frac{Z_{m_t}^2}{Z_{m_h^2}} \frac{\partial}{\partial \tau_t} \right)
 Z_g^2 \, Z_{\hat{\mathcal{G}}} \, \frac{Z^{1/2}_{\hat{h}}}{Z_v^{1/2}} Z_{HG}
\frac{128 (\alpha_s^{(r)})^2 \, \mu^{2 \epsilon}}{(\bar{v}_T^{(r)})^2} \left(1- \epsilon \, L_{\hat{m}_t} + \epsilon^2 \left(\frac{\pi^2}{12}+ \frac{1}{2}L_{\hat{m}_t}^2 \right)\right) \nonumber\\
&\times& \frac{(\hat{m}^8_h + s^4 + t^4 + u^4)(1-2\epsilon) + \frac 1 2 \epsilon\, (\hat{m}^4_h + s^2 + t^2 + u^2)^2}{s\, t\, u}  \tilde{C}_{HG}.
\eea

Dropping higher order terms and
using the BFM cancelation $\mu^{2 \epsilon} \, Z_g^2 Z_{\hat{\mathcal{G}}} \equiv 1$
simplifies the result. Multiplying by $d$ dimensional phase space,
\bea
d\Phi_2 \equiv \frac{1}{8 \pi} S^\epsilon \, e^{\epsilon \gamma_E} \frac{1}{\Gamma(1- \epsilon)} \left(\frac{1}{s}\right)^\epsilon \left[1- \frac{\hat{m}_h^2}{s}\right]^{1- 2 \epsilon}
\int_0^1 \omega^{- \epsilon} (1- \omega)^{- \epsilon} d \, \omega
\eea
and performing the color averaging and polarization sums yields
\bea
\sigma(\mathcal{G} \,\mathcal{G} \to h \mathcal{G}) = \frac{1}{512 \, s \, (1- \epsilon)^2} |\mathcal A(\mathcal{G} \,\mathcal{G} \to h \mathcal{G})|^2 d\Phi_2.
\eea
Explicitly, while using the definitions in Appendix~\ref{plusdist} for the plus distributions
we find
$\Delta^2 \delta \sigma(\mathcal{G} \,\mathcal{G} \to h \mathcal{G})_{IR}/\hat{\sigma}^{SM}_{LO, \epsilon \rightarrow 0}(\mathcal{G} \,\mathcal{G} \to h)$
is given by
\bea
&6& \! \! \! \!
\left[\frac{6}{\epsilon^2} -6 \frac{L_+}{\epsilon}
+ \frac{6}{\epsilon}  + 3 \, L_+^2 - 6 \, L_+ - \pi^2 + 6 \right] \delta(1-z) \tilde{C}_{HG} \nn
&+6&  \left[\left(12 \, f_1(z) \, (L_{\hat{m}_h} - \log(z)) - 11 f_1(z) + 11 z \right) f_1(z)
+ 11 \, (1-z)^2 \,z \right] \left(\frac{1}{1-z}\right)_+ \! \! \tilde{C}_{HG} \nn
&+& 144 \, f^2_1(z) \, \left(\frac{\log(1-z)}{1-z}\right)_+ \,  \tilde{C}_{HG} - 72 \,f^2_1(z) \left[\frac{1}{\epsilon} + 1 - L_{\hat{m}_t}\right] \left(\frac{1}{1-z}\right)_+ \! \! \! \tilde{C}_{HG}.
\eea

Here the distributions of the numerator have been included again that were suppressed in
Eqn.~\eqref{combinedresultUV}. Replacing the $1/(1-z)_+$ distribution in favor
of the Altarelli-Parisi (AP) splitting function via Eqn.~\eqref{APfunction} results in
\bea
&6& \! \! \! \!
\left[\frac{6}{\epsilon^2} + \frac{\beta_0}{\epsilon} -6 \frac{L_+}{\epsilon}
+ \frac{6}{\epsilon}  + 3 \, L_+^2 - 6 \, L_+  - \pi^2 + 6 \right] \delta(1-z) \tilde{C}_{HG}
-  72 \,f^2_1(z) \left[1 - L_{\hat{m}_t}\right] \left(\frac{1}{1-z}\right)_+ \, \tilde{C}_{HG}\nn
&+& 6 \left[\left(12 \, f_1(z) \, (L_{\hat{m}_h}  - \log(z)) - 11 f_1(z) + 11 z \right) f_1(z)
+ 11 \, (1-z)^2 \,z \right] \left(\frac{1}{1-z}\right)_+ \! \! \tilde{C}_{HG} \nn
&+& 144 \, f^2_1(z) \, \left(\frac{\log(1-z)}{1-z}\right)_+ \,  \tilde{C}_{HG} - 36 \, z \, p_{\mathcal{G}\mathcal{G}}(z) \, \left[\frac{1}{\epsilon}\right] \, \tilde{C}_{HG}.
\eea
We follow the splitting functions conventions of Ellis-Stirling-Webber \cite{ellis_stirling_webber_1996}
and introduce a counter term to remove the residual $1/\epsilon$
\bea
\Delta^2 \delta \sigma_{DR \, c.t}^{AP} &\equiv&36\, \Delta^2 \hat{\sigma}^{SMEFT}_{LO, \epsilon \rightarrow 0}(\mathcal{G} \,\mathcal{G} \to h) \left[\left(\frac{\mu^2}{\mu_F^2}\right)^\epsilon \right] (4 \pi)^\epsilon \frac{\Gamma(1+ \epsilon) \Gamma(1- \epsilon)^2}{\Gamma(1- 2 \epsilon)} \left[\frac{1}{\epsilon} \right]  \, z \, p_{\mathcal{G}\mathcal{G}}(z) \tilde{C}_{HG} \nn
\eea
where $\mu_F$ is a low renormalization scale for the Altarelli-Parisi splitting function, while
$\mu$ is the higher renormalization scale introduced for renormalizing the SMEFT perturbations.
Comparing to the literature, the Altarelli-Parisi function and counter term conventions differ between
references, in particular Ref.~\cite{Dawson:1990zj,ellis_stirling_webber_1996,Maltoni:2018dar,Deutschmann:2017qum}.
At times, conventions/schemes are unspecified.

The counter term is introduced proportional to the leading order SMEFT$\times$SM interference, as it must be proportional to $\tilde{C}_{HG}$. Formally,
the resulting splitting function is a SMEFT correction to the SM splitting function, since it
depends on the Wilson coefficient $\tilde{C}_{HG}$.
The introduction of the splitting function represents the factorization of
the long and short distance physics proportional to $\tilde{C}_{HG}$.
It is possible to modify the counterterm introduced via
the replacement $1/\epsilon \rightarrow 1/\epsilon + 1 - L_{\hat{m}_t}$. This choice
simplifies the final answer obtained, removing all $L_{\hat{m}_t}$ dependence.
As the evaluation of the resulting perturbation
of the cross section is done in fixed order perturbation theory with $\mu \sim m_h$,
the scale $\mu$ in the SMEFT AP counterterm is in the end set to a large renormalization scale.
Here we forgo this simplification of the final results, and retain an explicit factor of $1 - L_{\hat{m}_t}$.

We also note that an alternate calculational scheme convention for dipole subtraction to address $L_{\hat{m}_t}$ dependence is used in Ref.~\cite{Buchalla:2022igv},
based on Ref.~\cite{Gehrmann-DeRidder:2005btv}. Essentially, this is a rearrangement of $L_{\hat{m}_t}$ in intermediate steps of the calculation
Our use of results from Refs.~\cite{Anastasiou:2006hc,Anastasiou:2020qzk} to define
the SMEFT corrections to the cross section is similar to (but distinct from) Ref.~\cite{Gehrmann-DeRidder:2005btv,Buchalla:2022igv}
in intermediate steps, in that the Catani one loop IR operator multiplies
the full $\epsilon$ series with $L_{\hat{m}_t}$ dependence for the SM leading order cross section.
The net real emission result is consistent with some past literature, including Refs.~\cite{Djouadi:1991tka,Buchalla:2022igv}
once schemes and calculational conventions are appropriately accounted for.

The final result with $L_+$ is
\bea
&6& \! \! \! \!
\left[\frac{6}{\epsilon^2} + \frac{\beta_0}{\epsilon} -6 \frac{L_+}{\epsilon}
+ \frac{6}{\epsilon}  + 3 \, L_+^2 + \beta_0(1 - L_{\hat{m}_t}) - 6 \, L_+ - \pi^2 + 6 \right] \delta(1-z) \tilde{C}_{HG}\nn
&+& 6 \left[\left(12 \, f_1(z) \, (L_{\hat{m}_h}  - \log(z)) - 11 f_1(z) + 11 z \right) f_1(z)
+ 11 \, (1-z)^2 \,z \right] \left(\frac{1}{1-z}\right)_+ \! \! \tilde{C}_{HG} \nn
&+& 144 \, f^2_1(z) \, \left(\frac{\log(1-z)}{1-z}\right)_+  \tilde{C}_{HG}
+ 36 \, z  \, \log \left(\frac{\hat{\mu}^2}{\mu_F^2} \right) \,  p_{\mathcal{G}\mathcal{G}}(z) \, \tilde{C}_{HG}() \nn
&-&  72 \,f^2_1(z) \left[1 - L_{\hat{m}_t}\right] \left(\frac{1}{1-z}\right)_+ \, \tilde{C}_{HG} + \mathcal{O}(\epsilon).
\eea
\subsubsection{Combined NLO $\tilde{C}^{(6)}_{HG}$ result}

Combining the virtual and real emission, the poles in $\epsilon$ and the log squared terms
exactly cancel out. The final result is quite compact
\bea\label{SMEFTNLO}
\frac{\Delta^2 \delta \sigma^{SMEFT}}{\Delta^2 \,\hat{\sigma}^{SM}_{LO, \epsilon \rightarrow 0}} \frac{1}{{2} \,\tilde{C}^{(6)}_{HG}}
&=& 12\left[\pi^2 + \frac{11}{2}  \right] \, \delta(1-z)
-66 (1-z)^3  + 144 \, f^2_1(z)  \, \left(\frac{\log(1-z)}{1-z}\right)_+ \nn
&+&  72 \, f^2_1(z) \, \left[L_+ - \log \left(z\right)  - 1\right] \, \left(\frac{1}{1-z}\right)_+
+36 \, z \,  p_{\mathcal{G}\mathcal{G}}(z) \, \log \left(\frac{\hat{\mu}^2}{\mu_F^2} \right). \nn
\eea
This expression is understood to define the numerical rescaling required to generate the NLO result from the numerical value
of the SM cross section. The limit $\epsilon \rightarrow 0$ is thus already taken in determining the SM result, and the distribution in $z$
is averaged over the parton distribution functions in the SM result.

The full NLO results are different than those reported in Ref.~\cite{Corbett:2021cil}
and should be understood to supersede those results. The improvements of the calculation are multifold.
The full $\epsilon$ dependence results reported in Ref.~\cite{Anastasiou:2020qzk}
leads to modification of finite terms due to cross terms in the $1/\epsilon$ series and the top sub-loop
used in the $m_t \rightarrow \infty$ limit. The calculation in Ref.~\cite{Corbett:2021cil} used the
two loop matching calculation onto the heavy top quark effective operator, but this approximation
to the full NLO matrix element in the $m_t \rightarrow  \infty$ limit
cannot capture a full set of log terms (and descendent finite terms) that are numerically
relevant.
In addition, the calculation in  Ref.~\cite{Corbett:2021cil} neglected
the full two loop matrix element interference with the tree level SMEFT operator insertion,
only retaining a (poor) approximation of the one loop SM amplitude for interfering with
the one loop contribution to the QCD matrix element proportional to the
operator $C_{HG}$. Further, the cross section result
\bea
\Delta^2 \delta \sigma^{SMEFT}(\mathcal{G} \,\mathcal{G} \to h) \propto\left(\frac{\alpha_s}{4 \pi^2}\right)^2\tilde{C}_{HG},
\eea
has two contributions. The interference of
\bea
\langle \mathcal{G} \mathcal{G}|h \rangle^2_{SM} \, \times \, \langle \mathcal{G} \mathcal{G}|h \rangle^0_{\tilde{C}_{HG}},
\eea
and the interference of
\bea
\label{eq:NLOcHG}
\langle \mathcal{G} \mathcal{G}|h \rangle^1_{SM} \, \times \, \langle \mathcal{G} \mathcal{G}|h \rangle^1_{\tilde{C}_{HG}}.
\eea
In the $m_t \rightarrow \infty$ limit, the leading order result for
$\mathcal{G} \,\mathcal{G} \to h$ in the SM follows from the same local contact operator
that receives an additive SMEFT contribution from $\tilde{C}_{HG}$.
Each contribution to the cross section
can be built up with the full NLO virtual amplitude results in Ref.~\cite{Anastasiou:2020qzk}.
As the same local contact operator is present in the virtual NLO result, and the $\epsilon$
expansion of the SM leading order sub-diagram is properly accounted for, the two loop
result ``descends one loop order" in the $m_t \rightarrow \infty$ limit so that
\bea
\lim_{m_t \rightarrow \infty} \langle \mathcal{G} \mathcal{G}|h \rangle^2_{SM} \frac{\tilde{C}_{HG}}{\bar{v}_T^0 \, \Delta C^{SM}_{h \mathcal{G} \mathcal{G}}}
= \langle \mathcal{G} \mathcal{G}|h \rangle^1_{\tilde{C}_{HG}}.
\eea
Similarly,
\bea
\lim_{m_t \rightarrow \infty} \langle \mathcal{G} \mathcal{G}|h \rangle^1_{SM}
\equiv \langle \mathcal{G} \mathcal{G}|h \rangle^0_{\tilde{C}_{HG}}  \, \times \, \frac{\bar{v}_T^0 \,\Delta \, C^{SM}_{h \mathcal{G} \mathcal{G}}}{\tilde{C}_{HG}}.
\eea
The rescaling differences in each of these individual expressions cancel in the interference
of the virtual terms. The real emission result is determined at the
amplitude squared level in Eqn.~\eqref{eq:ggg2}, where the same rescaling relationship is present in the
$m_t \rightarrow \infty$ limit.
Combining the two sets of interference terms with their corresponding real emission
results leads to the overall factor of $2$ on the left hand side of Eqn.~\eqref{SMEFTNLO}.\footnote{
Note that the real emission result cancels the IR divergences in both of the interference terms
leading to the $1/2$ in Eqn.~\eqref{eq:ggg3}.}.
Such relationships between results is an example of the utility of the EFT approach.
Equation.~\eqref{SMEFTNLO} needs to be added to the terms in Appendix \ref{past}
taken (unchanged) from Ref.~\cite{Corbett:2021cil} and reiterated here for completeness
to build up the full NLO result.

The results of Ref.~\cite{Corbett:2021cil}, like most SMEFT literature, report results in a
mixed $\rm \overline{MS}$ like scheme with on-shell renormalization of $\alpha_s$ combined with BFM calculational scheme results.
Here we calculate in a consistent fashion in the BFM, and report the first complete
calculation of this form in the literature in the SMEFT for this process (to our knowledge)
in the $\rm \overline{MS}$ scheme.
The factorization of the results into a AP splitting function still requires
the introduction of a counterterm explicitly introducing dependence on the lower $\mu_F$
scale in the process. The $\beta_0$ dependent log proportional to $\delta (1-z)$ is absent
as the counterterms of the background field gluon wavefunction renormalization cancels against
the renormalization of the strong coupling. This scheme dependence, and the ability to rewrite distribution terms
using Eqn.~\eqref{usefulID} should be noted when comparing results in
differing schemes.

\subsection{$\Gamma(h \rightarrow \mathcal{G} \mathcal{G})$}
The matrix elements for $h \to \mathcal{G} \,\mathcal{G} $ and
$\mathcal{G} \,\mathcal{G} \to h$ are related by crossing symmetry.
As in the case of Higgs production, the $\mathcal O(\alpha^2_s)$ interference contributions for this decay are
\bea
\langle h |\mathcal{G} \mathcal{G} \rangle^2_{SM} \, \times \, \langle h| \mathcal{G} \mathcal{G}\rangle^0_{\tilde{C}_{HG}},
\eea
and
\bea
\langle h| \mathcal{G} \mathcal{G} \rangle^1_{SM} \, \times \, \langle h| \mathcal{G} \mathcal{G} \rangle^1_{\tilde{C}_{HG}}.
\eea
Also analogously to  $\mathcal{G} \,\mathcal{G} \to h$, there are IR divergences
cancelled by $h \to \mathcal{G} \,\mathcal{G} \,\mathcal{G}$ -- determined from the same
matrix elements reported in previous sections via crossing symmetry -- with additional contributions from $h \to \mathcal{G} \bar q q$ in the soft-collinear limit.

Some contributions to this decay are unchanged from the results in Ref.~\cite{Corbett:2021cil}
and are reiterated in Appendix \ref{pasthgg}. Here, we focus on presenting the differences compared to past
results due to the full two loop SM matrix elements now incorporated.
We uniform calculation conventions in our $\Gamma(h \rightarrow \mathcal{G} \,\mathcal{G})$
results with those in Section \ref{sigmaggh}. The two loop amplitudes explicitly presented in Ref.~\cite{Anastasiou:2020qzk}
are the key SM input, as in Section \ref{sigmaggh}.

The leading order results for the decay $\Gamma(h \rightarrow \mathcal{G} \,\mathcal{G})$
follows from $C_{h\mathcal{G} \,\mathcal{G}}^{SMEFT}$, with the decay width depending on
this Wilson coefficient as
\bea
\Gamma^{SMEFT}(h \to \mathcal{G} \,\mathcal{G}) \equiv \frac{2 \, \hat{m}_h^3}{\pi} \, |C_{h\mathcal{G} \,\mathcal{G}}^{SMEFT}|^2,
\eea
leading to \cite{Wilczek:1977zn,Georgi:1977gs}:
\begin{align*}
\Delta^2 \, \Gamma^{SM}_{m_t \rightarrow \infty}(h \to \mathcal{G} \,\mathcal{G}) &\equiv \frac{(\alpha_s^{(r)})^2 \, \hat{m}_h^3}{72 \, \pi^3 \hat{v}_T^2}, \quad &
\Delta \delta \, \Gamma(h \to \mathcal{G} \,\mathcal{G}) &\equiv \frac{\alpha_s^{(r)} \, \hat{m}_h^3 \, \tilde{C}_{HG}}{3 \, \pi^2 \hat{v}_T^2}.
\end{align*}

\subsubsection{$\Gamma(h \to \mathcal{G} \,\mathcal{G})$ Virtual terms}

We organize the NLO contributions as in the case of $ \mathcal{G} \mathcal{G} \rightarrow h$,
defining
\bea\label{twoloopmatch}
\langle h| \mathcal{G} \mathcal{G} \rangle^{2,F}_{SM} =
\frac{\alpha^{(r)}_s}{4 \pi}
\, \left[11 + c_1 \, \epsilon + (- \beta_0 + c_2 \, \epsilon) \log \left(- \frac{\hat{m}_h^2}{\hat{\mu}^2}\right) \right]
\langle h| \mathcal{G} \mathcal{G} \rangle^1_{SM,\epsilon \rightarrow 0},
\eea
leading to the contribution
\bea\label{polelimitNLOfinitedecay}
\frac{\Delta^2 \delta \Gamma(h \to \mathcal{G} \,\mathcal{G})_F}{\Delta^2 \hat{\Gamma}^{SM}_{LO}(h \to \mathcal{G} \,\mathcal{G})}
&=&
\frac{\alpha_s^{(r)}}{2 \pi}\left(11 - \beta_0 \, L_{\hat{m}_h}  \right)
\frac{\tilde{C}^{(6)}_{HG}}{\bar{v}^0_T \, \Delta C^{SM}_{h \mathcal{G} \mathcal{G}}}, \nonumber \\
&=& 6 \, \left(11 - \beta_0 \, L_{\hat{m}_h} \right)
\tilde{C}^{(6)}_{HG}.
\eea
The $\epsilon$ terms in $c_1, c_2$ again interfere and generate constant finite terms
\bea\label{polelimitNLO2finitedecay}
\frac{\Delta^3 \delta \Gamma(h \to \mathcal{G} \,\mathcal{G})_F}{\Delta^2 \, \hat{\Gamma}^{SM}_{LO}(h \to \mathcal{G} \,\mathcal{G})}
&=&
- \frac{3 \beta_0 \,\alpha_s^{(r)}}{2 \pi} \left({\rm Re}[c_1] + {\rm Re}[c_2] \, L_{\hat{m}_h}
+ \frac{3 \, \pi^2 \, \beta_0}{2}\right)
\tilde{C}^{(6)}_{HG}.
\eea
The net renormalization (using $\rm \overline{MS}$ for the top mass dependence)
again introduces a contribution to the cross section
\bea\label{polelimitrendecay}
\frac{\Delta^2 \delta \Gamma(h \to \mathcal{G} \,\mathcal{G})_{ren}}{\Delta^2 \, \hat{ \Gamma}^{SM}_{LO}(h \to \mathcal{G} \mathcal{G})}
&=& - 6 \beta_0 \left[\frac{1}{\epsilon} +1 - L_{\hat{m}_t} - L_{\hat{m}_h} \right] \, \tilde{C}^{(6)}_{HG}
+ 36 \, C_F \, \tilde{C}^{(6)}_{HG},
\eea
with an additional factor of $L_{\hat{m}_h}$ (compared to Eqn.~\eqref{polelimitren}) due to the $d$-dimensional two body phase space $d\Phi_2$.

\subsubsection{$\Gamma(h \to \mathcal{G} \,\mathcal{G})$ Real emission terms}

The interference of the two loop, scheme dependent terms with the tree level insertion of
$\tilde{C}_{HG}$ gives
\bea
\frac{\Delta^2 \delta \Gamma(h \to \mathcal{G} \,\mathcal{G})_{sch}}{\Delta^2 \, \hat{\Gamma}^{SM}_{LO}(h \to \mathcal{G} \mathcal{G})}
&=& 6 \, \tilde{C}_{HG} \, \left[-\frac{6}{\epsilon^2} + \frac{6 (L_+ + L_{\hat{m}_h}-1)}{\epsilon}
- 6 L_{\hat{m}_h}^2 -6 L_+^2 +3  L_{\hat{m}_t}^2 \right] \nonumber \\
&+& 6 \, \tilde{C}_{HG} \, \left[ 6  L_{\hat{m}_h} + 6  L_+ + \beta_0  L_{\hat{m}_h} + \frac{9 \pi^2}{2} - 20\right]
\eea
Leading to a net virtual interference result
\bea
\frac{\Delta^2 \delta \Gamma(h \to \mathcal{G} \,\mathcal{G})}{\Delta^2 \, \hat{\Gamma}^{SM}_{LO}(h \to \mathcal{G} \mathcal{G})}
&=& 12 \, \tilde{C}_{HG} \, \left[-\frac{6}{\epsilon^2} + \frac{6 (L_+ + L_{\hat{m}_h}-1) - \beta_0}{\epsilon}
- 6 L_{\hat{m}_h}^2 -6 L_+^2 +3  L_{\hat{m}_t}^2 \right] \nonumber \\
&+& 12 \, \tilde{C}_{HG} \, \left[ - \beta_0 + 6  L_{\hat{m}_h} + (6 + \beta_0) L_+  + \frac{9 \pi^2}{2} - 1
  \right].
\eea

The real emission contributions are a combination of $h \to \mathcal{G} \,\mathcal{G}\,\mathcal{G}$
in the soft limit and $h \to \mathcal{G} \,\bar{q} \,q$ in the collinear limit.
The former contribution is
\bea
\frac{\Delta^2 \delta \Gamma(h \to \mathcal{G} \,\mathcal{G} \,\mathcal{G})_{soft}}{\Delta^2 \, \hat{\Gamma}^{SM}_{LO}(h \to \mathcal{G} \mathcal{G})}
&=& 12 \, \tilde{C}_{HG} \, \left[\frac{6}{\epsilon^2} + \frac{6 (L_{\hat{m}_t} - 2 \, L_{+} +1) + 11}{\epsilon}
- \frac{9 \pi^2}{2} + \frac{119}{2} \right] \nonumber \\
&+& 12 \, \tilde{C}_{HG} \, \left[3 L_{\hat{m}_t}^2 + 17 L_{\hat{m}_t} - 12 L_{\hat{m}_t} L_++ 12 L_+^2 - 34 L_+ \right],
\eea
while the later is
\bea
\frac{\Delta^2 \delta \Gamma(h \to \mathcal{G} \, \bar{q} \,q)_{col.}}{\Delta^2 \, \hat{\Gamma}^{SM}_{LO}(h \to \mathcal{G} \mathcal{G})}
&=& 12 \, \tilde{C}_{HG} \, \left[ - \frac{2 \, N_F}{3 \epsilon} + \frac{N_F}{3} \left(4 L_+ - 2 L_{\hat{m}_t} -9\right)\right].
\eea
Combining all terms we find
\bea
\frac{\Delta^2 \delta \Gamma(h \to \mathcal{G} \, \mathcal{G})}{\Delta^2 \, \hat{\Gamma}^{SM}_{LO}(h \to \mathcal{G} \mathcal{G})}
&=& 12 \, \tilde{C}_{HG} \, \left[\frac{95}{2} - \frac{7 N_f}{3} - \beta_0 L_{\hat{m}_h}\right].
\eea
This result is consistent with the SM NLO result reported in Refs.~\cite{Inami:1982xt,Djouadi:1991tka,Chetyrkin:1997iv}.
See also Refs.~\cite{Contino:2014aaa,Buchalla:2022igv}.
\subsection{$\Gamma(h \rightarrow \mathcal{A} \mathcal{A})$ two loop QCD corrections}
The decay width is given by
\bea
\Gamma(h \rightarrow \mathcal{A} \mathcal{A}) &\simeq& \frac{\hat{m}_h^3}{4 \pi}
|\langle h|\mathcal{A} \mathcal{A}\rangle_{SM}^1+ \langle h|\mathcal{A} \mathcal{A}\rangle_{SM}^2
+ \langle h|\mathcal{A} \mathcal{A}\rangle^0_{\mathcal{O}(\bar{v}_T^2/\Lambda^2)} + \langle h|\mathcal{A}\mathcal{A} \rangle_{{\cal{O}}(\bar{v}_T^4/\Lambda^4)}^0 + \langle h|\mathcal{A} \mathcal{A}\rangle^1_{{\cal{O}}(\bar{v}_T^2/\Lambda^2)}|^2 \nn
\eea
All of the contributing terms except $\langle h|\mathcal{A} \mathcal{A}\rangle_{SM}^2$ were defined
in Ref.~\cite{Corbett:2021cil}. We reiterate these results in Appendix \ref{pasthaa} to make the paper self contained.
For example, the leading order result \cite{Ellis:1975ap,Shifman:1979eb,Bergstrom:1985hp}
is defined with the notation ($\tau_p = 4 m_p^2/\bar{m}_h^2$)
\begin{align}
&\langle h|\mathcal{A} \mathcal{A}\rangle_{SM}^1 = \frac{- \hat{g}_2 \, \hat{e}^2}{64 \, \pi^2 \, \hat{m}_W}
\Bigg(A_1(\tau_W)+ \sum_i \, N_c^i \,Q^2_i \, A_{1/2} (\tau_{\psi^i})\Bigg) \, \langle h \mathcal{A}^{\mu\nu} \mathcal{A}_{\mu \nu} \rangle^0,
\end{align}
The two loop QCD corrections we add in this work are reported in Refs.~\cite{Spira:1995rr,Fleischer:2004vb,Harlander:2005rq,Aglietti:2006tp}.
The QCD corrections are given by
\bea
\langle h|\mathcal{A} \mathcal{A}\rangle_{SM}^2 = \frac{- \hat{g}_2 \, \hat \alpha^{(r)}\,  \hat{e}^2}{64 \, \pi^3 \, \hat{m}_W} \sum_i \, N_c^i \,Q^2_i \, A_{1/2}(\tau_p) \, \left[C_1^H(\tau_p) + C_2^H(\tau_p) \log \left(\frac{4 \, \hat{\mu}^2}{\tau_p \hat m_h^2} \right)\right] \, \langle h \mathcal{A}^{\mu\nu} \mathcal{A}_{\mu \nu} \rangle^0, \nn
\eea
where
\bea
A_{1/2}(\tau_p) \, C_2^H(\tau_p) \equiv 4 \tau_p\, \left[1 + (1 - 2 \tau_p)f(\tau_p) + \tau_p \,(1- \tau_p)\, d\, f(\tau_p)/d \tau_p \right],
\eea
and $A_{1/2}(\tau_p) \, C_1^H(\tau_p)$ is lengthy and directly given in Ref.~\cite{Harlander:2005rq}.
Note that our definition of $\tau_p$ is the inverse of the definition used in Ref.~\cite{Harlander:2005rq}.
Numerically, we update the SM result including these corrections, thereby retaining
the corresponding $\propto \Delta^2 \delta$ interference terms
\bea
\langle h|\mathcal{A} \mathcal{A}\rangle_{SM}^2 \times \langle h|\mathcal{A} \mathcal{A}\rangle^0_{\mathcal{O}(\bar{v}_T^2/\Lambda^2)}
\eea
in the expression for $\Gamma^{SMEFT}(h \rightarrow \mathcal{A} \mathcal{A})/\Delta^2 \Gamma^{SM}(h \rightarrow \mathcal{A} \mathcal{A})$.
\subsection{$\Gamma(h \to \bar{\Psi} \,\Psi)$}
Defining the coupling of the Higgs to fermions with flavors $p,r$ as
\begin{equation}\label{hcouplings.def}
 \Lagr_{h,eff} =  - g_{\substack{h \psi \\ pr}} \, h \, \bar{\psi}_{\substack{R\\p}} \psi_{\substack{L\\r}} + h.c.
\end{equation}
the decays to $\psi = \{u,c,d,s,b,e,\mu,\tau\}$ are modified in the $\Delta, \delta$ expansions as
 \begin{align}
\bar{\Gamma} \left(h \to  \bar{\psi}_p \psi_p \right) &= \frac{\left|g_{\substack{h \psi\\pp}}^{SM} + \delta g_{\substack{h \psi\\pp}} +  \Delta g_{\substack{h \psi\\pp}} + \delta^2 g_{\substack{h \psi\\ppr}} + \delta \Delta g_{\substack{h \psi\\pp}} + \cdots \right|^2}{8 \pi \, |g_{\substack{h \psi\\pp}}^{SM}|^2} \, N_C^\psi \, \bar{M}_h \, \sqrt{2} \hat{G}_F \hat{M}_{\psi}^2
\beta^3,
 \end{align}
where $\beta \equiv \left(1- 4\bar M_\psi^2/\bar M_h^2\right)^{1/2}$.
The pole masses of quarks and leptons inferred from experimental results define input parameters $\hat{M}_\psi$
and determine the SM Yukawa couplings through the definition
\bea\label{Yhat}
 \hat Y_\psi &=& 2^{3/4}\hat M_\psi \sqrt{\hat G_F}.
\eea
When all SM parameters are defined via a particular input parameter scheme, we denote $\bar{\Gamma} \rightarrow \hat{\Gamma}$.
Known results are
\bea
g_{\substack{h \psi\\ pr}}^{SM} &=&  \delta_{pr} \hat Y_{\substack{\psi\\pr}}/\sqrt2,  \\
\d g_{\substack{h \psi\\pr}} &=&\frac{\hat Y_{\substack{\psi \\pr}}}{\sqrt{2}}\left[\ckin^{(6)} - \frac{\d G^{(6)}_F}{\sqrt{2}} \right]- \frac{1}{\sqrt2 } \,\tilde C^{*,(6)}_{\substack{\psi H \\ pr}} \,.
\eea
The geoSMEFT results in Ref.~\cite{Helset:2020yio,Hays:2020scx} lead directly to
\bea
\d^2 g_{\substack{h \psi\\pr}} &=&\frac{\hat Y_{\substack{\psi \\pr}}}{\sqrt{2}}\left[\ckin^{(8)} - \ckin^{(6)} \, \frac{\d G^{(6)}_F}{\sqrt{2}} + (\frac{\d G^{(6)}_F}{\sqrt{2}})^2 - \frac{\d G^{(8)}_F}{\sqrt{2}} \right]- \frac{1}{\sqrt2 } \,\tilde C^{*,(8)}_{\substack{\psi H \\ pr}} \nonumber \\
&-& \frac{1}{\sqrt2 } \, \left[\ckin^{(6)} - \frac{\d G^{(6)}_F}{\sqrt{2}} \right] \,\tilde C^{*,(6)}_{\substack{\psi H \\ pr}}.
\eea
Note that, in the $\rm U(3)^5$ limit, $\tilde C^{*,(6),(8)}_{\substack{\psi H \\ pr}}$ are proportional to $Y_{\substack{\psi\\ pr}}$.
The appearance of the shift in the measured value of the vev in muon decay, compared to the Lagrangian parameter is $\d G^{(6)}_F$,$\d G^{(8)}_F$.
The appearance of this shift at tree level is consistent in the dependence introduced due to the vev shift in the loop level SM decays
via Eqn.~\eqref{eq:cifi}.

For the SM decay at one loop (in QCD corrections), the results are given in Ref.~\cite{Braaten:1980yq}
in the limit $\beta \rightarrow 1$ (and neglecting subleading effects further suppressed by $m_{\psi}$). Specifically, \cite{Braaten:1980yq,Gauld:2016kuu}
\bea
\Delta g_{\substack{h \psi\\pp}} &\supset& g_{\substack{h \psi\\pp}}^{SM} \frac{\alpha_s^{(r)} \, C_F}{8 \pi} \left(17 + 6 \log \left(\frac{\hat{\mu}^2}{m_h^2}\right)\right), \\
\delta \Delta g_{\substack{h \psi\\pp}} &\supset& \Delta g_{\substack{h \psi\\pp}} \, \d g_{\substack{h \psi\\pp}}.
\eea
The universal EW corrections to the vev extraction are also given by $\Delta g_{\substack{h \psi\\pp}}\supset - g_{\substack{h \psi\\pp}}^{SM} \Delta G_F$.
$\Delta G_F$ is defined in Eqn.~\eqref{vevoneloop}.
This leads to the simple expression for $p=r$
\begin{align}
\frac{\delta {\Gamma}_{h \to  \bar{\psi} \psi}}{\hat{\Gamma}^{SM}_{h \to  \bar{\psi} \psi}}
&= 1+ 2 \, {\rm Re}\left(\delta g_{\substack{h \psi\\pp}} \right) + \frac{2 \, {\rm Re}\left(\delta^2 g_{\substack{h \psi\\pp}}\right)}{\Delta g_{\substack{h \psi\\pp}}} + \cdots
\end{align}
for the decays to $\psi = \{u,c,d,s,b,e,\mu,\tau\}$. Non-factorizable corrections are present in the last
term and also introduce $\delta \Delta$ effects through operator mixing. These corrections are
relatively suppressed by powers of $m_\psi$.

\begin{center}
\begin{table}
\centering
\tabcolsep 8pt
\begin{tabular}{|c|c|c|c|}
\hline
Input parameters&Value&onshell mass&Ref.\\
\hline
$\hat m_Z$ [GeV]&$91.1876\pm0.0021$&&\cite{Workman:2022ynf}\\
$\hat m_W$ [GeV]&$80.387\pm0.016$&&\cite{Aaltonen:2013iut}\\
$\hat m_h$ [GeV] &$125.15\pm0.15$&&\cite{Workman:2022ynf}\\
$\hat m_t$ (MC/onshell) [GeV] &$172.69\pm0.3$& &\cite{Workman:2022ynf}\\
$\hat{m}_b$ (msbar) [GeV]&     $4.18\pm0.03$ & $4.92$& \cite{Workman:2022ynf,Bauer:2004ve, LHCHiggsCrossSectionWorkingGroup:2016ypw}\\
$\hat{m}_c$ (msbar)[GeV]&     $1.27\pm0.02$   &$1.51$ & \cite{Workman:2022ynf,Bauer:2004ve,LHCHiggsCrossSectionWorkingGroup:2016ypw}\\
$\hat{m}_d$ (curr.-msbar)[MeV]&     $4.67\pm0.48$   &  & \cite{Workman:2022ynf}\\
$\hat{m}_s$ (curr.-msbar)[MeV]&     $93.4\pm8.6$   & 100 &  \cite{Workman:2022ynf,LHCHiggsCrossSectionWorkingGroup:2016ypw}\\
$\hat{m}_u$ (curr.-msbar)[MeV]&     $2.16\pm 0.49$   &  & \cite{Workman:2022ynf}\\
$\hat{m}_\tau$ (pole) [GeV]&  $1.77686\pm0.00012$   &  &   \cite{Workman:2022ynf}\\
$\hat{m}_\mu$ (pole) [MeV]&  $105.6583755\pm0.0000023$   & & \cite{Workman:2022ynf}\\
$\hat{m}_e$ (pole) [MeV]&  $0.510-\pm1.5\times 10^{-10}$   & & \cite{Workman:2022ynf}\\
$\hat{G}_F$ [GeV$^{-2}$] & 1.166 $\cdot 10^{-5}$&&  \cite{Olive:2016xmw,Mohr:2012tt} \\
$\hat \alpha_{EW}$&1/137.03599084(21)&&\cite{Workman:2022ynf}\\
$\nabla\alpha$&$0.0590\pm0.0005$&&\cite{Dubovyk:2019szj}\\
$\hat \alpha_s$&$0.1179\pm0.0010$&&\cite{Workman:2022ynf}\\
\hline
$m_W^{\hat\alpha}$&$80.36\pm0.01$&&--\\
$\nabla\alpha^{\hat m_W}$&$0.0576\pm0.0008$&&--\\
\hline
\end{tabular}
\caption{Input parameter values used.
$m_W^{\hat\alpha}$ is the value of $m_W$ inferred in the $\{\hat\alpha,\hat m_Z,\hat G_F\}$ scheme
using the interpolation formula of Refs.~\cite{Freitas:2014hra,Awramik:2003rn,Awramik:2006uz,Dubovyk:2019szj},
while $\Delta\alpha^{\hat m_W}$ is the shift in the
value of alpha due to hadronic effects for the $\{\hat m_W,\hat m_Z,\hat G_F\}$ scheme.
The on-shell masses used for the numerical evaluations to be consistent with past
literature conventions are also listed.}
\label{tab:inputs}
\end{table}
\end{center}

\section{Scheme choice and Numerics}
We report numerical results for $\sigma(\mathcal{G} \mathcal{G} \rightarrow h)$,
$\Gamma(h \rightarrow \mathcal{A} \mathcal{A})$, and $\Gamma(h \rightarrow \mathcal{G} \mathcal{G})$.
As SMEFT corrections are determined to higher orders in the operator and
perturbative expansions, scheme dependence becomes a more relevant issue of concern
for numerical accuracy.
Scheme dependence comes in three forms in the SMEFT: operator basis dependence,
perturbative/renormalization scheme dependence, and input parameter dependence.
\begin{table}[t]\centering
\renewcommand{\arraystretch}{1.2}
 \begin{tabular}{cc|cc|cc}\hline
 $\Delta R^{\hat{m}_W}_{\mathcal{A}}$ & 0.12 & $\Delta R^{\hat{\alpha}_{ew}}_{\mathcal{A}}$ & 0.12 & $\Delta R^{\hat{\alpha}_{ew}(0)}_{\mathcal{A}}$ & 0.13 \\
 $\Delta G_F^{\hat{m}_W}$ & 0.024 & $\Delta G_F^{\hat{\alpha}_{ew}}$ & 0.024 & $\Delta G_F^{\hat{\alpha}_{ew}(0)}$ & 0.024\\
 $\Delta R_{M^2_W}^{\hat{m}_W}$ & -0.041& $\Delta R_{M^2_W}^{\hat{\alpha}_{ew}}$ &-0.041 & $\Delta R_{M^2_W}^{\hat{\alpha}_{ew}(0)}$ &-0.041\\
 $\Delta R_{M^2_Z}^{\hat{m}_W}$ & -0.055  & $\Delta R_{M^2_Z}^{\hat{\alpha}_{ew}}$ &-0.055 & $\Delta R_{M^2_Z}^{\hat{\alpha}_{ew}(0)}$ &-0.055\\
 $\frac{\Delta R_{\phi_4}^{\hat{m}_W}}{2}+ \frac{\Delta v}{v}$ & -0.003 & $\frac{\Delta R_{\phi_4}^{\hat{\alpha}_{ew}}}{2}+ \frac{\Delta v}{v}$  & -0.003 & $\frac{\Delta R_{\phi_4}^{\hat{\alpha}_{ew}(0)}}{2}+ \frac{\Delta v}{v}$  & -0.003\\
 $\Delta M_1^{\hat{m}_W}$ & -0.010 & $\Delta M_1^{\hat{\alpha}_{ew}}$ & -0.0096 & $\Delta M_1^{\hat{\alpha}_{ew}(0)}$ & -0.0098\\
 $\Delta g_1^{\hat{m}_W}$ & -0.014 & $\Delta g_1^{\hat{\alpha}_{ew}}$ & -0.096 & $\Delta g_1^{\hat{\alpha}_{ew}(0)}$ & -0.097\\
 $\Delta g_2^{\hat{m}_W}$ & -0.0054 & $\Delta g_2^{\hat{\alpha}_{ew}}$ & 0.039 & $\Delta g_2^{\hat{\alpha}_{ew}(0)}$ & 0.033 \\
 \hline
 \end{tabular}\caption{Numerical values of the one loop corrections to various Lagrangian parameters
 and matrix element corrections in both input schemes, updated to new input parameter values
 in Table~\ref{tab:inputs}.
 We only report gauge independent combinations of parameters. We have chosen $\mu= \hat{m}_h$ in these evaluations
 for the scale dependence associated with the one loop improvement of input parameters and
 finite on shell renormalization conditions in the LSZ formula. For operator mixing effects, we
set $\mu = \Lambda$.}\label{tab:inputs2}
  \end{table}
There is operator basis dependence
at each order in the $\mathcal{O}(1/\Lambda)$ expansion, and higher orders in
$\mathcal{O}(1/\Lambda)$ also depend on the scheme choice made at lower orders in
$\mathcal{O}(1/\Lambda)$ \cite{Barzinji:2018xvu}. We address this scheme dependence
by using the Warsaw basis \cite{Grzadkowski:2010es}, and the geoSMEFT formalism
\cite{Corbett:2019cwl,Helset:2020yio,Hays:2020scx}
for higher order corrections in $\mathcal{O}(1/\Lambda)$.

For perturbative/renormalization scheme dependence, we renormalize in a mixed on shell-$\rm \overline{MS}$
scheme, use the BFM for gauge fixing, and a FJ tadpole scheme \cite{Fleischer:1980ub}.
This approach is consistent with the background field
independence of the geoSMEFT formalism. For numerical evaluations we use the on shell
masses given in Table~\ref{tab:inputs}.

\subsection{$\alpha_{ew}$ and the Hadronic resonance region}
A significant numerical effect, larger than some of the two loop QCD corrections added in this work, is the treatment of
the hadronic resonance region for the running of $\alpha_{EW}(0)$ measured in the $p^2 \rightarrow 0$
Thompson limit. As discussed in Ref.~\cite{ALEPH:2005ab,Agashe:2014kda,Mohr:2015ccw,Hartmann:2016pil,Corbett:2021cil} this effect is numerically significant
in the SM and in the numerical coefficients of SMEFT perturbations.
Including this effect leads to the numerical difference \cite{Workman:2022ynf}
\begin{align}
1/\alpha_{ew}(p^2 \sim \hat{m}_Z^2) = 128.951 \pm 0.009,\quad {\rm while} \quad 1/\alpha_{ew}(p^2 \rightarrow 0) = 137.035 999139(31). \nonumber \\
\end{align}
In \texttt{Hdecay} \cite{Djouadi:1997yw,Djouadi:2018xqq}, a modified $\rm \overline{MS}$ subtraction scheme is used,
motivated by this large numerical effect,
consistent with results developed in Ref.~\cite{Passarino:2007fp,Actis:2008ts}. As this scheme choice
is more numerically significant compared to the size of the two loop corrections we incorporate here
to $\Gamma (h \rightarrow \mathcal{A}\mathcal{A})$, we adjust our numerical results to this convention.

Essentially, the scheme choice used in \cite{Passarino:2007fp,Actis:2008ts,Djouadi:1997yw,Djouadi:2018xqq} is
to use a $\alpha_{ew}(0)$ input, instead of $\alpha_{ew}(\hat{m}_Z)$.
This choice is made to exploit that the hadronic resonance region from bound states in QCD, preserves $\rm QED$, and
hence naive QED Ward identities relate the wavefunction and charge renormalization.
This is the case if a suitable renormalization scheme and gauge fixing term is used.
As a result, the nonperturbative corrections from the hadronic resonance region are not present in the SM prediction of $\Gamma(h \rightarrow \mathcal{A}\, \mathcal{A})$,
but are shifted to  other observables.

To uniform the SMEFT perturbations to this scheme choice \cite{Passarino:2007fp,Actis:2008ts,Djouadi:1997yw,Djouadi:2018xqq}, we modify our finite terms as follows.
As verified in Ref.~\cite{Corbett:2021cil}, the finite terms of the charge and wavefunction renormalization
are related by the preserved $\rm QED$ Ward identity to be
\bea\label{BFMphoton}
\Delta Z_e  &= - \frac{1}{2} \Delta Z_{\hat{\mathcal{A}}}, \nonumber \\
\Delta R_e  &= - \frac{1}{2} \Delta R_{\hat{\mathcal{A}}}.
\eea
We extend $\Delta R_e$ and
$\Delta R_{\hat{\mathcal{A}}}$ by finite terms to cancel the effect of the
running through the hadronic resonance region. Explicitly, $\Delta R_e$
is defined at one loop to be \cite{Sirlin:1980nh,Dekens:2019ept,Corbett:2021cil}
\bea
\Delta R_e = \frac{\bar{g}_1^2 \bar{g}_2^2}{(\bar{g}_1^2 + \bar{g}_2^2)} \,
\left[\frac{7}{32 \pi^2} \log \left(\frac{\mu^2}{m_W^2} \right)
- \frac{N^{f}_c Q^2_{f}}{24 \pi^2}
\,  \log \left(\frac{\mu^2}{\bar{m}_f^2} \right) + \frac{1}{48 \pi^2}\right],
\eea
and the charge renormalization is related to the Thompson limit measured value by
\bea
- i \, \left[\frac{4 \, \pi \, \hat{\alpha}(q^2)}{q^2}\right]_{q^2 \rightarrow 0}
\equiv \frac{- i \, (e_0 + \Delta R_{e})^2}{q^2} \left[1 + {\rm Re} \frac{\Sigma^{AA}(m_Z^2)}{m_Z^2} + \nabla \alpha \right].
\eea
Here $\nabla \alpha$ includes corrections form QCD bound states (see Table.~\ref{tab:inputs})
\cite{Wells:2005vk,Agashe:2014kda,Baikov:2012zm,PhysRevD.22.971} and $\Sigma^{AA}(\bar{m}_Z^2)$ is given explicitly in
Ref.~\cite{Corbett:2021cil}. Now, redefining
\bea
\Delta R_e \rightarrow \Delta R_e + e_0 \left({\rm Re} \frac{\Sigma^{AA}(m_Z^2)}{m_Z^2} + \nabla \alpha\right) + \cdots
\eea
numerically absorbs the effect of running through the Hadronic resonance region into the finite renormalization
of the electric charge. So long as the Ward identity derived relation for finite terms
$\Delta R_e  = - \frac{1}{2} \Delta R_{\hat{\mathcal{A}}}$ is imposed, this leads to the cancelation of the
numerical effects of running through the hadronic resonance region in $\Gamma(h \rightarrow \mathcal{A}\mathcal{A})$
in the (so-defined) $\left\{\alpha(0),\hat{M}_W, \hat{G}_F\right\}$ input scheme. For further discussion see Refs.~\cite{Denner:2019vbn,Dittmaier:2021loa}.

The SM predictions from \texttt{Hdecay} are produced in the effective $\left\{\alpha(0),\hat{M}_W, \hat{G}_F\right\}$
scheme. While the $\left\{\hat{M}_Z,\hat{M}_W, \hat{G}_F\right\}$ scheme is used in
in global studies \cite{Ellis:2020unq,Almeida:2021asy,Ethier:2021bye,ATLAS:2022xyx} for SMEFT perturbations.

This leads to an important numerical shift in the central value of the SM prediction compared to a
$\left\{\alpha(\hat{M}_Z),\hat{M}_W, \hat{G}_F\right\}$ input scheme.
This numerical difference should be noted given that, at leading order,
$\Gamma(h \rightarrow \mathcal{A}\mathcal{A}) \propto \alpha_{ew}^2$, and
\begin{align}
(\alpha_{ew}^{\alpha(0)})^2 &= 5.33 \times 10^{-5}, \quad & (\alpha_{ew}^{\alpha(\hat{m}_Z)})^2 &= 6.01 \times 10^{-5}, \quad & (\alpha_{ew}^{\hat{m}_W})^2 &= 5.72 \times 10^{-5}.
\end{align}
As the perturbations (or lack of perturbations) in $\Gamma(h \rightarrow \mathcal{A}\mathcal{A})$ numerically
is quite dominant in global SMEFT fits, numerical consistency on this issue is critical for precise constraints.
In what follows we present results in the $\left\{\alpha(0),\hat{M}_W, \hat{G}_F\right\}$,
$\left\{\alpha(\hat{M}_Z),\hat{M}_W, \hat{G}_F\right\}$
and $\left\{\hat{M}_Z,\hat{M}_W, \hat{G}_F\right\}$ schemes for $\Gamma(h \rightarrow \mathcal{A}\mathcal{A})$.
Scheme dependence is minimal in the observables $\sigma(\mathcal{G} \mathcal{G} \rightarrow h)$
and $\Gamma(h \rightarrow \mathcal{G}\mathcal{G})$.

\subsection{Uniforming Quark Masses}
We uniform the fermion mass inputs
to a common $\rm \overline{MS}$ convention, consistent with Refs.~\cite{Brooijmans:2016vro,Djouadi:1997yw,Djouadi:2018xqq,LHCHiggsCrossSectionWorkingGroup:2016ypw}.
The top mass is taken as an on shell mass, related to the $\rm \overline{MS}$
at one loop via
\bea
M^i_{os,t} = m^i(\mu) \left(1 + \frac{\alpha_s(M^i_{os,t})}{\pi}\left(\log \frac{\mu^2}{(m^i)^2} + \frac{4}{3}\right) \right).
\eea
For the on-shell charm quark mass used for numerical evaluations,
we determine this value from the relationship \cite{Bauer:2004ve} free of renormalons at leading order
in the $\rm 1S$ scheme
\bea
m_b- m_c = 3.41 \, {\rm GeV}.
\eea
Numerical dependence on the light quark masses is negligible. The masses used are listed in
Table \ref{tab:inputs}.

In the case of results reported in Ref.~\cite{Harlander:2005rq} we note that, the running masses are related
to the pole mass via the convention in Ref.~\cite{Spira:1995rr}
\bea
M^i_{os,\mathcal{A}\mathcal{A}} = m^i(\mu) \left(1 + \frac{\alpha_s(m^i)}{\pi}\, \log \frac{\mu^2}{(m^i)^2} \right).
\eea

Finally, for the lepton pole masses the relationship to the $\rm \overline{MS}$ masses
is \cite{PhysRevD.46.3945}
\bea
M^i_{os,lep} = m^i(\mu) \left(1 + \frac{\alpha_{ew}(m^i)}{\pi}\, \left(1+ \frac{3}{4} \log \frac{\mu^2}{(m^i)^2} \right)\right).
\eea

\subsection{$\sigma(\mathcal{G} \mathcal{G} \rightarrow h)$}

To numerically evaluate $\sigma(\mathcal{G} \mathcal{G} \rightarrow h)$, we use NNPDF3.0 NLO parton distribution functions~\cite{Hartland_2013, Ball_2015} and $\alpha_s = 0.118$. We set all $\mu$ scales to $\hat m_h$, with the exception of scales associated with operator mixing, following Ref.~\cite{Corbett:2021cil}. For these choices, and taking the $m_t \to \infty$ limit, the NLO SM cross section for $\sigma(\mathcal{G} \mathcal{G} \rightarrow h)$, $\sqrt s = 13\, \rm{TeV}$ is (for all EW input schemes):
\begin{align}
\hat{\sigma}_{{\rm SM}, m_t \to \infty}( \mathcal G \mathcal G \to h) = \Delta^2 \sigma^{\rm SM}_{m_t \to \infty}( \mathcal G \mathcal G \to h) +  \Delta^3 \sigma^{\rm SM}_{m_t \to \infty}( \mathcal G \mathcal G \to h) = 31.6\, {\rm pb},
\end{align}
where the analytic expressions for the LO ($\Delta^2$) and NLO ($\Delta^3$) pieces are given in Eqn.~\eqref{eq:SMgghLO} and Eqn.~\eqref{SMNLO} respectively.

Adding up the full set of SMEFT contributions to the inclusive  $\sigma(\mathcal{G} \mathcal{G} \rightarrow h)$ cross section and dividing by the SM result, we find:
\begin{align}
\frac{ \sigma^{\hat{\alpha}}_{\rm SMEFT}(\mathcal G\mathcal G \to h)}{\hat \sigma_{{\rm SM}, m_t \to \infty}(\mathcal G\mathcal G \to h)}\simeq 1
& +  289\, \tilde C^{(6)}_{HG} \nn &+ 289\, \tilde C^{(6)}_{HG}\Big(\tilde C^{(6)}_{H\Box} - \frac 1 4 \tilde C^{(6)}_{HD} \Big) + 4.68\times10^4\, (\tilde C^{(6)}_{HG})^2 + 289\, \tilde C^{(8)}_{HG} \nn
& + 0.85\, \Big(\tilde C^{(6)}_{H\Box} - \frac 1 4 \tilde C^{(6)}_{HD} \Big) + 369\, \tilde C^{(6)}_{HG} -0.91\, \tilde C^{(6)}_{uH} - 7.26\, {\rm Re }\, \tilde C^{(6)}_{uG} \nn
& - 0.60\,\delta G^{(6)}_F - 4.42\, {\rm Re }\, \tilde C^{(6)}_{uG}\,\log\Big(\frac{\hat m^2_h}{\Lambda^2} \Big) - 0.126\,{\rm Re}\,\tilde C^{(6)}_{dG}\,\log\Big(\frac{\hat m^2_h}{\Lambda^2} \Big)  \nn
&-0.057\,{\rm Re}\,\tilde C^{(6)}_{dG} + 2.06\, \tilde C^{(6)}_{dH},
\label{eq:gghrationumeric}
\end{align}
where coefficient $\delta G_F^{(6)}$ stands for the combination
\begin{align}
\label{eq:deltaGF}
 \delta G_F^{(6)} &= \frac{1}{\sqrt2} \left(\tilde C^{(3)}_{\substack{Hl \\ee}}+\tilde C^{(3)}_{\substack{Hl \\ \mu \mu}} - \frac{1}{2}(\tilde C'_{\substack{ll \\ \mu ee \mu}}+\tilde C'_{\substack{ll \\ e \mu \mu e}})\right), \nonumber
\end{align}

The superscript $\hat \alpha$ on the left hand side of the result indicates we used the $\hat \alpha(m_Z)$ scheme, though we find the result is identical for the other two schemes, at least to the order of accuracy presented. The right hand side of Eqn.~\eqref{eq:gghrationumeric} is grouped according to the $\bar v_T/\Lambda$ and loop order of the terms. Specifically, the first line is the $\mathcal O(\bar v^2_T/\Lambda^2)$ interference, the second line is the $\mathcal O(\bar v^4_T/\Lambda^4)$ contribution coming from dimension six operators squared and the interference of dimension eight effects with the SM, and the last three lines are the one loop times $\mathcal O(v^2_T/\Lambda^2)$ contributions. Not surprisingly, the largest loop contribution is the $\mathcal O(\tilde C_{HG}\, \alpha^2_s)$ correction, which is split roughly evenly between the $\delta(1-z)$ term and the $z>1$ contribution.

These results are different than what was presented in Ref.~\cite{Corbett:2021cil}. One cause for the difference is that we are dividing by full NLO SM result in Eqn.~\eqref{eq:gghrationumeric}, while in Ref.~\cite{Corbett:2021cil} we retained only a part of the $\mathcal O(\alpha^3_s)$ SM in the denominator. The difference, $31.6\, {\rm pb}$ here versus $18.15\, {\rm pb}$ in Ref.~\cite{Corbett:2021cil}, explains the approximate halving of all the numbers multiplying the Wilson coefficients. The other main differences is that Eqn.~\eqref{eq:gghrationumeric} has the complete $\mathcal O(\tilde C_{HG}\alpha^2_s)_{m_t \to \infty}$ dependence, consistently calculated in the BFM with the $\rm \overline{MS}$ scheme, while the result in Ref.~\cite{Corbett:2021cil} was incomplete and used an ad hoc combination of different schemes.

To compare our result, the obvious candidate is \texttt{SMEFT@NLO} \cite{Degrande:2020evl}, a recently advanced (NLO) SMEFT Monte Carlo operating within the \texttt{MadGraph} \cite{Alwall:2014hca} framework. However, a direct comparison of our full, analytic result with \texttt{SMEFT@NLO} is complicated by several subtleties. First, the internal \texttt{MadGraph} classification of processes into tree versus loop-level complicates scenarios like $\mathcal G\mathcal G \to h$, where the SM and SMEFT contributions fall into different categories. Second, the counterterm for operator $\tilde C_{HG}$ is not part of the current \texttt{SMEFT@NLO} suite, so terms such as the interference between the lowest order (loop level) SM amplitude and the NLO $\tilde C_{HG}$ amplitude (Eqn.~\eqref{eq:NLOcHG}) cannot be generated.

A further comparison is potentially possible between a subset of terms in this result
and Ref.~\cite{Deutschmann:2017qum}, Table 2. However, the operators
in Ref.~\cite{Deutschmann:2017qum} are, in fact, distinct from ours due to the choice to subtract $\bar v_T^2$ in the operator definition.
Further, the results in Ref.~\cite{Deutschmann:2017qum} have rescaled Wilson coefficient with factors of $\alpha_s$ being introduced.
These differences complicate compensating for different scale and PDF choices between this work and Ref.~\cite{Deutschmann:2017qum}.
As no result equivalent to Eqn.~\eqref{SMEFTNLO} is given in Ref.~\cite{Deutschmann:2017qum}, an analytic parton-level comparison is not possible, so only proton level results can be compared. With these caveats in mind, the central values do differ, though the order of magnitude of the subset of numerical coefficients is consistent within errors and
after rescaling of the coefficients to uniform conventions. A more thorough error analysis on the PDF and scale uncertainty is beyond this work.

\subsection{$\Gamma(h \rightarrow \mathcal{G} \mathcal{G})$}
Using inputs in Table~\ref{tab:inputs} and the SM result for $\Gamma(h \to \mathcal G \mathcal G)$
in the $m_t \rightarrow \infty$ limit at two loop order we have
\bea
\Gamma^{SM}_{m_t \rightarrow \infty}(h \to \mathcal{G} \,\mathcal{G}) = \Delta^2 \, \Gamma^{SM}_{m_t \rightarrow \infty}(h \to \mathcal{G} \,\mathcal{G})
+ \Delta^3 \, \Gamma^{SM}_{m_t \rightarrow \infty}(h \to \mathcal{G} \,\mathcal{G}),
\eea
where \cite{Wilczek:1977zn,Georgi:1977gs,Inami:1982xt,Djouadi:1991tka,Chetyrkin:1997iv}
\begin{align*}
\Delta^2 \, \Gamma^{SM}_{m_t \rightarrow \infty}(h \to \mathcal{G} \,\mathcal{G}) &\equiv \frac{(\alpha_s^{(r)})^2 \, \hat{m}_h^3}{72 \, \pi^3 \hat{v}_T^2},\\
\Delta^3 \, \Gamma^{SM}_{m_t \rightarrow \infty}(h \to \mathcal{G} \,\mathcal{G}) &\equiv
\frac{(\alpha_s^{(r)})^2 \, \hat{m}_h^3}{72 \, \pi^3 \hat{v}_T^2} \left( \frac{\alpha_s^{(r)}}{\pi} \right) \left(\frac{95}{4} - \frac{7 \, n_F}{6} - \frac{\beta_0}{2} \log \frac{\hat{m}_h^2}{\hat{\mu}^2} \right).
\end{align*}
Numerically, this evaluates to $2.01 + 1.35 = 3.37 \times 10^{-4}\,\text{GeV}$.

Including SMEFT contributions, we have the result
\begin{align}
\frac{\Gamma_{SMEFT}(h \to \mathcal G \mathcal G)}{\hat{\Gamma}_{SM,m_t \rightarrow \infty}(h \to \mathcal G \mathcal G)} &\simeq 1
+ \frac{24 \pi}{\alpha_s^{(r)}} \tilde{C}^{(6)}_{HG}
+\frac{4 \pi}{\alpha_s^{(r)} \,  \kappa_{h\mathcal{G} \,\mathcal{G}} } \, \left(12 + \frac{36 \pi}{\alpha_s^{(r)}}\right) (\tilde{C}^{(6)}_{HG})^2 \\
&+\frac{24 \pi}{\alpha_s^{(r)} \, \kappa_{h\mathcal{G} \,\mathcal{G}}}\times \left(\left[\Delta G_F + \Delta M_1 + \Delta R_\mathcal{G}\right] \, \tilde C^{(6)}_{HG} + \sum_{i} \,  \frac{{\rm Re} \, \tilde C_i^{(6)} \Delta f_i^{(6)}}{16 \pi^2} \right) \nonumber \\
&+\frac{24 \pi}{ \alpha_s^{(r)}} \, \left[\langle \sqrt{h}^{44}\rangle_{\mathcal O(v^2/\Lambda^2)} \tilde C^{(6)}_{HG} +  \tilde C^{(8)}_{HG}\right],\nonumber
\end{align}
where we have defined
\bea
\kappa_{h\mathcal{G} \,\mathcal{G}} \equiv 1+ \Delta^3 \, \Gamma^{SM}_{m_t \rightarrow \infty}(h \to \mathcal{G} \,\mathcal{G})/\Delta^2 \, \Gamma^{SM}_{m_t \rightarrow \infty}(h \to \mathcal{G} \,\mathcal{G})
\equiv 1.67.
\eea
In the $m_t \rightarrow \infty$ limit, the SM QCD correction cancels against the same overall
correction for the $\tilde{C}^{(6)}_{HG}$ linear term.
The rescaling of the local contact operator forms present in the last term is also the same, leading to another
cancelation of $\kappa_{h\mathcal{G} \,\mathcal{G}}$.
The remaining terms have non-factorizable corrections that are not included here,
so only the SM two loop normalization is present. See Ref.~\cite{Asteriadis:2022ras} for recent work on these effects.

Only the second line is input parameter scheme dependent, so scheme effects on the SMEFT
perturbations are quite small. Numerically (using the same inputs and scales as Eq.~\eqref{eq:gghrationumeric}), the SMEFT result is
\begin{align}
\frac{\Gamma_{SMEFT}}{\hat{\Gamma}_{SM,m_t \rightarrow \infty}} \simeq 1 & +  640\, \left[ \tilde C^{(6)}_{HG}\left(1 + \Big(\tilde C^{(6)}_{H\Box} - \frac 1 4 \tilde C^{(6)}_{HD} \Big)\right)+ \tilde C^{(8)}_{HG}\right] + S_1 \, \tilde C^{(6)}_{HG} + 6.20\times10^4\, (\tilde C^{(6)}_{HG})^2 \nn
& + 1.24 \, \Big(\tilde C^{(6)}_{H\Box} - \frac 1 4 \tilde C^{(6)}_{HD} \Big) - 0.87 \,\delta G^{(6)}_F
- 1.24 \, \tilde C^{(6)}_{tH} +2.73\, \tilde C^{(6)}_{bH} \\
&- 7.86\, {\rm Re }\, \tilde C^{(6)}_{uG}
-4.85\,{\rm Re }\,  \tilde C^{(6)}_{uG}\,\log\Big(\frac{\hat m^2_h}{\Lambda^2} \Big) -0.14\,{\rm Re}\,\tilde C^{(6)}_{dG}\,\log\Big(\frac{\hat m^2_h}{\Lambda^2} \Big)
- 0.06\,{\rm Re}\,\tilde C^{(6)}_{dG}. \nonumber
\end{align}
The input parameter scheme dependence of the numerical coefficients is negligible,
with the largest dependence being
\bea
\label{eq:S1hgg}
\left(S_1^{\hat{m}_W},S_1^{\hat{\alpha}_{ew}(\hat{m}_Z)},S_1^{\hat{\alpha}_{ew}(0)}\right) = \left(-26.8,-26.6,-26.7\right).
\eea

\subsection{$\Gamma(h \rightarrow \mathcal{A} \mathcal{A})$}

For these numeric, we again use the input parameters in Table~\ref{tab:inputs} and the related results
in Table~\ref{tab:inputs2}.
Including the two loop QCD SM results at the amplitude level in this manner
gives the following SM $h \rightarrow \mathcal{A} \mathcal{A}$ partial widths for the SM
with out chosen numerical input parameters:
\bea
\Gamma^{ \hat{m}_W}_{\rm SM}(h \rightarrow \mathcal{A} \mathcal{A}) &=& 1.10 \times 10^{-5}\, {\rm GeV}, \\
\Gamma^{\hat{\alpha}_{ew}(\hat{m}_Z)}_{\rm SM}(h \rightarrow \mathcal{A} \mathcal{A}) &=& 1.16 \times 10^{-5}\, {\rm GeV}, \\
\Gamma^{\hat{\alpha}_{ew}(0)}_{\rm SM}(h \rightarrow \mathcal{A} \mathcal{A}) &=& 1.01 \times 10^{-5}\, {\rm GeV}.
\eea
where here we retain the
two loop squared contribution to the decay width. Interference corrections of
three loop order interfering with the SM one loop amplitude are the same order, but numerically neglected
in the normalization. We include the
two loop QCD interference effects with the tree level operator (leading) interference results in the SMEFT.
We neglect these two loop SM interference effects in the other interference terms. The result is
\begin{align}
\frac{\Gamma_{SMEFT}}{\hat{\Gamma}_{\rm SM}}
&\simeq 1 + S_1 \left[ f_1 +  \left(\tilde C_{H\Box}^{(6)} - \frac{\tilde C_{HD}^{(6)}}{4}\right)  \, f_1+ f_2\right]
+ S_2 \, f_1^2 + S_3 \, (\tilde{C}_{HW}^{(6)} - \tilde{C}_{HB}^{(6)})^2  + S_4 \, \delta G_F^{(6)} \, \tilde{C}_{HB}^{(6)} \nonumber \\
&+ S_5 \, \delta G_F^{(6)} \, \tilde{C}_{HW}^{(6)} + S_6 \, \delta G_F^{(6)} \, \tilde{C}_{HWB}^{(6)} + S_7 \, \tilde{C}_{HD}^{(6)} \, \tilde{C}_{HB}^{(6)}
 + S_8 \, \tilde{C}_{HD}^{(6)} \, \tilde{C}_{HW}^{(6)}  + S_9 \, \tilde{C}_{HD}^{(6)} \, \tilde{C}_{HWB}^{(6)} \nonumber \\
  &+ S_{10} \, \tilde{C}_{HWB}^{(6)} \, \tilde{C}_{HB}^{(6)} + S_{11} \, \tilde{C}_{HWB}^{(6)} \, \tilde{C}_{HW}^{(6)}
  + S_{12} \, (\tilde{C}_{HWB}^{(6)})^2 + S_{13} \, \tilde{C}_{HB}^{(6)} + S_{14} \, \tilde{C}_{HW}^{(6)} \nonumber \\
& + \left[S_{15} + S_{16} \log \left(\frac{\hat{m}_h^2}{\Lambda^2}\right)\right] \, \tilde{C}_{HWB}^{(6)}
+ \left[S_{17} + S_{18} \log \left(\frac{\hat{m}_h^2}{\Lambda^2}\right)\right]\, \tilde{C}_{W}^{(6)} \nn
&+ \left[S_{19} + S_{20} \log \left(\frac{\hat{m}_h^2}{\Lambda^2}\right)\right]\, {\rm{Re} \, \tilde{C}_{\substack{uB \\ 33}}^{(6)}}
+ \left[S_{21} + S_{22} \log \left(\frac{\hat{m}_h^2}{\Lambda^2}\right)\right]\, {\rm{Re} \, \tilde{C}_{\substack{uW \\ 33}}^{(6)}}
+ S_{23} \, {\rm{Re} \, {\tilde{C}_{\substack{uH \\ 33}}^{(6)}}} \nn
&+ S_{24}\, {\rm Re} \,{\tilde{C}_{\substack{dH \\ 33}}^{(6)}} + S_{25} \, (\tilde{C}_{H \Box}^{(6)} - \frac{\tilde{C}_{HD}^{(6)}}{4}) + S_{26} \, \tilde{C}_{HD}^{(6)} + S_{27} \, \tilde{C}_{HWB}^{(6)} + S_{28} \,\delta G_F^{(6)}. \nonumber
\end{align}
The input scheme dependent numerical results are given in Table~\ref{tab:schemeresults5}.
\begin{table}[t]\centering
\renewcommand{\arraystretch}{1.2}
 \begin{tabular}{|c|c|c|c|c|c|c|c|c|c|c|}\hline
& $S_1$ & $S_2$ & $S_3$ & $S_4$ & $S_5$ & $S_6$ & $S_7$ & $S_8$ & $S_9$ & $S_{10}$   \\
\hline
$\hat{M}_W$ &$-753$ &  $1.41 \times 10^5$ & $-321$ & $2041$ & $586$ & $-1093$ & $897$ & $721$ & $-914$ & $1880$   \\
$\hat{\alpha}_{ew}^{(\hat{M}_Z)}$ & $-724$ & $1.31 \times 10^5$ & $-320$ & $1402$ & $-126$ & $-269$ & $149$ & $-149$ & $95.0$ & $297$  \\
$\hat{\alpha}_{ew}^{(0)}$ & $-794$ & $1.56 \times 10^5$ & $-317$ & $1447$ & $-105$ & $-274$ & $138$ & $-138$ & $97.0$ & $227$  \\
 \hline
 \end{tabular}
  \renewcommand{\arraystretch}{1.2}
   \begin{tabular}{|c|c|c|c|c|c|c|c|c|c|c|c|c|c|c|c|}\hline
  &  $S_{11}$ & $S_{12}$ &$S_{13}$ & $S_{14}$ & $S_{15}$ & $S_{16}$ & $S_{17}$ & $S_{18}$ & $S_{19}$ & $S_{20}$ & $S_{21}$ \\
  \hline
  $\hat{M}_W$ & $1587$ &$-1843$ & $-91$ & $-26.1$ & $52.3$ & $1.87$ & $-0.51$ & $3.28$ & $24.4$ & $-25.6$ & $13.1$   \\
  $\hat{\alpha}_{ew}^{(\hat{M}_Z)}$ & $-297$  & $320$ & $-198$ & $31.4$ & $-15.3$ & $1.80$ & $-0.55$ & $3.25$ & $23.9$ & $-25.0$ & $43.6$ \\
  $\hat{\alpha}_{ew}^{(0)}$& $-227$ & $317$ & $-203$ & $26.5$ & $-16.9$ & $1.95$ & $-0.42$ & $3.10$ & $23.5$ & $-24.6$ & $45.2$ \\
   \hline
   \end{tabular}
     \renewcommand{\arraystretch}{1.2}
     \begin{tabular}{|c|c|c|c|c|c|c|c|}\hline
    & $S_{22}$ & $S_{23}$ & $S_{24}$ & $S_{25}$ & $S_{26}$ & $S_{27}$ & $S_{28}$ \\
    \hline
    $\hat{M}_W$ & $-13.7$ & $0.51$ & $-0.28$  & $2$ &$-3.49$ & $-7.5$ & $-3 \sqrt{2}$  \\
    $\hat{\alpha}_{ew}^{(\hat{M}_Z)}$& $-45.7$& $0.51$ &  $-0.28$ & $2$ & $0$ & $0$ & $- \sqrt{2}$ \\
    $\hat{\alpha}_{ew}^{(0)}$& $-47.3$ & $4.71$ & $-1.14$ & $2$ & $0$ & $0$ & $- \sqrt{2}$  \\
     \hline
     \end{tabular}\caption{Numerical coefficients for SMEFT perturbations to
     $\Gamma(h \rightarrow \mathcal{A} \mathcal{A})$ in three input parameter schemes, including two loop QCD interference effects.}\label{tab:schemeresults5}
      \end{table}
Several numerically small corrections compared to the retained terms are neglected here. These neglected corrections are generally further suppressed by small
(SM) Yukawa couplings. Here the short hand functions $f^{\hat{m}_W}_i \simeq f^{\hat{\alpha}_{ew}}_i$ for $i=1,2$ are approximately scheme independent,
\begin{align}
f^{\hat{m}_W}_1 &=  \left[\tilde{C}_{HB}^{(6)} +0.29 \, \, \tilde{C}_{HW}^{(6)} -0.54  \, \tilde{C}_{HWB}^{(6)}\right],\\
f^{\hat{m}_W}_2 &=   \left[\tilde{C}_{HB}^{(8)} +0.29 \, \, (\tilde{C}_{HW}^{(8)}+ \tilde{C}_{HW,2}^{(8)}) -0.54  \, \tilde{C}_{HWB}^{(8)}\right].
\end{align}

The above result can be compared to Eq.~(5.6) and (5.11) of Ref.~\cite{Corbett:2021cil}. The new result fixes minor mistakes in the old result and should be taken to supersede it. In addition, a few inputs have shifted slightly, leading to small changes in a few of the $\Delta M, \Delta R$ in Table~\ref{tab:inputs2}. More significantly, we have included the two-loop squared contribution to $\hat \Gamma_{\rm SM}$, which increases it by $\mathcal O(10\%)$.

\subsection{$\delta \Gamma_{h,full}^{SMEFT}$}
The total width of the SMEFT was calculated systematically in Ref.~\cite{Brivio:2019myy} including
all corrections $\mathcal{O}(1/\Lambda^2)$ interfering with SM amplitudes in the $\rm U(3)^5$ limit for
$\tilde{C}_i^{(6)}$. In this section we discuss how this result is surprisingly robust against the leading QCD corrections.
The dependence of the total inclusive width on the $\mathcal{L}^{(6)}$ Wilson coefficients of the SMEFT was found to be \cite{Brivio:2019myy}
\bea
\begin{aligned}
\frac{\d\Gamma_{h,full}^{SMEFT}}{\Gamma_h^{SM}}
\simeq \,  1  &
- 1.50    \,\tilde{C}_{HB}^{(6)}
- 1.21    \,\tilde{C}_{HW}^{(6)}
+ 1.21    \,\tilde{C}_{HWB}^{(6)}
+ 50.6    \,\tilde{C}_{HG}^{(6)}
\\ &
+ 1.83    \,\tilde{C}_{H\square}^{(6)}
- 0.43    \,\tilde{C}_{HD}^{(6)}
+ 1.17    \,\tilde{C}_{ll}'^{(6)}
\\&
- 7.85  \,  \hat{Y}_{\substack{u \\ cc}} \, \re\tilde{C}_{uH}^{(6)}
- 48.5 \,  \hat{Y}_{\substack{d \\ bb}} \, \re\tilde{C}_{dH}^{(6)}
- 12.3   \,  \hat{Y}_{\substack{\ell \\ \tau \tau}} \, \re\tilde{C}_{eH}^{(6)}
\\&
+ 0.002   \,\tilde{C}_{Hq,(1)}^{(6)}
+ 0.06    \,\tilde{C}_{Hq,(3)}^{(6)}
+ 0.001 \,\tilde{C}_{Hu}^{(6)}
- 0.0007  \,\tilde{C}_{Hd}^{(6)}
\\&
- 0.0009   \,\tilde{C}_{Hl,(1)}^{(6)}
- 2.32    \,\tilde{C}_{Hl,(3)}^{(6)}
- 0.0006   \,\tilde{C}_{He}^{(6)},
\end{aligned}
\eea
using the $\{\hat M_W,\hat M_Z,\hat G_F,\hat M_h\}$ input scheme.
Here, we have pulled out the explicit Yukawa factor from the Wilson coefficient.
Using the $\{\hat \alpha_{ew},\hat M_Z,\hat G_F,\hat M_h\}$ input scheme, the result is
\bea
\begin{aligned}
\frac{\d\Gamma_{h,full}^{SMEFT}}{\Gamma_h^{SM}}
\simeq \,  1  &
- 1.40    \,\tilde{C}_{HB}^{(6)}
- 1.22    \,\tilde{C}_{HW}^{(6)}
+ 2.89    \,\tilde{C}_{HWB}^{(6)}
+ 50.6    \,\tilde{C}_{HG}^{(6)}
\\ &
+ 1.83    \,\tilde{C}_{H\square}^{(6)}
+ 0.34    \,\tilde{C}_{HD}^{(6)}
+ 0.70    \,\tilde{C}_{ll}'^{(6)}
\\&
- 7.85  \,  \hat{Y}_{\substack{u \\ cc}} \, \re\tilde{C}_{uH}^{(6)}
- 48.5 \,  \hat{Y}_{\substack{d \\ bb}} \, \re\tilde{C}_{dH}^{(6)}
- 12.3   \,  \hat{Y}_{\substack{\ell \\ \tau \tau}} \, \re\tilde{C}_{eH}^{(6)}
\\&
+ 0.002   \,\tilde{C}_{Hq,(1)}^{(6)}
+ 0.06    \,\tilde{C}_{Hq,(3)}^{(6)}
+ 0.001 \,\tilde{C}_{Hu}^{(6)}
- 0.0008  \,\tilde{C}_{Hd}^{(6)}
\\&
- 0.0008   \,\tilde{C}_{Hl,(1)}^{(6)}
- 1.38    \,\tilde{C}_{Hl,(3)}^{(6)}
- 0.0007   \,\tilde{C}_{He}^{(6)}.
\end{aligned}
\eea
In Ref.~\cite{Brivio:2019myy}, loop effects (outside the SM loop suppressed decays
to  $\mathcal{G} \, \mathcal{G}$, $\mathcal{A} \, \mathcal{A}$, $\mathcal{Z} \, \mathcal{A}$) were neglected.
As such, the $\{\hat \alpha_{ew}(\hat M_Z),\hat M_Z,\hat G_F,\hat M_h\}$ and $\{\hat \alpha_{ew}(0),\hat M_Z,\hat G_F,\hat M_h\}$
scheme are identified. We have used the $\{\hat \alpha_{ew}(\hat M_Z),\hat M_Z,\hat G_F,\hat M_h\}$
scheme.\footnote{The use of the results of the Higgs cross section working group for branching ratios
effectively shifts some the numerical results to that of the $\{\hat \alpha_{ew}(0),\hat M_Z,\hat G_F,\hat M_h\}$
scheme. This correction to the presentation of Ref.~\cite{Brivio:2019myy} should be noted.}

Numerically important loop contributions to $\d\Gamma_{h,full}^{SMEFT}/\Gamma_h^{SM}$
come about from decays to $\bar{b} \, b$,  $\mathcal{G} \, \mathcal{G}$ and $\mathcal{A} \, \mathcal{A}$.
QCD corrections to $\mathcal{A} \, \mathcal{A}$ decay are small.
The leading $\Gamma(h \rightarrow \bar{b} \, b)$ QCD corrections factorize and are the same as in the SM in the EFT (neglecting $m_\psi$ the small known IR mass parameters),
thus they cancel in the SMEFT width expression. Therefore, the $\Delta$  corrections to the decay
$\Gamma(h \rightarrow \mathcal{G} \mathcal{G})$ dominate the dependence of the total width on $\tilde{C}_{HG}^{(6)}$.
This correction can be incorporated by adding the term
\bea
- \frac{0.33}{\Gamma^{SM}_h} \times 619 \, \tilde{C}_{HG}^{(6)} + \frac{0.337}{\Gamma^{SM}_h} \left(640 + S_1 \right) \, \tilde{C}_{HG}^{(6)},
\eea
to $\frac{\d\Gamma_{h,full}^{SMEFT}}{\Gamma_h^{SM}}$, where $S_1$ refers to the quantity in Eq.~\eqref{eq:S1hgg}. Using $\Gamma^{SM}_h = 4.100 \,{\rm MeV}$, this
leads to the partial QCD-improved result of the SMEFT width reported in Ref.~\cite{Brivio:2019myy}
\bea
\frac{\d\Gamma_{h,full}^{SMEFT}}{\Gamma_h^{SM}} + \left(0.58, 0.59\right) \tilde{C}_{HG}^{(6)},
\eea
in the $\{\hat M_W,\hat M_Z,\hat G_F,\hat M_h\}$,$\{\hat \alpha_{ew}(0),\hat M_Z,\hat G_F,\hat M_h\}$
schemes respectively.
This correction is only partial, it neglects many other QCD correction in the partial decay width.
Nevertheless it is the leading correction for the operator $C_{HG}^{(6)}$ dependence in the total width.
Due to the numerical dominance of the decay to $\mathcal{G} \, \mathcal{G}$
for the operator $C_{HG}^{(6)}$ in the SMEFT, this is a relevant numerical improvement.

\section{Conclusions}
In this paper we have advanced the results in the geoSMEFT formulation of the SMEFT for $\sigma(\mathcal{G} \,\mathcal{G}\rightarrow h)$,
$\Gamma(h \rightarrow \mathcal{G} \,\mathcal{G})$, $\Gamma(h \rightarrow \mathcal{A} \mathcal{A})$, $\Gamma(h \rightarrow \bar{\psi} \, \psi)$, and the total Higgs width. Previous literature \cite{Alonso:2013hga,Brivio:2017vri,Brivio:2017btx,Helset:2020yio,Brivio:2020onw,Corbett:2021cil,Brivio:2019myy} has provided terms in the SMEFT $\times$ loop expansion of orders $\mathcal O(\bar v^2_T/\Lambda^2), \mathcal O(\bar v^2_T/\Lambda^2 (16\pi^2))$ and $\mathcal O(\bar v^4_T/\Lambda^4)$. This work extends the expansion to $\mathcal O(\bar v^2_T/\Lambda^2 (16\pi^2)^2)$ by consistently including the interference of two-loop (NLO in QCD) SM amplitudes with $\mathcal O(\bar v^2_T/\Lambda^2)$ SMEFT terms. Additionally, we have incorporated a set of QCD loop corrections determined previously in \cite{Braaten:1980yq,Gauld:2016kuu}
into the characterization of $\Gamma(h \rightarrow \bar{\psi} \, \psi)$. Combining these updated results, we determine the leading loop correction to the Higgs total width.
We have also characterized a more consistent numerical treatment of input parameter choices and effects,
updating past numerical results.

\acknowledgments
We thank B. Anastasiou, G. Buchalla, L. Dixon, C. Duhr,  A. Helset, A. Manohar, K. Mimasu, B. Mistlberger, M. Spira
and E. Vryonidou for helpful discussion and correspondence.
M.T. acknowledges support from the Villum Fund, project number 00010102, Caltech and Mark Wise
for financial support.
The work of A.M. was supported in part by the National Science Foundation under Grant Number PHY-2112540.

\appendix
\section{SMEFT/geoSMEFT notation and conventions}\label{setup}

The SM Lagrangian \cite{Glashow:1961tr,Weinberg:1967tq,Salam:1968rm} notation is fixed to be
\bea\label{sm1}
\mathcal{L} _{\rm SM} &=& -\frac14 G_{\mu \nu}^A G^{A\mu \nu}-\frac14 W_{\mu \nu}^I W^{I \mu \nu} -\frac14 B_{\mu \nu} B^{\mu \nu}  + \sum_{\psi} \overline \psi\, i \slashed{D} \, \psi \\
&\,&\hspace{-0.75cm} + (D_\mu H)^\dagger(D^\mu H) -  \lambda \left(H^\dagger H -\frac12 v^2\right)^2 -  \left[H^{\dagger j} \, \overline d\, Y_d\, q_{j}
+ \widetilde H^{\dagger j} \overline u\, Y_u\, q_{j} + H^{\dagger j} \overline e\, Y_e\,  \ell_{j} + \hbox{h.c.}\right]. \nonumber
\eea
The chiral projectors have the convention $\psi_{L/R} = P_{L/R} \, \psi$ where
$P_{R} = \left(1 + \gamma_5 \right)/2$, and
the gauge covariant derivative is defined with a positive sign convention
\bea
D_\mu = \partial_\mu + i g_3 T^A A^A_\mu + i g_2  \sigma^I W^I_\mu/2 + i g_1 {\bf \hyp_i} B_\mu,
\eea
with $I=\{1,2,3\}$, $A=\{1\dots 8\}$ , $\sigma^I$ denotes the Pauli matrices and
${\bf \hyp_i}$ the $\rm U_Y(1)$ hypercharge generator with charge normalization ${\bf \hyp_i}= \{1/6,2/3,-1/3,-1/2,-1,1/2\}$ for $i =\{q,u,d,\ell,e,H \}$.
The SMEFT Lagrangian is
\begin{align}
 \Lagr_{\textrm{SMEFT}} &= \Lagr_{\textrm{SM}} + \Lagr^{(d)}, &   \Lagr^{(d)} &= \sum_i \frac{C_i^{(d)}}{\Lambda^{d-4}}\mathcal{Q}_i^{(d)}
 \quad \textrm{ for } d>4.
\end{align}
The SM Lagrangian notation and conventions are consistent with
Refs.~\cite{Grzadkowski:2010es,Alonso:2013hga,Brivio:2017vri,Brivio:2017btx,Helset:2020yio,Brivio:2020onw,Corbett:2021cil}
with some slight variations.
The operators $\mathcal{Q}_i^{(d)}$ are labelled with a mass dimension $d$ superscript
and multiply unknown Wilson coefficients $C_i^{(d)}$; while $\bar{v}_T \equiv \sqrt{\langle 2 H^\dagger H \rangle}$ and
$\tilde{C}^{(d)}_i \equiv C^{(d)}_i \bar{v}_T^{d-4}/\Lambda^{d-4}$.
Due to strong constraints from low energy CP violating observables~\cite{Cirigliano:2016njn},
we restrict our study to CP even operators.
\subsection{geoSMEFT}
The geoSMEFT \cite{Corbett:2019cwl,Helset:2020yio,Hays:2020scx} is a organization of the physics of the SMEFT
in terms of field-space connections $G_i$ depend on the group indices $I,A$ of the (non-spacetime) symmetry groups
and multiplying composite operator forms $f_i$ (which include powers of $D^\mu H$).
The re-organization is represented schematically by
\bea\label{basicdecomposition}
\Lagr_{\textrm{SMEFT}} &=& \sum_i G_i(I,A,\phi \dots) \, f_i , \nn
 &=& \frac{1}{2} h_{IJ}(\phi) (D_{\mu} \phi)^{I} (D^{\mu} \phi)^{J}
- \frac{1}{4} g_{AB}(\phi) {\mathcal{W}}^{A}_{\mu\nu} {\mathcal{W}}^{B\mu\nu}
- \frac{1}{4} k_{\mathpzc{AB}}(\phi) G^{\mathpzc{A},\mu \nu} G_{\mathpzc{B},\mu \nu}
+ \cdots.
\eea
Our notation is such that the covariant derivative acting on the bosonic fields of the SM in the doublet,
using real scalar field coordinates, is given by \cite{Helset:2018fgq}
\bea
(D^{\mu}\phi)^I &=& (\partial^{\mu}\delta_J^I - \frac{1}{2}\mathcal{W}^{A,\mu}\tilde\gamma_{A,J}^I)\phi^J,
\eea
with symmetry generators/structure constants ($\tilde{\epsilon}^{A}_{\, \,BC},\tilde{\gamma}_{A,J}^{I}$).
See Refs.~\cite{Helset:2018fgq,Helset:2020yio} for the generators/structure constants for the real scalar representation.
The real scalar field co-ordinates ($\phi_I $) of the Higgs scalar doublet are introduced as
\begin{align}
\hat{H}(\hat{\phi}_I) &= \frac{1}{\sqrt{2}}\begin{bmatrix} \hat{\phi}_2+i\hat{\phi}_1 \\ \hat{\phi}_4 + \bar{v}_T - i \hat{\phi}_3\end{bmatrix},
& \quad  H(\phi_I) &= \frac{1}{\sqrt{2}}\begin{bmatrix} \phi_2+i\phi_1 \\ \phi_4 - i\phi_3\end{bmatrix}.
\end{align}

The field-space connections (or metrics) $h_{IJ},g_{AB},k_{\mathpzc{AB}}$ are functions of $\phi_I$
and depend on $I$, the indicies of the generalized canonically normalised Yang Mills ($\mathcal{W}^A$)
or the gluon fields $(G^{\mathpzc{A}})$.
The mass eigenstate fields are $\Phi^L, \mathcal{A}^A$ and the mass eigenstate ghost field is defined as $c^A$.
Explicitly, the field sets are
\begin{align*}
\phi_I &= \{\phi_1,\phi_2,\phi_3,\phi_4\}, & \quad \quad \mathcal{W}^A &= \{W^1,W^2,W^3,B\}, \\
\Phi^L &= \{\Phi^+, \Phi^-, \chi, H \}, &\quad \quad \mathcal{A}^A &= \{\mathcal{W}^+,\mathcal{W}^-,\mathcal{Z},\mathcal{A}\}, \\
c^A &= \{c_{{W}^+},c_{{W}^-},c_{Z},c_{A}\}.
\end{align*}
Here $\mathpzc{A}=\{1 \cdots 8 \}$, $A,L,I =\{1 \cdots 4\}$ and the EW
couplings are $\alpha_A = \{g_2, g_2, g_2, g_1\}$.

The weak/mass eigenstate field and coupling transformations
at all orders in the $\bar{v}_T/\Lambda$ expansion are given by
\begin{align*}
\phi^{J} &= \sqrt{h}^{JK} V_{KL} \Phi^{L}, &\quad \quad \mathcal{W}^{A,\nu} &=  \sqrt{g}^{AB} U_{BC} \mathcal{A}^{C,\nu}, \\
u^A &= \sqrt{g}^{AB} U_{BC} \, c^C, & \quad \quad \alpha^{A} &= \sqrt{g}^{AB} U_{BC} \mathcal{\beta}^{C}, \\
G^{A,\nu} &=  \sqrt{\kappa} \, \mathcal{G}^{A,\nu}, & \quad \quad
\bar{g}_3 &= g_3 \, \sqrt{\kappa}.
\end{align*}

$k_{\mathpzc{AB}}(\phi) \rightarrow \kappa(\phi) \, \delta_{AB}$
and $\mathcal{\beta}^{C}$ is obtained directly
from $\alpha^{A}$ and $U_{BC}$. Note that $\alpha_A \mathcal{W}^{A,\nu}$ and $g_3 ^{A,\nu}$ linear terms in the
covariant derivative are unchanged by these transformations at all orders in
the $\bar{v}_T/\Lambda$ expansion.\footnote{The matrix square roots of these field space connections are $\sqrt{g}_{AB} = \langle g_{AB} \rangle^{1/2}$.
$\langle \rangle$ indicates
a background field expectation value. The inverses are defined via
$\sqrt{g}^{AB} \sqrt{g}_{BC} \equiv \delta^A_C$ .
The field-space connections are positive semi-definite matrices,
with unique positive semi-definite square roots.
We also use the hat notation for the background field expectation values at times. These conventions apply to $h^{IJ}, k_{\mathpzc{AB}}$.}

The matrices $U,V$ are unitary rotations; i.e. orthogonal matrices whose transpose is equal to the matrix inverse,
and given by
\begin{align*}
 U_{BC} &= \begin{bmatrix}
   \frac{1}{\sqrt{2}} & \frac{1}{\sqrt{2}} & 0 & 0 \\
   \frac{i}{\sqrt{2}} & \frac{-i}{\sqrt{2}} & 0 & 0 \\
   0 & 0 & c_{\overline{\theta}} & s_{\overline{\theta}} \\
   0 & 0 & -s_{\overline{\theta}} & c_{\overline{\theta}}
 \end{bmatrix},& \quad
 V_{JK} &= \begin{bmatrix}
   \frac{-i}{\sqrt{2}} & \frac{i}{\sqrt{2}} & 0 & 0 \\
   \frac{1}{\sqrt{2}} & \frac{1}{\sqrt{2}} & 0 & 0 \\
 0 & 0 & -1 & 0 \\
 0 & 0 & 0 & 1
 \end{bmatrix}.
\end{align*}
Here the angle is defined via the generalized Yang Mills field space metric
\bea
s_{\bar{\theta}}^2 &=& \frac{(g_1 \sqrt{g}^{44} - g_2 \sqrt{g}^{34})^2}{g_1^2 [(\sqrt{g}^{34})^2+ (\sqrt{g}^{44})^2]+ g_2^2 [(\sqrt{g}^{33})^2+ (\sqrt{g}^{34})^2]
- 2 g_1 g_2 \sqrt{g}^{34} (\sqrt{g}^{33}+ \sqrt{g}^{44})}.
\eea

The geoSMEFT  masses and couplings are consistent with Ref.~\cite{Alonso:2013hga} and used
(at leading order) in SMEFTsim, see Refs.~\cite{Brivio:2017btx,Brivio:2019myy}.
For completeness, the canonically normalised (geometric) masses at $\mathcal{O}(\bar{v}_T^2/\Lambda^2)$ are
\begin{align}
\bar{M}_{W}^2 &= \frac{\bar g_2^2 \bar{v}_T^2}{4}, \\
\bar{M}_{Z}^2 &= \frac{\bar{v}_T^2}{4}(\bar g_1^2+\bar g_2^2)+\frac{1}{8} \, \bar{v}_T^2 \, (\bar g_1^2+\bar g_2^2) \, \tilde{C}_{HD} +\frac{1}{2} \bar{v}_T^2 \, \gcb \, \gcw \, \tilde{C}_{HWB},\\
\bar{m}_{h}^2 &= 2 \lambda \bar{v}_T^2 \left[1- 3 \frac{\tilde{C}_H}{2 \, \lambda} + 2 \left(\tilde{C}_{H \Box} - \frac{\tilde{C}_{HD}}{4}\right)\right].
\end{align}
The geometric SMEFT couplings with $\mathcal{L}^{(6)}$ corrections are
\begin{align}
\bar{e} &= \frac{\gcb \, \gcw}{\sqrt{\bar g_1^2+\bar g_2^2}} \left[1- \frac{\gcb \, \gcw}{\bar g_1^2+\bar g_2^2} \, \tilde{C}_{HWB} \right],
& \quad
\bar{g}_Z &= \sqrt{\bar g_1^2+\bar g_2^2}+ \frac{\gcb \, \gcw}{\sqrt{\bar g_1^2+\bar g_2^2}} \, \tilde{C}_{HWB}, \\
\gcb &= g_1(1+ \tilde{C}_{HB}),& \quad
\gcw &= g_2(1+ \tilde{C}_{HW}) \\
\bar{g}_3 &= g_s(1+ \tilde{C}_{HG}).
\end{align}
Bowing to past notational conventions we define $\phi_4 = h$ and use the later symbol in the bulk of this work.

Our gauge fixing is given by Ref.~\cite{Helset:2018fgq} in the BFM for the SMEFT.
For the EW sector it is
\begin{align}\label{gaugefixing1}
	\Lagr^{EW}_{\textrm{GF}} &= -\frac{\hat{g}_{AB}}{2 \, \xi} \mathcal{G}^A \, \mathcal{G}^B, &
\mathcal{G}^X &\equiv \partial_{\mu} \mathcal{W}^{X,\mu} -
		\tilde\epsilon^{X}_{ \, \,CD}\hat{\mathcal{W}}_{\mu}^C \mathcal{W}^{D,\mu}
    + \frac{\xi}{2}\hat{g}^{XC}
		\phi^{I} \, \hat{h}_{IK} \, \tilde\gamma^{K}_{C,J} \hat{\phi}^J,
\end{align}
for the QCD coupling we have analogously the BFM gauge fixing term \cite{Corbett:2020bqv}
\begin{align}\label{gaugefixing2}
	\Lagr^{QCD}_{\textrm{GF}} &= -\frac{\hat{\kappa}}{2 \, \xi_G} \, \mathcal{G}^{\mathpzc{A}} \, \mathcal{G}_{\mathpzc{A}},
  & \mathcal{G}^{\mathpzc{A}} &\equiv \partial_{\mu} \mathcal{G}^{\mathpzc{A},\mu} - \frac{\bar{g}_3}{\sqrt{\kappa}} \, f^{\mathpzc{ABC}} \,
  \hat{\mathcal{G}}_{\mu,\mathpzc{B}} \, \mathcal{G}_{\mu, \mathpzc{C}}.
\end{align}

\subsection{Combining SMEFT and SM on shell renormalizations}\label{combiningschemes}

The manner in which the ultraviolet (UV) divergences of the SMEFT combine with those of the SM is subtle.
The counterterm induced modifications in results depend on the renormalization scheme used.
The different schemes at use in the literature
mean that results cannot be casually combined without introducing inconsistent scheme dependence,
that can rise to level of the deviations being searched for and interpreted.
We specify our scheme for combining SMEFT and SM counterterms in some detail here, along with modifications of SM results.

For UV divergences, one has to define a subtraction scheme for the SM and the SMEFT effects.
The SM is renormalized in a combined on shell/$\rm \overline{MS}$ subtraction scheme in ($d= 4 - 2 \epsilon$) dimensional regularization, following \cite{Sirlin:1980nh,Denner:1991kt,Denner:1994xt,Corbett:2020ymv}.
Renormalization constants
\bea
Z_{\hat{h}},Z_{\hat{\mathcal{A}}},Z_{\hat{\mathcal{G}}},Z_{e},Z_{g},Z_v, Z_{m_{W}^2}, Z_{m_{Z}^2},Z_{m_{f}^2}, Z_{m_{h}^2},
\eea
are introduced for the background fields and the couplings
(here a $0/r$ superscript means a bare/renormalized parameter) via
\bea\label{onshell1}
\hat{h}^0 &=& Z^{1/2}_{\hat{h}} \, \hat{h}^{(r)}, \\
\hat{\mathcal{A}}_\mu^0 &=& Z^{1/2}_{\hat{\mathcal{A}}} \, \hat{\mathcal{A}}_\mu^{(r)}, \\
\hat{\mathcal{G}}_\mu^0 &=& Z^{1/2}_{\hat{\mathcal{G}}} \, \hat{\mathcal{G}}_\mu^{(r)}, \\
\bar{e}^0 &=& Z_{e} \, \bar{e}^{(r)} \, \mu^{\epsilon}, \\
\bar{g}_3^0 &=& Z_{g} \, \bar{g}_3^{(r)} \, \mu^{\epsilon}, \\
\bar{v}_T^{0} &=& Z_v^{1/2}  \, \bar{v}_T^{(r)} \label{onshell2},
\eea
and the masses
\bea\label{onshell3}
(\bar{m}_{W}^{(0)})^2 &= Z_{m_{W}^2} \, (\bar{m}_{W}^{(r)})^2,  \quad (\bar{m}_{Z}^{(0)})^2 &= Z_{m_{Z}^2} \,(\bar{m}_{Z}^{(r)})^2, \\
(\bar{m}_{f}^{(0)})^2 &=  Z_{m_{f}^2} \,(\bar{m}_{f}^{(r)})^2,  \quad (\bar{m}_{h}^{(0)})^2 &=  Z_{m_{h}^2} \,(\bar{m}_{h}^{(r)})^2,
\label{onshell4}
\eea
with $m_{f}$ is a mass of fermion field $f$. Here we restrict our results to renormalization factors
relevant to two loop improving $\sigma(\mathcal{G} \,\mathcal{G}\rightarrow h)$,
$\Gamma(h \rightarrow \mathcal{G} \,\mathcal{G})$ and $\Gamma(h \rightarrow \mathcal{A} \mathcal{A})$.\footnote{The CKM entries and massive gauge fields are also
renormalized, see Ref.~\cite{Denner:1994xt} for details.}
In addition a tadpole scheme must be defined. We use an FJ tadpole scheme \cite{Fleischer:1980ub}.
The one loop correction ($\Delta v$) to the vacuum expectation value is fixed
by the condition that the one point function of the Higgs field vanishes, including the factor of $\Delta v$.
As in Ref.~\cite{Corbett:2021cil}, each of the renormalization constants is expanded as $ Z_i = 1 + \Delta Z_i + \cdots$.
Our notation is to use $\Delta Z_i$ for the divergence
chosen to cancel in a $\rm \overline{MS}$ subtraction. The notation $\Delta R_i$ is reserved
for the finite renormalization factors. Again, we generally use $\Delta$ to indicate a loop correction to a Lagrangian parameter
while $\delta$ is used to indicate a SMEFT perturbation $\propto 1/\Lambda^n$.

The full one loop renormalization of $\mathcal{L}^{(6)}$ is only
systematically defined and known for the Warsaw basis \cite{Grzadkowski:2010es},
and is given in Refs.~\cite{Grojean:2013kd,Jenkins:2013zja,Jenkins:2013wua,Alonso:2013hga,Alonso:2014zka}.
These renormalization results are reported in the unbroken phase of the theory with manifest $\rm SU(2) \times U(1)_Y$ symmetry.
The counter terms map consistently to the broken phase of the theory \cite{tHooft:1971qjg,tHooft:1971akt,tHooft:1972tcz}.
This is well known in the SM and also the case in the SMEFT. There are some differing results
due to renormalization scheme dependence in the literature.\footnote{In addition, there is some confusion in the literature on how the
SMEFT counter terms combine with the SM counter terms due to the effects of dimension six operators mixing down.
It is instructive to compare our discussion with Refs.~\cite{Degrande:2014vpa,Deutschmann:2017qum,Degrande:2020evl}.}

Mass terms in the SMEFT can compensate for powers of $1/\Lambda$
in the numerator of divergent terms, even when dimensional regularization is
used. This means that SMEFT UV counter terms can redefine
the running of the SM parameters at the one loop level.
 In the case of SMEFT $\mathcal{L}^{(6)}$ running, the only dimensionful parameter in the SM in the unbroken phase that
can appear in the numerator is the Higgs vacuum expectation value $\bar{v}_T$, or equivalently, the
Higgs mass. These ``mixing down" results are given completely in Ref.~\cite{Jenkins:2013zja}
for the full set of $i$ SM parameters and $\mathcal{L}^{(6)}$. We denote the full set of these renormalization
factors for the $i$ SM parameters by $ \Delta \delta Z^{SM,i}$.

We defined the set of SM parameters as $P_i$, and these parameters are renormalized in an on shell scheme as in Eqns.~\ref{onshell1}-\ref{onshell4}.
This renormalization is denoted as
\bea
\hat{P}_i^{(0)} = Z_{ij}^{SM} \, \hat{P}_j^{(r)},
\eea
and in the SM at one loop $Z_{ij}^{SM} \propto \delta_{ij}$ for the parameters of interest in this calculation.
The leading tree level mapping of the $\Delta \delta Z^{SM}_{ij}$
renormalization factors to the on shell $i$ SM parameters is given at one loop by
\bea
\hat{P}_i^{(0)} = Z_{ij}^{SM} \,  L^{SMEFT}_{jk} \hat{P}_k^{(r)},
\eea
where $L^{SMEFT}_{jk} = \delta_{ij} + \Delta \delta Z^{SM}_{ij} + \cdots$ is a function of the $\Delta \delta Z^{SM}_{ij}$
and follows from a simple set of linear algebra based transformations
between the unbroken/broken phase SM parameters at tree level.
For example, in the simple case of the Higgs mass, from Ref.~\cite{Jenkins:2013zja} one has
\bea
L^{SMEFT}_{m_h^2 m_h^2} = \left(1 + \frac{1}{16 \pi^2 \epsilon} \, \frac{m_h^2}{\Lambda^2} \, \left[C_{HD} - 2 \, C_{H \Box}\right] \cdots \right)
\eea
so that in the BFM with SMEFT $\xi$ gauge fixing \cite{Helset:2018fgq} one has (to one loop order)
\bea
(\bar{m}_{h}^{(0)})^2 &=&  Z^{SM}_{m_{h}^2} \, L^{SMEFT}_{m_h^2 m_h^2}\,(\bar{m}_{h}^{(r)})^2, \\
&=& (\bar{m}_{h}^{(r)})^2 \left(1 +
\frac{(3+ \xi)(\bar{g}_1^2 + 3 \,\bar{g}_2^2)}{64 \pi^2 \epsilon} - \frac{Y}{16 \pi^2 \epsilon}
+
\frac{\left[C_{HD} - 2 \, C_{H \Box}\right]}{16 \pi^2 \epsilon} \, \frac{(\bar{m}_{h}^{(r)})^2}{\Lambda^2} \right),
\eea
with
\begin{align}
Y &= \text{Tr}\left[N_c Y_u^\dagger Y_u + N_c Y_d^\dagger Y_d + Y_e^\dagger Y_e\right].
\end{align}
Our notation for Yukawa matricies is defined in Appendix \ref{setup}.

A more involved example is the top quark. Combining an on-shell renormalization with the SMEFT corrections
\bea
(\bar{m}_{t}^{(0)})^2 &=&  Z^{SM}_{m_{t}^2} \, L^{SMEFT}_{m_t^2 m_t^2} \, (\bar{m}_{t}^{(r)})^2, \\
&=& Z^{SM}_{m_{t}^2} (Z^{SMEFT}_{[Y_u]_{tt}[Y_u]_{tt}})^2  \, (\bar{m}_{t}^{(r)})^2.
\eea
The contributions from the SMEFT only follow from the renormalized top Yukawa and are given
in Eqn.~4.3 of Ref.~\cite{Jenkins:2013zja}. SMEFT running effects on SM parameters
could exist in principal for $Z_v$, but these corrections are related to $Z_{\hat{h}}$ in the BFM.
The latter does not have $\Delta \delta Z_{\hat{h}}$ corrections in the SMEFT \cite{Jenkins:2013zja},
in the unbroken phase of the theory.
In fact, a stronger statement can be made about the lack of such mixing down effects
for $\Delta \delta^n Z_{\hat{h}}$ based on the geoSMEFT. See Appendix \ref{wavefunction} for a short proof to this effect.

In a similar manner as to the $\bar{m}_{t},\bar{m}_{h}$ examples, all SMEFT corrections to the running of the
SM parameters relevant for renormalizing the one loop SM amplitudes for $\sigma(\mathcal{G} \,\mathcal{G}\rightarrow h)$,
$\Gamma(h \rightarrow \mathcal{G} \,\mathcal{G})$ and $\Gamma(h \rightarrow \mathcal{A} \mathcal{A})$
descend from the results in Ref.~\cite{Jenkins:2013zja}.
The SMEFT corrections to the SM amplitude in the background field method,
and $\rm \overline{MS}$, of interest here are to the gluon coupling and the top quark mass (via the top Yukawa).

The corrections in the SMEFT to these counter terms due to $\mathcal{L}^{(6)}$ $\propto \bar{g}_3$ are
\bea\label{firstdelta}
\Delta \delta Z_g &=& - \frac{\bar{m}_h^2}{16 \pi^2 \, \epsilon} (4 \pi)^{\epsilon} e^{- \epsilon \, \gamma_E} \frac{\tilde{C}^{(6)}_{HG}}{\Lambda^2},
\eea
Combining SMEFT and SM corrections in a consistent fashion strongly depends on the
normalization choice for the parameters introduced in $\mathcal{L}^{(6)}$.
If the choice is made that $\tilde{C}^{(6)}_{HG} \propto g_3$ then the corrections in $\Delta \delta Z_g$ shown scale as $g_3$
and should be retained when considering a NLO calculation restricted to QCD couplings. Alternatively
if $\tilde{C}^{(6)}_{HG}$ is defined with a normalization not $\propto g_3$, then including these corrections means improving a calculation
to include mixed terms in the SM couplings and SMEFT Wilson coefficients. Any reasonable choice can be made in terms
of what terms to include and what normalization to choose, so long as consistency
is maintained in the calculation. In the case of the BFM, the corrections in $\Delta Z_g$ in the SM and the SMEFT
cancel against the corrections introduced to the gluon field strength renormalization as
the identity $Z_G \, Z_{g}^2= 1$ is maintained for the counter terms, including finite corrections.

$\Delta Z_{m_t}$ can be extended with a further correction in the SMEFT
that is not due to the gluon correction to the mass renormalization. SMEFT corrections are also
present in $\Delta \delta Z_{m_h^2}^{SMEFT}$.
if one retains terms that are expected to be sizable due to known SM couplings (i.e. known IR
physics of the SM). It is reasonable to choose to retain all of these sets of terms
via a correction $Z_{m_t}^2/Z_{m_h^2} = 1 + 2 \Delta \delta Z_{m_t} - \Delta \delta Z_{m_h^2} + \cdots$.
In this case, we note
\bea
\Delta \delta Z_{m_t} &=&  \frac{1}{32\pi^2 \,\epsilon} \frac{\bar{m}_H^2}{\Lambda^2}  \biggl[3 \, C_{\substack{uH \\ tt}}^* - C_{H \Box} [Y_u]_{tt} + \frac12 C_{H D} [Y_u]_{tt} -
[Y_u]_{tt} \left( C^{(1)}_{\substack{H q \\ tt}} +3 C^{(3)}_{\substack{H q \\ tt}} \right) \nn
&& \hspace{3cm}+ C_{\substack{H u \\ tt}} [Y_u]_{tt} -2 \left( C^{(1)*}_{\substack{qu \\ tttt}} + c_{F,3} C^{(8)*}_{\substack{qu \\ tttt}} \right) [Y_u]_{tt} + \cdots  \biggr]\,, \\
\Delta \delta Z_{m_h^2} &=& \frac{1}{16\pi^2 \,\epsilon} \, \frac{\bar{m}_H^2}{\Lambda^2}  \, \left(C_{H D} - 2 \,C_{H \Box} \right),
\eea
here $c_{F,3} = 4/3$ and $[Y_u]_{tt}$ is the SM top Yukawa.
There are more $\Delta \delta Z^{SMEFT}_{m_t}$ terms that we have suppressed here, as these terms
multiply the known small Yukawa couplings of the SM.

\subsection{Higgs Wavefunction Renormalization $\Delta \delta^n$ effects in geoSMEFT}\label{wavefunction}
Using heat kernel techniques, at one loop order, the Higgs wavefunction renormalization and mass renormalization can be
defined to subtract the divergences present in the theory. The corresponding divergent terms can be written
in this approach at one loop order geometrically
as \cite{tHooft:1973bhk,Alonso:2016oah}
\begin{eqnarray}\label{div}
\mathcal{L}_{div} = \frac{1}{64 \pi^2 \epsilon} \left[- 2 (\nabla^I \nabla^J \mathcal{I}) R_{IKJL} (D \phi)^K (D \phi)^L - (\nabla^I \nabla_J \mathcal{I}) (\nabla^J \nabla_I \mathcal{I}) \right] + \cdots
\end{eqnarray}
Here the dimensional regularization is given by $d= 4 - 2 \epsilon$, $R_{IKJL}$ is the Riemann curvature tensor for the scalar metric $h_{IJ}$
and $\mathcal{I}$ is an invariant scalar density on the scalar manifold.
Note that
\begin{eqnarray}
R_{IKJL} &=& h_{IM} R^M_{KJL}, \\
&=& h_{IM} \left[\partial_J \Gamma^M_{LK}-\partial_L \Gamma^M_{JK} +  \Gamma^M_{JN} \Gamma^N_{LK}
- \Gamma^M_{LN} \Gamma^N_{JK} \right],
\end{eqnarray}
and
\begin{eqnarray}
\Gamma^I_{JK} &=& \frac{1}{2} h^{IL}\left(h_{LJ,K}+ h_{LK,J}- h_{JK,L}\right), \\
\nabla_I \nabla_J \mathcal{I} &=& \frac{\partial^2 \mathcal{I}}{\partial \phi^I \partial \phi^J} - \Gamma^K_{IJ} \frac{\partial \mathcal{I}}{\partial \phi^K}.
\end{eqnarray}
The first term in Eqn.~\eqref{div} with indices $I=J=K=L=4$ corresponds to Higgs wavefunction renormalization in the geoSMEFT,
and in particular possible effects of dimension $d$ higher dimensional operators mixing down modifying the
Higgs wavefunction renormalization proportional to $(v^2/\Lambda^2)^{d-4}$. It is easy to verify that
$R_{4444} \equiv 0$. As a result, this tower of higher dimensional operator mixing down effects exactly vanish at one loop.
In the background field method, this has an important consequence. As a result of the identity
\begin{eqnarray}
\left(\sqrt{Z_v}+ \frac{\Delta^n \delta^m v}{\bar{v}_T^0} \right)_{div} \equiv \Delta^n \delta^m Z_{\hat{h}},
\end{eqnarray}
such corrections to the tadpole corrected vev also vanish for all $n,m \geq 1$. This argument is an example
of the utility of the geoSMEFT and thinking in terms of field space geometries.
Using an operator approach, at each order,
two point functions and four point functions would have to be laboriously
and explicitly evaluated for divergences for each operator, at each mass dimension in the SMEFT
to draw the same conclusion.

The geoSMEFT also makes clear how the mass renormalization
of the Higgs is modified by mixing down effects, introduced to cancel the second term in Eqn.~\eqref{div}.
All of these effects are proportional to the Higgs mass, as this is only dimensionful scale
in the unbroken phase of the theory where the renormalization of the SM and SMEFT corrections can
be carried out \cite{tHooft:1971qjg,tHooft:1971akt,tHooft:1972tcz}.

\section{One Loop Functions}\label{functions}

We define the standard function ($\tau_p = 4 m_p^2/\bar{m}_h^2$)
\begin{align}
A_{1/2}(\tau_p)=- 2 \tau_p \left[1+ (1- \tau_p)f(\tau_p)\right],
\label{eq:A12}
\end{align}
taking $m_t \to \infty$, $A_{1/2}(\tau_f \gg 1) \to -\frac 4 3 + \mathcal O(1/\tau_f)$.
Similarly, we also define
\begin{align}
A_1(\tau_p) &=2 + 3 \tau_p \left[1+ (2- \tau_p)f(\tau_p)\right],
\label{eq:A1}
\end{align}
We also note
\bea
f (\tau_p) =
\begin{cases}
   \arcsin^2 \sqrt{1/\tau_p} ,&  \tau_p \ge 1\\
    - \frac{1}{4} \left[\ln \frac{1 + \sqrt{1- \tau_p}}{1 - \sqrt{1- \tau_p}} - i \pi \right]^2,              & \tau_p <1.
\end{cases}
\eea
Also used are
\begin{align}
\mathcal{I}[m^2] & \equiv   \int_0^1 dx \,\log \left( \frac{m^2-\bar{m}_h^2\, x \, (1-x)}{\bar{m}_h^2} \right)
& \quad
\mathcal{J}_{x}[m^2] & \equiv  \int_0^1 dx \, \frac{x \, m^2}{m^2 - \bar{m}_h^2\, x\, (1-x)}, \\
\mathcal{I}_y[m^2] & \equiv \int_0^{1-x} dy \int_0^1 dx \, \frac{m^2}{m^2 - m_h^2\, x\, (1-x-y)}.
\end{align}
$\mathcal{I}, \mathcal{I}_y, \mathcal{I}_{xx} $ for $\tau \ge 1$ (while restricting our results to top loops) are
\bea
\mathcal{I}[m_p] & \equiv& \log(\frac{\tau_p}{4}) + 2 \sqrt{\tau_p -1} \, \arctan \left(\frac{1}{\sqrt{\tau_p -1}} \right) - 2, \\
\mathcal{I}_y [m_p] &\equiv& \frac{\tau_p}{2} \arcsin^2 (1/\sqrt{\tau_p}), \\
\mathcal{I}_{xx}[m_p] & \equiv& \frac{\tau_p}{\sqrt{\tau_{p} - 1}} \,  \arctan \left(\frac{1}{\sqrt{\tau_{p} - 1}}\right).
\eea

\section{Endpoint regulation} \label{plusdist}
Regulation of the $z=1$ singularity is done with
\bea
(1-z)^{-1 - 2 \epsilon} = \left(\frac{1}{1-z}\right)_+ - 2 \epsilon \left(\frac{\log (1-z)}{1-z}\right)_+ - \frac{1}{2\epsilon} \delta(1-z),
\eea
with plus function definitions
\bea
\int^1_0 d \, x \frac{f(x)}{(x)_+} &=& \int_0^1 d x \frac{f(x)-f(0)}{x}, \\
\int^1_0 d \, f(x) \left(\frac{log(x)}{x}\right)_+ &=& \int_0^1 d x \frac{(f(x)-f(0)) log(x)}{x}.
\eea
The Altarelli-Parisi \cite{Altarelli:1977zs} splitting function is defined as
\bea\label{APfunction}
p_{\mathcal{G} \mathcal{G}}(z) = 2 \, z \, \left(\left(\frac{1}{1-z}\right)_+ - z + \frac{f_1(z)}{z^2} \right) + \frac{\beta_0}{6} \delta(1-z).
\eea
A common function of $z$ is $f_1(z) = z^2-z+1$. A useful distribution identity is
\bea\label{usefulID}
2 \left(\frac{1}{1-z}\right)_+ f_1(z)^2 \equiv z \, p_{\mathcal{G} \mathcal{G}}(z) - \frac{\beta_0}{6} \delta(1-z).
\eea

\section{Common One loop results}
$\hat{v}^2_T$ corresponds
to an experimentally measured extraction of the vacuum expectation value
\bea
\bar{v}_T = \hat{v}_T \left(1 + \Delta G_F + \frac{\delta G_F^{(6)}}{\sqrt{2}}  \right),
\label{eq:deltaGFdefn}
\eea
Here
\bea
\delta G_F^{(6)} \equiv  \frac{1}{\sqrt{2}}
\left(\tilde{C}^{(3)}_{\substack{Hl \\ ee }} +  \tilde{C}^{(3)}_{\substack{Hl \\ \mu\mu }}
- \frac{1}{2}\left(\tilde{C}'_{\substack{ll \\ \mu ee \mu}} +  \tilde{C}'_{\substack{ll \\ e \mu\mu e}}\right)\right).
\eea
The one loop corrections to the vev are \cite{Corbett:2021cil}
\bea\label{vevoneloop}
\Delta G_F = - \frac{\bar{v}_T^2}{4} \Delta L^{V,LL} + \frac{\Delta L_{ew}^{V,LL}}{2},
\eea
with \cite{Kallen:1968gaa,Green:1980bd}
giving
\bea
\Delta L_{ew}^{V,LL} &=& -\frac{\alpha_{ew}}{4 \pi}\left(\pi^2 - \frac{25}{4}\right),
\eea
and the remaining term has been determined in Ref.~\cite{Dekens:2019ept} to be\footnote{Here we have set the evanescent scheme parameter in this result
(bEvan$=1$) to be consistent with naive tree level Fierz identities used in the matching. Note the correction posted in the erratta to Ref.~\cite{Dekens:2019ept} dealing with this issue.}
\bea
\bar{v}_T^2 \, \Delta L^{V,LL} &=&
\frac{\left(7 \bar{m}_{h}^4+\bar{m}_{h}^2
   \left(2 \bar m_t^2 \, N_c -5 \left(2 \bar{m}_{W}^2+\bar{m}_{Z}^2\right)\right)+4
   \left(-4 \, \bar m_t^4 \, N_c +2 \,\bar{m}_{W}^4+ \bar{m}_{Z}^4 \right)\right)}{16 \pi ^2 \,
  \bar{m}_{h}^2 \, \bar{v}_T^2} \nn
&+&\frac{3 (\bar{m}_{h}^4 - 2 \bar{m}_{h}^2 \, \bar{m}_{W}^2)}{8 \pi^2 \, \bar{v}_T^2 (\bar{m}_{h}^2 -\bar{m}_{W}^2)}
  \log \left(\frac{\hat \mu^2}{\bar{m}_{h}^2} \right)
+ \frac{\bar m_t^2 \,N_c \,(\bar{m}_{h}^2 - 4 \bar m_t^2)}{4 \pi^2 \bar{m}_{h}^2 \, \bar{v}_T^2}
\log \left(\frac{\hat \mu^2}{m_t^2} \right) \nn
&+&\frac{3((\bar{m}_{h}^2(\bar{m}_{Z}^4-2 \bar{m}_{W}^2 \bar{m}_{Z}^2) + 2 \bar{m}_{Z}^4 (\bar{m}_{W}^2-\bar{m}_{Z}^2))}
{8 \pi^2 \bar{m}_{h}^2  \bar{v}_T^2 (\bar{m}_{W}^2-\bar{m}_{Z}^2)}
\log \left(\frac{\hat \mu^2}{\bar{m}_{Z}^2} \right) \\
&-& \frac{3 \, \bar{m}_{W}^2 \left(\bar{m}_{h}^4 \left(\bar{m}_{W}^2 -2 \bar{m}_{Z}^2\right)+\bar{m}_{h}^2
   \left(7 \,\bar{m}_{W}^2  \bar{m}_{Z}^2-6 \bar{m}_{W}^4\right)+4 \bar{m}_{W}^4
   ( \bar{m}_{W}^2 - \bar{m}_{Z}^2 )\right)}{8 \pi ^2\bar{m}_{h}^2 \bar{v}_T^2
   (\bar{m}_{h}^2-\bar{m}_{W}^2)(\bar{m}_{W}^2-\bar{m}_{Z}^2)}
\log \left(\frac{\hat \mu^2}{\bar{m}_{W}^2} \right). \nonumber
\eea

The one loop function $\Delta M_1$ is given by \cite{Hartmann:2015oia}
\begin{align}
\Delta M_1 &\equiv
\left(\frac{\Delta R_h}{2} + \frac{\Delta v}{v}
+ \frac{(\sqrt{3} \pi - 6) \lambda}{ 16 \pi^2}
+ \frac{1}{16 \pi^2}\left(\frac{\bar{g}_1^2}{4} + \frac{3 \bar{g}_2^2}{4} + 6 \lambda \right)
\log\left[\frac{\bar{m}_h^2}{\hat \mu^2}\right]\right) \nn
&+ \frac{1}{16 \pi^2}
\left(\frac{\bar{g}_1^2}{4} \mathcal{I}[\bar{m}_Z] + (\frac{\bar{g}_2^2}{4} + \lambda) (\mathcal{I}[\bar{m}_Z] + 2 \mathcal{I}[\bar{m}_W])\right).
\label{eq:M1defn}
\end{align}
This expression is formally dependent in individual terms on a gauge fixing parameter which cancels in the
common sum of terms present in $ \Delta M_1$. See Refs.~\cite{Hartmann:2015oia,Hartmann:2015oia} for details.
We have set $\xi = 1$ in this expression for brevity of presentation.
$\Delta v$ is defined by the condition $T=0$ on \cite{Hartmann:2015oia}
(with $\xi =1$)
\bea
T &=&   \, \bar{m}_{h}^2 \, h \,\bar{v}_T \, \frac{1}{16\pi^2}  \left[- 16 \pi^2 \, \frac{\Delta v}{\bar{v}_T}   + 3  \, \lambda \left(1+ \log \left[\frac{\hat \mu^2}{\bar{m}_{h}^2} \right] \right) + \frac{1}{4} \,  \bar{g}_2^2 \,
  \left(1+ \log \left[\frac{\hat \mu^2}{\bar{m}_{W}^2} \right] \right)  \right.  \\
&\,& \hspace{2.3cm} \left. + \frac{1}{8}(\bar{g}_1^2+ \bar{g}_2^2) \, \left(1+ \log \left[\frac{\hat \mu^2}{\bar{m}_{Z}^2} \right] \right)
- 2 \sum_\psi y_\psi^2  \, N_c \frac{\bar m_\psi^2}{\bar{m}_{h}^2}\left(1+ \log \left[\frac{\hat \mu^2}{\bar m_\psi^2} \right] \right)  \right. \nn
&\,&  \hspace{2.3cm} \left.  + \frac{\bar{g}_2^2}{2} \frac{\bar{m}_{W}^2}{\bar{m}_{h}^2} \left(1+ 3\log \left[\frac{\hat \mu^2}{\bar{m}_{W}^2} \right] \right)
 +\frac{1}{4} (\bar{g}_1^2 + \bar{g}_2^2) \frac{\bar{m}_{Z}^2}{\bar{m}_{h}^2} \left(1+ 3\log \left[\frac{\hat \mu^2}{\bar{m}_{Z}^2} \right] \right) \right]. \nonumber
\eea

The finite results for the Higgs wavefunction renormalization
in the BFM are \cite{Hartmann:2015oia}
\begin{align}
16 \, \pi^2 \,  \Delta R_{h} &=   2 \,  \lambda \, \Big(6  - \sqrt{3} \pi - \mathcal{J}_x[\bar{m}_Z^2] - 2 \, \mathcal{J}_x[\bar{m}_W^2] \Big)
+2 \, \bar{g}_2^2 \Bigg(
  \Big( \mathcal{J}_x[\bar{m}_W^2]- \frac{1}{2} \Big)  \,  \left(1 -\frac{3 \bar{m}_W^2}{\bar{m}_h^2} \right)  - \mathcal{I}[\bar{m}_W^2]\Bigg)
\nn
 & +\left(\sum_\psi \, y_\psi^2 \, N_c - \bar{g}_1^2-3  \bar{g}_2^2\right) \log
    \left(\frac{\bar{m}_h^2}{\hat \mu ^2}\right)  +\left( \bar{g}_1^2+ \bar{g}_2^2\right) \, \Bigg( \Big( \mathcal{J}_x[\bar{m}_Z^2] - \frac{1}{2}\Big)\left(1 - \frac{3
   \bar{m}_Z^2}{\bar{m}_h^2}\right) -\mathcal{I}[\bar{m}_Z^2] \Bigg)
   \nn
   & +\sum_\psi \, y_\psi^2 \, N_c \Bigg(1+
   \left(1+\frac{2 \bar m_\psi^2}{\bar{m}_h^2}\right)\, \mathcal{I}[\bar m_\psi^2] -\frac{2 \bar m_\psi^2}{\bar{m}_h^2} \,\log
   \left(\frac{\bar m_\psi^2}{\bar{m}_h^2}\right) \Bigg).
\end{align}
We also use
\bea
\Delta R_{\hat{\mathcal{A}}} = \frac{\bar{g}_1^2 \bar{g}_2^2}{(\bar{g}_1^2 + \bar{g}_2^2)} \,
\left[-\frac{7}{16 \pi^2} \log \left(\frac{\hat \mu^2}{\bar{m}_W^2} \right)
+ \sum_{\psi} \frac{N^\psi_c Q^2_\psi}{12 \pi^2}
\,  \log \left(\frac{\hat \mu^2}{\bar{m}_\psi^2} \right) - \frac{1}{24 \pi^2}\right].
\eea
This result was successfully verified comparing to the explicit calculation reported in
in Ref.~\cite{Dekens:2019ept}.
The Ward identities in the SMEFT in the BFM \cite{Corbett:2019cwl}
have been validated at one loop \cite{Corbett:2020ymv,Corbett:2020bqv}.
These identities also give
\bea
\Delta Z_e  &= - \frac{1}{2} \Delta Z_{\hat{\mathcal{A}}}, \nonumber \\
\Delta R_e  &= - \frac{1}{2} \Delta R_{\hat{\mathcal{A}}}.
\eea
In the $\{\hat{m}_{W}, \hat{m}_Z, \hat{G}_F \}$ scheme one has \cite{Corbett:2021cil}
\begin{align}
\frac{\Delta g_1}{\hat{g}_1} &= \frac{\Delta G_F}{2}+ \frac{\Delta R_{m_W} \, \hat{m}_W^2- \Delta R_{m_Z} \, \hat{m}_Z^2}{\hat{m}_W^2-\hat{m}_Z^2}, \\
\frac{\Delta g_2}{\hat{g}_2} &= \frac{\Delta G_F}{2} + \Delta R_{m_W},
\end{align}
while the $\{\hat{\alpha}_{ew}, \hat{m}_Z, \hat{G}_F \}$ scheme defines \cite{Corbett:2021cil}
\begin{align}
\frac{\Delta g_1}{\hat{g}_1} &= \frac{\Delta e}{\hat{e}} -\frac{\Delta c_\theta}{c_{\hat{\theta}}}, \\
\frac{\Delta g_2}{\hat{g}_2} &= \frac{\Delta e}{\hat{e}} -\frac{\Delta s_\theta}{s_{\hat{\theta}}},
\end{align}
where
\bea
\frac{\Delta s_\theta}{s_{\hat{\theta}}} = \frac{1 - s_{\hat{\theta}}^2}{2 \, (1 - 2 s_{\hat{\theta}}^2)} \left[ \frac{\Delta \alpha}{\alpha}
-  \Delta G_F -2 \, \Delta R_{M_Z} \right].
\eea
The BFM expressions for $\Delta R_{m_W,m_Z}$ are somewhat lengthy and given in the Appendix Ref.~\cite{Corbett:2021cil}.

\section{$\Delta \delta \sigma (\mathcal{G} \mathcal{G} \rightarrow h)$ and quadratic $\delta^2 \sigma (\mathcal{G} \mathcal{G} \rightarrow h)$ results}\label{past}
Explicitly, in the $m_t \rightarrow \infty$ limit,
the leading results for the interference with the SM one loop amplitude are \cite{Georgi:1977gs,Gounaris:1998ni,Manohar:2006gz}
\bea
\label{eq:SMgghLO}
\Delta^2 \hat{\sigma}^{SM,m_t \rightarrow \infty}_{LO}(\mathcal{G} \mathcal{G} \rightarrow h)&\equiv&
\frac{\pi}{4} \lim_{\epsilon \rightarrow 0}
\left|\Delta C^{SM,m_t \rightarrow \infty}_{h \mathcal{G} \mathcal{G}}\right|^2, \nn
&=& \frac{(\alpha_s^{(r)})^2}{576 \, \pi \, \bar{v}_T^2},
\eea
while
\bea
\frac{\Delta \delta \hat{\sigma}(\mathcal{G} \mathcal{G} \rightarrow h)[\tilde C^{(6)}_{HG}]}
{\Delta^2 \hat{\sigma}^{SM,m_t \rightarrow \infty}_{LO}(\mathcal{G} \mathcal{G} \rightarrow h)} &=&
\frac{24 \, \pi}{\alpha_s^{(r)}} \, \tilde{C}_{HG},
\eea
and a contribution at $\delta^2$ order (in the $m_t \rightarrow \infty$ limit) is
\bea
\frac{\delta^2 \hat{\sigma}(\mathcal{G} \mathcal{G} \rightarrow h)[(\tilde C^{(6)}_{HG})^2]}{\Delta^2 \hat{\sigma}^{SM}_{LO}(\mathcal{G} \mathcal{G} \rightarrow h)}&=&
9 \left(\frac{4 \, \pi}{\alpha_s^{(r)}}\right)^2 (\tilde{C}^{(6)}_{HG})^2.
\eea
The results for these ratios reported
in in Ref.~\cite{Corbett:2021cil} were further scaled by a correction factor
of $(1 + 11 \, \alpha_s^{(r)}/2 \pi)^{-1} \simeq 1.21$, using $\alpha_s \simeq 0.118$
due to the inclusion of the partial NLO result easily
retaining by the two loop matching correction to the SM result.
Using Ref.~\cite{Anastasiou:2020qzk} we can improve this rough approximation
(while still in the $m_t \rightarrow \infty$ limit) using
\bea\label{SMNLO}
\frac{\Delta^3 \sigma^{SM}}{\Delta^2 \, \hat{\sigma}^{SM}_{LO, \epsilon \rightarrow 0}}
&=& \frac{\alpha_s^{(r)}}{4 \pi}
\left[4 \pi^2 + 22\right] \, \delta(1-z) + \frac{6 \, \alpha_s^{(r)} \, f^2_1(z)}{\pi} \log \left(\frac{s}{\hat \mu^2} \right)\left(\frac{1}{1-z}\right)_+
- 11 \frac{\alpha_s^{(r)}}{2 \pi} \, (1-z)^3 \nn
&+& \frac{3 \, \alpha_s^{(r)}}{\pi}  \, z \,  p_{\mathcal{G}\mathcal{G}}(z) \, \log \left(\frac{\hat{\mu}^2}{\mu_F^2} \right)
+ \frac{12 \, \alpha_s^{(r)}}{\pi} \, f^2_1(z)  \, \left(\frac{\log(1-z)}{1-z}\right)_+.
\eea
Here we used the AP counterterm that accounts for $L_{\hat m_t}$ dependence
\bea
\Delta^2 \delta \sigma_{DR \, c.t}^{AP} &\equiv& \Delta^2 \hat{\sigma}^{SM}_{LO, \epsilon \rightarrow 0}(\mathcal{G} \,\mathcal{G} \to h) \frac{3 \alpha^{(r)}_s}{2 \, \pi} \left[\left(\frac{\mu^2}{\mu_F^2}\right)^\epsilon \right] \nn
&\times& (4 \pi)^\epsilon \frac{\Gamma(1+ \epsilon) \Gamma(1- \epsilon)^2}{\Gamma(1- 2 \epsilon)} \left[\frac{1}{\epsilon} + 1 - 2 L_{\hat{m}_t}\right]\,z \, p_{\mathcal{G}\mathcal{G}}(z).
\eea

\subsection{$\Delta \delta \sigma (\mathcal{G} \mathcal{G} \rightarrow h)$}
The contributions to $\langle \mathcal{G} \mathcal{G}|h \rangle^1_{\mathcal O(v^2/\Lambda^2)}$
that need to be added to Eqn.~\eqref{SMEFTNLO} follow from the following perturbations.
We express these various terms as \cite{Corbett:2021cil}
\bea
\langle \mathcal{G} \mathcal{G}|h\rangle^1_{\mathcal{O}(v_T^2/\Lambda^2)} =
- 4[\Delta G_F + \Delta M_1]\, \frac{\tilde C_{HG}^{(6)}}{\hat v_T} \, K_{ab}
- 4 \left(\frac{\tilde C_i \, \Delta f_i}{16 \, \pi^2 \, \hat v^2_T}\right) \, \hat{v}_T  \, K_{ab}.
\eea
where $\tilde C_i\, \Delta f_i$ contains all corrections -- from operator mixing and  $\mathcal O(\bar v^2_T/\Lambda^2)$ corrections to the SM
-- that are not proportional to $\tilde C^{(6)}_{HG}$.
The $\Delta f_i$ are \cite{Hartmann:2015aia,Deutschmann:2017qum,Grazzini:2016paz} (using $\tau_p = 4 m_p^2/\bar{m}_h^2$)
\bea
\label{eq:cifi}
\Delta f_{H\Box} &=& -\sum_f \, \alpha_s \pi \,  A_{1/2}(\tau_f), \\
\Delta f_{HD} &=& \frac{1}{4} \, \sum_f \, \alpha_s \pi \,  A_{1/2}(\tau_f), \\
\Delta f_{\delta G_F} &=& \frac{1}{\sqrt{2}} \, \sum_f \, \alpha_s \pi \, A_{1/2}(\tau_f), \\
\left[Y_{\psi'_{ff}}\right] \, \Delta f_{\substack{\psi' H \\ ff}} &=&  \sum_f \, \alpha_s \pi \,  A_{1/2}(\tau_f).
\eea
In practice, contributions from light fermions to the $\Delta f_i$ are suppressed since $A_{1/2}(\tau_f \ll 1) \sim \tau \sim \left[Y_{\psi'_{ff}}\right]^2 $, so we will only include effects from the top and bottom quarks.
The dipole operators enter at one loop \cite{Hartmann:2015aia,Deutschmann:2017qum,Grazzini:2016paz}, the only term which enter at $\mathcal O(v^2_T/16\pi^2\Lambda^2)$ (again retaining only the $Y_{t,b}$ terms)
are the $\mathcal{L}^{(6)}$ operators $\tilde{C}_{\substack{uG\\ tt}}$ and $\tilde{C}_{\substack{dG\\ bb}}$:
\bea\label{oneloopfirstterm}
\langle \mathcal{G} \mathcal{G}|h \rangle^1_{\mathcal O(v^2/\Lambda^2)} \supset -  \frac{\sqrt{\kappa} \, \bar{g}_3}{8\pi^2 \, \bar{v}_T}
 \, \left(\Delta f_{uG}\, \tilde{C}_{\substack{uG\\ tt}}[Y]_{tt}+ \Delta f_{dG}\, \tilde{C}_{\substack{dG\\ bb}}[Y]_{bb} + h.c. \right) K_{ab},
\eea
where
\bea
\Delta f_{uG}  &=&  \left[-1 +  2 \, \log \left(\frac{\hat \mu^2}{\hat{m}_h^2} \right) +  \, \log \left(\frac{4}{\tau_{t}} \right)
 \right] - 2 \, \mathcal{I}_{y} [m_t^2] -  \mathcal{I}[m_t^2], \nn
 \Delta f_{dG} & = & \left[-1 + 2 \, \log \left(\frac{\hat \mu^2}{\hat{m}_h^2} \right) +  2\, \log \left(\frac{4}{\tau_{b}} \right)\right] - \tau_b \, f(\tau_b)
  - 4 \, i \,\sqrt{1-\tau_b} \, f^{1/2}(\tau_b).
  \eea

\noindent This set of $\Delta^2 \delta$ corrections in the $m_t \to \infty$ limit are \cite{Corbett:2021cil}
\bea
  \frac{\Delta^2 \delta \sigma(\mathcal{G} \,\mathcal{G} \to h)}{\Delta^2 \, \hat{\sigma}^{SM}_{LO, \epsilon \rightarrow 0}(\mathcal{G} \,\mathcal{G} \to h)} =
  \frac{24 \pi}{ \alpha_s^{(r)}} \,
\left[[\Delta G_F + \Delta M_1]\, \tilde C_{HG}^{(6)}
 + {\rm{Re}} \left(\frac{\tilde C_i \, \Delta f_i}{16 \, \pi^2}\right)\right] \delta(1-z).
\eea

\subsection{$\delta^2 \sigma (\mathcal{G} \mathcal{G} \rightarrow h)$ geoSMEFT terms}

The $\mathcal O(v^4_T/\Lambda^4)$ terms are denoted as $\delta^2$ terms.
There are two sets of terms of this order. One that follows from
the ``self-square" or ``quadratic" term of the tree level $\tilde{C}_{HG}$ dependence, and
a further set of terms that are obtained consistently expanding to $\delta^2$ order.
In this subsection we report the later set of terms.

For the $\mathcal{G} \,\mathcal{G} \to h$ amplitude these corrections are \cite{Corbett:2021cil}
\begin{align}
 \langle \mathcal{G}\mathcal{G}|\phi_4 \rangle^0_{\mathcal O(v^4/\Lambda^4)} &=  \langle \sqrt{h}^{44}\rangle_{\mathcal O(v^2/\Lambda^2)} \langle \mathcal{G}\mathcal{G}|\phi_4 \rangle^0_{\mathcal O(v^2/\Lambda^2)} \\ \nonumber
&\quad\quad\quad\quad\quad\quad  + 2\,\frac{ v_T [\langle \mathcal{G}\mathcal{G}|\phi_4 \rangle^0_{\mathcal O(v^2/\Lambda^2)}]^2}{\langle \mathcal G\, \mathcal G | \phi_4 \rangle^0}  + \, (\langle \mathcal{G}\mathcal{G}|\phi_4 \rangle^0_{\mathcal O(v^2/\Lambda^2)})\Big|_{\tilde C^{(6)}_{HG} \to \tilde C^{(8)}_{HG}}
\end{align}
where $\langle \sqrt{h}^{44}\rangle_{\mathcal O(v^2/\Lambda^2)} = \tilde C^{(6)}_{H\Box} - \frac 1 4 \tilde C^{(6)}_{HD}$. A term from
the redefinition of $\bar{v}_T$ in its relation to input observables is formally present but suppressed as it cancels
when the SM amplitude is interfered with, which is $\propto 1/\bar{v}_T$.
These dimension eight interference corrections in the $m_t \to \infty$ limit are \cite{Corbett:2021cil}
\begin{align}
\frac{\Delta \delta^2 \sigma(\mathcal{G} \,\mathcal{G} \to h)}{\Delta^2 \, \hat{\sigma}^{SM}_{LO, \epsilon \rightarrow 0}(\mathcal{G} \,\mathcal{G} \to h)} =
\frac{24 \pi}{ \alpha_s^{(r)}} \, \left[\left(\langle \sqrt{h}^{44}\rangle_{\mathcal O(v^2/\Lambda^2)} + 2 \tilde C^{(6)}_{HG}\right) \tilde C^{(6)}_{HG} +  \tilde C^{(8)}_{HG}\right] \delta(1-z).
\end{align}

Note that taking the ``quadratic" dependence on $\tilde{C}_{HG}^2$ (the square of the $\delta$ correction due to this operator)
does not generate all terms dependent on $\tilde{C}_{HG}^2$ in the observable. See Ref.~\cite{Hays:2020scx}
for more discussion.

\section{$\Delta \delta \Gamma (h \rightarrow \mathcal{G} \mathcal{G})$ and quadratic $\delta^2 \Gamma (h \rightarrow \mathcal{G} \mathcal{G})$ results}\label{pasthgg}
The results unchanged from Ref.~\cite{Corbett:2021cil} are as follows.
The leading order result in the $m_t \rightarrow \infty$ limit is
\bea
\Delta^2 \Gamma(h \to \mathcal{G} \,\mathcal{G})_{SM} \equiv
\frac{2}{\pi} \, \hat{m}_h^3  \lim_{\epsilon \rightarrow \infty} |\Delta C^{SM,m_t \rightarrow \infty}_{h \mathcal{G} \mathcal{G}}|^2,
\eea
leading to
\bea
\frac{\Delta \delta \Gamma(h \to \mathcal{G} \,\mathcal{G})}{\Delta^2 \Gamma(h \to \mathcal{G} \,\mathcal{G})_{SM}}
\equiv \frac{24 \pi}{\alpha_s^{(r)}} \tilde{C}^{(6)}_{HG},
\eea
and
\bea
\frac{\delta^2 \Gamma(h \to \mathcal{G} \,\mathcal{G})}{\Delta^2 \Gamma(h \to \mathcal{G} \,\mathcal{G})_{SM}}
 \equiv 9 \left(\frac{4 \pi}{\alpha_s^{(r)}}\right)^2 \, (\tilde{C}^{(6)}_{HG})^2.
\eea

\subsection{$\Delta \, \delta \Gamma(h \to \mathcal{G} \,\mathcal{G})$}
As previously reported in Ref.~\cite{Corbett:2021cil}, the EW correction is identical
to the case of $\sigma(\mathcal{G} \mathcal{G} \rightarrow h)$,
\begin{align}
\frac{\Delta \delta \Gamma(h \to \mathcal{G} \,\mathcal{G})^{m_t\to \infty}_{EW}}{\Delta^2 \Gamma(h \to \mathcal{G} \,\mathcal{G})_{SM}} =
\frac{24 \pi}{\alpha_s}\times \left(\left[\Delta G_F + \Delta M_1 + \Delta R_\mathcal{G}\right] \, \tilde C^{(6)}_{HG} + \sum_{i} \,  \frac{{\rm Re} \, \tilde C_i^{(6)} \Delta f_i^{(6)}}{16 \pi^2}
\right).
\end{align}
In this expression we also include the
BFM wavefunction renormalization finite  factor of the final state gluons
\bea
\Delta R_\mathcal{G} = \frac{1}{24 \pi^2} \sum_f \log \left(\frac{m_f^2}{\hat{\mu}^2}\right),
\eea
as the $\tilde C^{(6)}_{HG}$ operator was not redefined to
rescale it by $g_3^2$. Note that in the BFM this has the result
of the $\Delta R_\mathcal{G}$ contribution not canceling against a corresponding finite term for $g_3^2$, but contributing.

\subsection{$\delta^2 \Gamma(h \to \mathcal{G} \,\mathcal{G})$ geoSMEFT terms}
The $\Delta \delta^2$ terms for this decay that follow from the geoSMEFT
and added to the naive Quadratic terms are
\begin{align}
\frac{\Delta \delta^2 \Gamma(h \to \mathcal{G} \,\mathcal{G})}{\Delta^2 \Gamma(h \to \mathcal{G} \,\mathcal{G})_{SM}} =
\frac{24 \pi}{ \alpha_s^{(r)}} \, \left[\left(\langle \sqrt{h}^{44}\rangle_{\mathcal O(v^2/\Lambda^2)} + 2 \tilde C^{(6)}_{HG}\right) \tilde C^{(6)}_{HG} +  \tilde C^{(8)}_{HG}\right].
\end{align}

\section{$\Delta \delta \Gamma (h \rightarrow \mathcal{A} \mathcal{A})$ and quadratic $\delta^2 \Gamma (h \rightarrow \mathcal{A} \mathcal{A})$ results}\label{pasthaa}
We define \cite{Corbett:2021cil}
\begin{align}
&\langle h|\mathcal{A} \mathcal{A}\rangle_{SM}^1 = \frac{- \hat{g}_2 \, \hat{e}^2}{64 \, \pi^2 \, \hat{m}_W}
\Bigg(A_1(\tau_W)+ \sum_i \, N_c^i \,Q^2_i \, A_{1/2} (\tau_{\psi^i})\Bigg) \, \langle h \mathcal{A}^{\mu\nu} \mathcal{A}_{\mu \nu} \rangle^0,
\end{align}
with $\psi^i$ a mass eigenstate fermion and the loop functions are reported in Appendix~\ref{functions}.
For notational convenience we define a short hand notation
\bea
A_{\mathcal{A} \mathcal{A}} \equiv \Bigg(A_1(\tau_W)+ \sum_i \, N_c^i \,Q^2_i \, A_{1/2} (\tau_{\psi^i})\Bigg).
\eea
For later convenience we define
\bea
A^{\mathcal{L}^{(6)}}_{\mathcal{A} \mathcal{A}} \equiv \left[\frac{\hat{g}_2^2 \,   \tilde{C}_{HB}^{(6)}
  + \hat{g}_1^2 \,  \tilde{C}_{HW}^{(6)} - \hat{g}_1 \, \hat{g}_2 \,  \tilde{C}_{HWB}^{(6)}}{(\hat{g}^{\rm SM}_Z)^2} \right].
\eea
Directly one has
\begin{align}
	\label{eq:hgamgamDim6}
\langle h|\mathcal{A} \mathcal{A}\rangle^0_{\mathcal{O}(\bar{v}^2/\Lambda^2)} &= \frac{A^{\mathcal{L}^{(6)}}_{\mathcal{A} \mathcal{A}}}{\hat{v}_T} \, \langle h \mathcal{A}^{\mu\nu} \mathcal{A}_{\mu \nu} \rangle^0,
\end{align}
leading to the $\Delta \delta$ contribution
\bea
\frac{\Delta \delta \Gamma(h \rightarrow \mathcal{A} \mathcal{A})}{\Delta^2 \Gamma_{SM}(h \rightarrow \mathcal{A} \mathcal{A})}
= - 64 \pi^2 \,  \frac{{\rm Re}A_{\mathcal{A} \mathcal{A}}}{|A_{\mathcal{A} \mathcal{A}}|^2}
\left[\frac{\tilde{C}_{HB}^{(6)}}{\hat{g}_1^2}
  + \frac{\tilde{C}_{HW}^{(6)}}{\hat{g}_2^2} - \frac{\tilde{C}_{HWB}^{(6)}}{\hat{g}_1 \, \hat{g}_2}\right],
\eea
and a $\delta^2$ contribution
\bea
\frac{\delta^2 \Gamma(h \rightarrow \mathcal{A} \mathcal{A})}{\Delta^2 \Gamma_{SM}(h \rightarrow \mathcal{A} \mathcal{A})}
=  \frac{1024 \pi^4}{|A_{\mathcal{A} \mathcal{A}}|^2}
\left[\frac{\tilde{C}_{HB}^{(6)}}{\hat{g}_1^2}
  + \frac{\tilde{C}_{HW}^{(6)}}{\hat{g}_2^2} - \frac{\tilde{C}_{HWB}^{(6)}}{\hat{g}_1 \, \hat{g}_2}\right]^2.
\eea
\subsection{$\Delta^2 \delta \Gamma (h \rightarrow \mathcal{A} \mathcal{A})$}
\bea
\langle h|\mathcal{A} \mathcal{A}\rangle^1_{\mathcal{O}(v_T^2/\Lambda^2)} &=&
\langle h | [C^{(6)}_{i}] |\mathcal{A} \mathcal{A}\rangle^1 \frac{\langle h \mathcal{A}^{\mu\nu} \mathcal{A}_{\mu \nu} \rangle_0}{\hat v_T}
+ \left(\frac{A^{\mathcal{L}^{(6)}}_{\mathcal{A} \mathcal{A}}}{\bar{v}_T^2} \, \Delta M_1
 +  \frac{\tilde{C}_i \, \Delta f_i}{16 \, \pi^2 \,\hat{v}_T^2}\right) \, \hat{v}_T  \, \langle h \mathcal{A}^{\mu\nu} \mathcal{A}_{\mu \nu} \rangle^0. \nn
\eea
Here we have redefined notation slightly from Ref.~\cite{Corbett:2021cil} and explicitly
\bea
\langle h | [C^{(6)}_{i}] |\mathcal{A} \mathcal{A}\rangle^1 &=&
  \frac{\hat{e}^2}{32 \, \pi^2} \, A_{\mathcal{A} \mathcal{A}}
  \left[\frac{\delta G_F}{\sqrt{2}} -\frac{\delta \alpha}{\alpha}\right]
 + \frac{2.1 \, \hat{e}^2}{16 \, \pi^2 } \, \frac{\delta M_W^2}{\hat{M}_W^2} \\
&+& \left(\Delta R_A + \Delta G_F + \frac{2 \hat{g}_1 (\Delta g_2 \hat{g}_1 - \Delta g_1 \hat{g}_2)}
 {\hat{g}_2 \, (\hat{g}^{\rm SM}_Z)^2} \right)
 \frac{\hat{g}_2^2 \, \tilde{C}^{(6)}_{HB}}{(\hat{g}^{\rm SM}_Z)^2}\nn
&+&
  \left(\Delta R_A + \Delta G_F
 + \frac{2 \hat{g}_2 (\Delta g_1 \hat{g}_2 - \Delta g_2 \hat{g}_1)}
 {\hat{g}_1 \, (\hat{g}^{\rm SM}_Z)^2} \right)
 \frac{\hat{g}_1^2 \, \tilde{C}^{(6)}_{HW}}{(\hat{g}^{\rm SM}_Z)^2} \nn
&-& \left(\Delta R_A + \Delta G_F
 + \frac{(\hat{g}_1^2-\hat{g}_2^2) (\Delta g_2 \hat{g}_1-\Delta g_1 \hat{g}_2)}{\hat{g}_1 \hat{g}_2 (\hat{g}^{\rm SM}_Z)^2}\right)
 \frac{\hat{g}_1 \, \hat{g}_2 \, \tilde{C}^{(6)}_{HWB}}{(\hat{g}^{\rm SM}_Z)^2}. \nonumber
\eea
The remaining $\Delta f_i$'s are in Refs.~\cite{Hartmann:2015oia,Hartmann:2015aia}
in terms of the one loop functions are
\bea
\frac{\hat{g}_1 \,\hat{g}_2}{\hat{e}^2} \Delta f_{HWB} &=&  \left(- 3 \, \hat{g}_2^2 +4 \, \lambda  \right) \,  \log \left( \frac{\bar{m}_h^2}{\hat \mu^2}\right) + (4 \lambda - \hat{g}_2^2) \, \mathcal{I} [\bar{m}_W^2]
- 4 \, \hat{g}_2^2 \, \mathcal{I}_y [\bar{m}_W^2] - 2 \, \hat{g}_2^2 \left[1+ \log \left( \frac{\tau_{W}}{4}\right) \right] \nonumber \\
&+& \hat{e}^2 \, (2+ 3 \, \tau_{W})+ 6 \, \hat{e}^2 (2 - \tau_{W}) \, \, \mathcal{I}_y [\bar{m}_W^2],
\eea
\begin{align}
\frac{\hat{g}_2^2}{\hat{e}^2} \Delta f_{HW} &= - \hat{g}_2^2 \Bigg[3 \, \tau_{W} + \left(16 - \frac{16}{\tau_{W}}- 6 \, \tau_{W}\right) \, \mathcal{I}_y [\bar{m}_W^2] \Bigg],  \\
\frac{\hat{g}_2^3}{\hat{e}^2} \Delta f_{W} &= - 9 \, \hat{g}_2^4 \,  \log \left( \frac{\bar{m}_h^2}{\hat \mu^2}\right) - 9 \, \hat{g}_2^4 \,  \mathcal{I} [\bar{m}_W^2] - 6 \, \hat{g}_2^4 \,  \mathcal{I}_y [\bar{m}_W^2]
+ 6  \, \hat{g}_2^4 \,  \mathcal{I}_{xx} [\bar{m}_W^2] \, \left(1-1/\tau_{W}\right) - 12 \, \hat{g}_2^4, \\
\frac{\hat{g}_1}{\hat{e}^2} \Delta f_{\substack{eB \\ ss}}  &= 2 \, Q_\ell \, [Y_\ell]_{ss}  \left[-1 +  2 \, \log \left(\frac{\hat \mu^2}{\bar{m}_h^2} \right) +  \, \log \left(\frac{4}{\tau_{s}} \right) \right]
- 2 \, Q_\ell \, [Y_\ell]_{ss} \, \Big[ 2 \, \mathcal{I}_{y} [m_s^2] + \mathcal{I}[m_s^2] \Big], \\
\frac{\hat{g}_1}{\hat{e}^2} \Delta f_{\substack{uB \\ ss}}  &= 2 \, N_c \, Q_u \, [Y_u]_{ss}  \left[-1 +  2 \, \log \left(\frac{\hat \mu^2}{\bar m_h^2} \right) +  \, \log \left(\frac{4}{\tau_{s}} \right) \right]
- 2 \, Q_u \, [Y_u]_{ss} \, \Big[ 2 \, \mathcal{I}_{y} [m_s^2] + \mathcal{I}[m_s^2] \Big], \\
\frac{\hat{g}_1}{\hat{e}^2} \Delta f_{\substack{dB \\ ss}}  &= 2 \, N_c \, Q_d \, [Y_d]_{ss}  \left[-1 +  2 \, \log \left(\frac{\hat \mu^2}{\bar{m}_h^2} \right) +  \, \log \left(\frac{4}{\tau_{s}} \right) \right]
- 2 \, Q_d \, [Y_d]_{ss} \, \Big[ 2 \, \mathcal{I}_{y} [m_s^2] + \mathcal{I}[m_s^2] \Big].
\end{align}
Note $\hat{g}_2 \Delta f_{\substack{eW \\ ss}} \rightarrow  - \hat{g}_1 \, \Delta f_{\substack{eB \\ ss}}$.
In the case of up quarks $\hat{g}_2 \Delta f_{\substack{uW \\ ss}} \rightarrow  \hat{g}_1 \Delta f_{\substack{uB \\ ss}}$, while in the case
of down quarks $\hat{g}_2 \Delta f_{\substack{dW \\ ss}} \rightarrow  - \hat{g}_1 \Delta f_{\substack{dB \\ ss}}$.
Here, $s=\{1,2,3\}$ sums over the flavors. The Wilson coefficients
are summed with their Hermitian conjugates, and the normalization is such that $\Delta f_{\substack{eB \\ ss}}$ multiplies
${\rm Re} \, C_{\substack{eB \\ ss}}$.
The remaining contributions proportional to the SM loop functions are
\begin{align}
[Y_e]_{ss} \, \Delta f_{\substack{eH \\ ss}} &= \hat{e}^2\, \frac{Q_\ell^2}{2} A_{1/2}(\tau_s), \nonumber \\
[Y_u]_{ss} \, \Delta f_{\substack{uH \\ ss}} &= N_c \, \hat{e}^2\, \frac{Q_u^2}{2} A_{1/2}(\tau_s),  \nonumber \\
 [Y_d]_{ss} \, \Delta f_{\substack{dH \\ ss}} &= N_c \, \hat{e}^2\, \frac{Q_d^2}{2} A_{1/2}(\tau_s),  \nonumber \\
\Delta f_{H \Box} &= -\hat{e}^2 \frac{Q_\ell^2}{2} A_{1/2}(\tau_p) - N_c \, \hat{e}^2\, \frac{Q_u^2}{2} A_{1/2}(\tau_r)
 -  N_c \, \hat{e}^2 \,\frac{Q_d^2}{2} A_{1/2}(\tau_s) - \frac{1}{2} \, \hat{e}^2\, A_1(\tau_{W}),
\end{align}
and $\Delta f_{HD}  = -\Delta f_{H \Box} /4$. Here $p,r,s$ run over $1,2,3$ as flavor indices. Several of these results
have been cross checked against Ref.~\cite{Ghezzi:2015vva}.

These input parameter scheme dependent corrections perturb $\Gamma(h \rightarrow \mathcal{A} \mathcal{A})$ as
\bea
\frac{\Delta^2 \delta \Gamma(h \rightarrow \mathcal{A} \mathcal{A})}{\Delta^2 \Gamma_{SM}(h \rightarrow \mathcal{A} \mathcal{A})}
\simeq \frac{- 16 \pi^2}{\hat \alpha_{ew} \, |A_{\mathcal{A} \mathcal{A}}|^2} {\rm Re} A_{\mathcal{A} \mathcal{A}} \left[
\langle h | [C^{(6)}_{i}] |\mathcal{A} \mathcal{A}\rangle^1
+ A^{\mathcal{L}^{(6)}}_{\mathcal{A} \mathcal{A}} \, \Delta M_1  + \frac{\tilde{C}_i \, \Delta f_i}{16 \, \pi^2}\right].
\eea
\subsection{$\delta^2 \Gamma(h \to \mathcal{A} \,\mathcal{A})$ geoSMEFT terms}
The  ${\cal{O}}(v^4/\Lambda^4)$ terms in the full three-point function are \cite{Hays:2020scx}
\begin{align}
	\label{eq:hggDim8}
	\langle h|\mathcal{A}\mathcal{A} \rangle_{{\cal{O}}(v^4/\Lambda^4)}^0 &=
   \langle \sqrt{h}^{44}\rangle_{{\cal{O}}(v^2/\Lambda^2)} \, \langle h|\mathcal{A} \mathcal{A}\rangle^0_{\mathcal{O}(\bar{v}^2/\Lambda^2)}
  + 2  \, \frac{\bar{v}_T \, [\langle h|\mathcal{A}\mathcal{A}\rangle_{\mathcal{O}(\bar{v}^2/\Lambda^2)}^0]^2}{\langle h \mathcal{A}^{\mu\nu} \mathcal{A}_{\mu \nu} \rangle^0} \\
   &+ 2 \, \langle h|\mathcal{A}\mathcal{A}\rangle_{\mathcal{O}(\bar{v}^2/\Lambda^2)}^0|_{C_i^{(6)} \rightarrow C_i^{(8)}}.\nonumber
\end{align}
Leading to the $\Delta \delta^2$ interference term
\bea
\frac{\Delta \delta^2 \Gamma(h \rightarrow \mathcal{A} \mathcal{A})}{\Delta^2 \Gamma_{SM}(h \rightarrow \mathcal{A} \mathcal{A})}
\simeq \frac{- 16 \pi^2}{\hat \alpha_{ew} \, |A_{\mathcal{A} \mathcal{A}}|^2} {\rm Re} A_{\mathcal{A} \mathcal{A}} \left[
\langle \sqrt{h}^{44}\rangle_{{\cal{O}}(v^2/\Lambda^2)} \, A^{\mathcal{L}^{(6)}}_{\mathcal{A} \mathcal{A}}
+ 2 (A^{\mathcal{L}^{(6)}}_{\mathcal{A} \mathcal{A}})^2 + 2 A^{\mathcal{L}^{(6)}}_{\mathcal{A} \mathcal{A}}|_{C_i^{(6)} \rightarrow C_i^{(8)}}\right]. \nn
\eea
Here we have used the short-hand notation
\begin{align}
	\langle \sqrt{h}^{44}\rangle_{{\cal{O}}(v^2/\Lambda^2)}
	&= \tilde C_{H\Box}^{(6)} - \frac{1}{4}\tilde C_{HD}^{(6)}, & \quad C_{HB}^{(6)} &\rightarrow \frac{1}{2} C_{HB}^{(8)},  \\
	C_{HW}^{(6)} &\rightarrow \frac{1}{2} \left(  C_{HW}^{(8)}+C_{HW,2}^{(8)}\right), &  \quad
	C_{HWB}^{(6)} &\rightarrow \frac{1}{2} C_{HWB}^{(8)}.
\end{align}
\section{Past literature results}
We have included a significant set of numerical and analytic detail in this work to aid reproducibility.
However, a complete reproduction of the numerical results requires the following additional literature expressions:
\begin{itemize}
\item{The two loop QCD corrections for $C_1^H(\tau_p)$ is lengthy and directly given in Ref.~\cite{Harlander:2005rq}.
Specifically Eqn.~2.8 in this work.}
\item{The explicit expression for $\delta G_F^{(8)}$ can be derived from Appendix C, Eqn.~C.12 in Ref.~\cite{Hays:2020scx}.}
\item{The explicit expression for $C_{H,kin}^{(8)}$ can be derived from Eqn.~3.10 in Ref.~\cite{Helset:2020yio}.}
\item{The one loop corrections to the $W,Z$ masses in the BFM reported in Appendix A, Eqns.~A.1, A.2 in Ref.~\cite{Corbett:2021cil}.}
\end{itemize}

\bibliographystyle{JHEP}
\bibliography{bibliography.bib}

\end{document}